\documentclass[prb,eqsecnum,twocolumn,showpacs,superscriptaddress]{revtex4-1}

\usepackage{amsmath,amssymb,bm}
\usepackage[driverfallback=dvipdfm]{hyperref}
\usepackage{xcolor}
\usepackage{graphicx}

\pdfoutput=1

\begin{document}

\title{
Nonlocal Andreev transport through a quantum dot in a magnetic field:\\
Interplay between 
Kondo, Zeeman, and Cooper-pair correlations
}

\author{Masashi Hashimoto}
\affiliation{
Department of Physics, Osaka City University, Sumiyoshi-ku, 
Osaka, 558-8585, Japan}

\author{Yasuhiro Yamada}
\affiliation{
NTT Basic Research Laboratories, NTT Corporation, Atsugi, Kanagawa, 243-0198, Japan}

\author{Yoichi Tanaka}
\affiliation{
Advanced Simulation Technology  
Of 
Mechanics R$\&$D Co., Ltd., Bunkyo-ku, Tokyo, 112-0002, Japan.}

\author{Yoshimichi Teratani}
\affiliation{
Department of Physics, Osaka City University, Sumiyoshi-ku, 
Osaka, 558-8585, Japan}
\affiliation{ 
NITEP, Osaka Metropolitan University, Sumiyoshi-ku, Osaka, 558-8585, Japan
}

\author{Takuro Kemi}
\affiliation{
Department of Physics, Osaka City University, Sumiyoshi-ku, 
Osaka, 558-8585, Japan}

\author{Norio Kawakami}
\affiliation{
Department of Physics, Ritsumeikan University, Kusatsu, Shiga, 525-8577, Japan}
\affiliation{
Department of Materials Engineering Science, Osaka University, Toyonaka, Osaka, 
560-8531, Japan}


\author{Akira Oguri}
\affiliation{
Department of Physics, Osaka City University, Sumiyoshi-ku, 
Osaka, 558-8585, Japan}
\affiliation{ 
NITEP, Osaka Metropolitan University, Sumiyoshi-ku, Osaka, 558-8585, Japan
}

\date{\today}

\begin{abstract}
We study the nonlocal 
magnetotransport 
through a strongly correlated quantum dot, 
connected to multiple terminals consisting of two normal and 
one superconducting (SC) leads.
Specifically, we present a comprehensive view on 
the interplay between the crossed Andreev reflection (CAR), 
the Kondo effect, and the Zeeman splitting 
 at zero temperature in the large SC gap limit.  
The ground state of this network shows an interesting variety,  
which varies continuously with the system parameters, such as 
the coupling strength $\Gamma_S^{}$ between the SC lead and the quantum dot, 
the Coulomb repulsion $U$, the impurity level $\varepsilon_d^{}$, 
and the magnetic field $b$. 
We show, using the many-body optical theorem 
which is derived from the Fermi-liquid theory,  
 that the nonlocal conductance is determined by the transmission rate of the Cooper pairs 
$\mathcal{T}_{\mathrm{CP}}^{}  =  
\frac{1}{4} \sin^2 \Theta\, 
\sin^2 \bigl(\delta_{\uparrow}+ \delta_{\downarrow})$ 
and that of the Bogoliubov particles  
$\mathcal{T}_{\mathrm{BG}}^{}
= \frac{1}{2}\sum_{\sigma} \sin^2 \delta_{\sigma}^{}$.  
Here, $\delta_\sigma^{}$ is 
the phase shift  of  the renormalized Bogoliubov particles, 
and 
 $\Theta \equiv \cot^{-1} (\xi_d^{}/ \Gamma_S^{})$ 
is the Bogoliubov-rotation angle 
 in the Nambu pseudo spin space,  
with $\xi_d^{} =\varepsilon_d^{}+U/2$.
It is also demonstrated, using Wilson's numerical renormalization group approach,    
that the CAR is enhanced
in the crossover region between  
the Kondo regime and the SC-proximity-dominated regime at zero magnetic field.
The magnetic fields induce another crossover
between the Zeeman-dominated regime and the SC-dominated regime, 
which occurs when the renormalized Andreev resonance
 level of majority spin crosses the Fermi level. 
We find that the CAR is enhanced and becomes less sensitive 
to magnetic fields in the SC-dominated regime 
 close to the crossover region spreading 
over the angular range of $\pi/4 \lesssim \Theta \lesssim 3\pi/4$.  
 At the level crossing point, 
a spin-polarized current flows 
between the two normal leads,  
and it is significantly enhanced in the directions of 
 $\Theta \simeq 0$ and $\Theta \simeq \pi$ 
where the SC proximity effect  is suppressed.   

\end{abstract}

\maketitle

\section{Introduction}

Quantum dots (QD) connected to multi terminal networks 
consisting of normal and superconducting (SC) leads  
is one of the active fields of current research.
In such networks, the quantum coherence and entanglements 
can be probed through the Andreev reflection
\cite{Hofstetter_2009,Schindele_2012,Das_2012,PhysRevB.89.045422,PhysRevB.90.235412,PhysRevLett.114.096602,Borzenets2016,Tan2020,Golubev_2007,Eldridge_2010,PhysRevB.98.241414,Wrzeinfmmode_2017,PhysRevB.85.035419,PhysRevLett.120.087701,PhysRevB.99.075429,PhysRevB.99.045120,PhysRevB.101.205422,HLee2013,Bordoloi2022}
and Josephson effect.
\cite{PhysRevB.62.13569,PhysRevLett.106.037002,Deacon2015,PhysRevB.95.205437,PhysRevB.92.155437,Ueda2019}

In particular, 
the crossed Andreev reflection (CAR) is 
one of the most interesting processes 
caused by a Cooper-pair tunneling
in which an incident electron entering from a normal lead forms 
a Cooper pair with another electron from 
the other normal leads to tunnel into the SC leads, 
leaving a hole in the normal lead where the second electron came from.
The time-reversal process of the CAR 
corresponds to  
a splitting of a Cooper pair that is emitted from the SC lead    
 into two entangled electrons penetrating the different normal leads.
The CAR and  the Cooper-pair splitting have also been studied 
in the multi-terminal systems  
 without quantum dots.
\cite{PhysRevLett.97.237003,refId0,PhysRevLett.95.027002,Yeyati_2007,Zimansky_2009,Wei_2010,PhysRevB.99.144504,PhysRevB.99.115127,Ranni_2021}

Quantum dots give a variety 
to the transport properties of multi-terminal systems, 
through the tunable parameters such as electron correlations, 
 resonant-level positions, 
and local magnetic fields which can polarize the spins of electrons. 
The strong electron correlations induce 
an interesting  
crossover between the Kondo singlet and the Cooper-pair singlet. 
\cite{Yoshioka_2000,YoichTanaka_2007,Buizert_2007,Deacon_2010,Deacon_2010Rapid,Yamada_2011,Governale_2008,Oguri_2013,PhysRevB.87.115409,Domanski_2017,Vecino_2003,Oguri_2004, YoshihideTanaka_2007, PhysRevLett.129.207701,Zitko2013,Zonda2017,Zonda2023}
Furthermore,  the  magnetic field  induces 
a crossover occurring between the Kondo singlet state and 
the spin-polarized state due to the Zeeman splitting of 
discrete energy levels of quantum dots,    
which has recently been revisited to find that  
the three-body Fermi-liquid corrections play 
an essential role  in the crossover region.
\cite{PhysRevB.98.075404,ao2017_3_PRB}

The CAR contributions can be probed through 
the nonlocal conductance for the current 
flowing from the QD towards one of the normal drain electrode  
when the bias voltage is applied to the source electrode.
\cite{Hofstetter_2009, PhysRevB.89.045422, PhysRevB.90.235412, PhysRevLett.114.096602}
However, 
the nonlocal current also includes the contributions  
of the single-electron-tunneling process,  
 in which an incident electron from the source electrode 
 transmits directly towards the drain electrode through the QD. 
In order to observe the CAR contributions, 
it is important to find some sweet spots in the parameter space, 
at which the superconducting proximity effect dominates the 
nonlocal current  and enhances the Cooper-pair tunneling   
by reconciling it with the other effects 
from electron correlations and magnetic fields.

The CAR in a single correlated quantum dot has theoretically 
been studied over a decade, particularly for 
 a three-terminal QD connected to two normal and one superconducting leads. 
In the early stage,   
Futterer {\it et al}.\cite{Futterer2009} and  Micha\l ek {\it et al}.\cite{Michalek_2013,Michalek_2015}
 demonstrated some behaviors of  the nonlocal transport conductance 
typical to this three-terminal configuration,\cite{Michalek_2015} 
 taking also into account the Coulomb interaction 
with a generalized master equation\cite{Futterer2009}
 or the equation-of-motion method.\cite{Michalek_2013}
It has been extended to the configuration 
in which the normal leads are replaced by ferromagnetic metals
and has been investigated intensively, using also the methods 
such as the real-time diagrammatic method and  
the numerical renormalization group (NRG).
\cite{Futterer2009,Trocha_2018,Weymann2014,Wojcik2014,Weymann2015}

Effects of the Zeeman splitting 
induced by the external magnetic field applied to quantum dots 
have also been theoretically investigated,  
 mainly for two-terminal systems in which a quantum dot is connected 
to a single paramagnetic normal and a SC lead so far.\cite{PhysRevB.64.094505,Gorski2020,YAMADA2007265,PhysRevB.78.144515,Domanski_2015,PhysRevB.91.045441}  
Specifically, 
these theories addressed such subjects as
the field dependence of the Andreev transport,\cite{PhysRevB.64.094505,Gorski2020}
the role of the Coulomb interaction in this configuration, 
\cite{YAMADA2007265,PhysRevB.78.144515,Domanski_2015} 
 and the quantum phase transition between the spin-singlet 
and -doublet ground states.\cite{PhysRevB.91.045441}
However, it is still not fully clarified how 
the CAR contributions evolve at low energies  
in a wide parameter space of the multi-terminal networks, 
with and without magnetic fields.

\begin{figure}
\includegraphics[width=\linewidth]{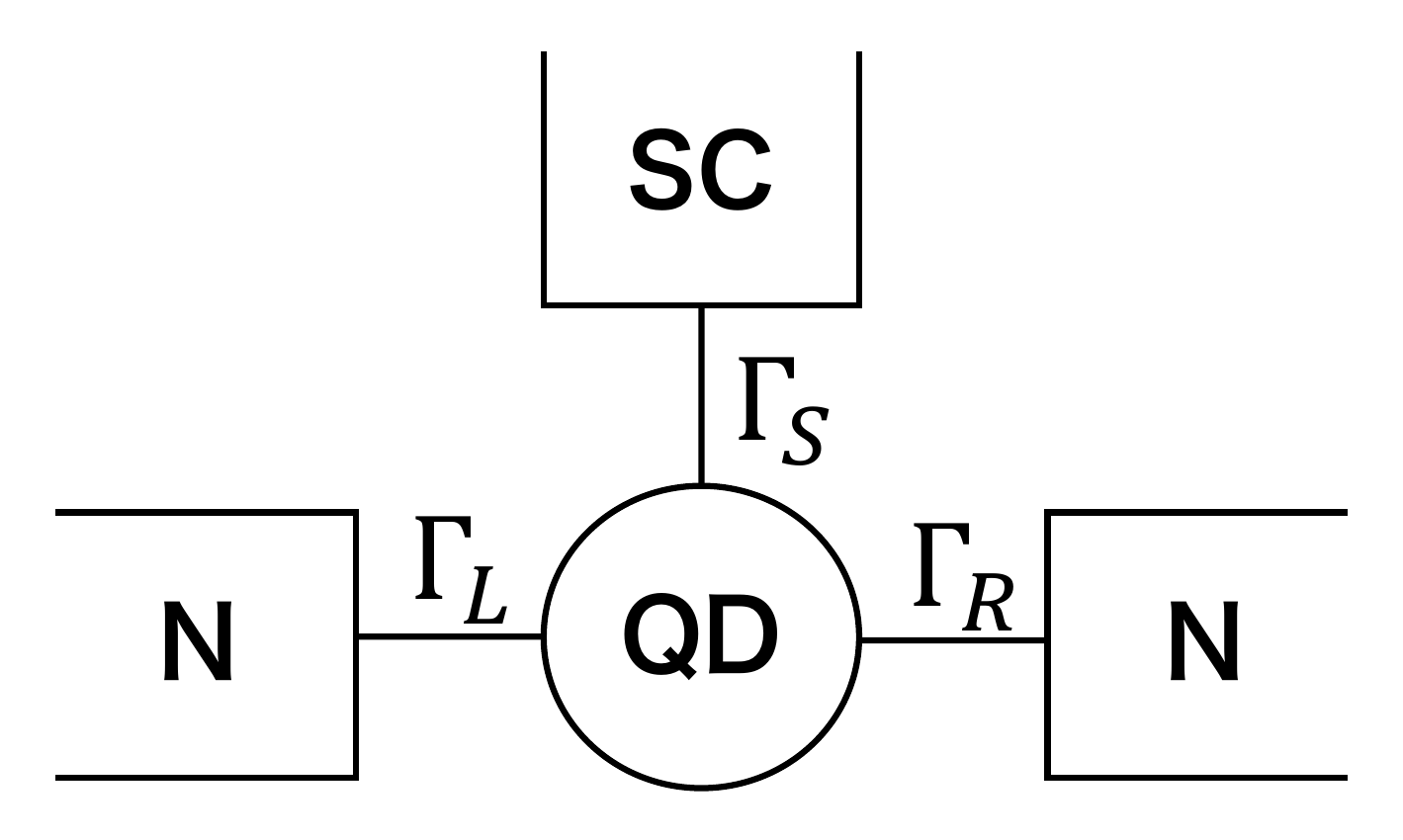}
\caption{Single quantum dot (QD) connected to two normal leads (N) 
and one superconductor lead (SC). 
$\Gamma_L^{}$, $\Gamma_R^{}$, and $\Gamma_S^{}$ 
represent the coupling strengths 
of the QD with the left ($L$),  the right ($R$), and the SC leads, 
 respectively.  
The contributions of the normal tunnelings 
are given by $\Gamma_N^{}=\Gamma_L^{}+\Gamma_R^{}$. 
}
\label{fig:NRGOneDot}
\end{figure}

\begin{figure}
\includegraphics[width=\linewidth]{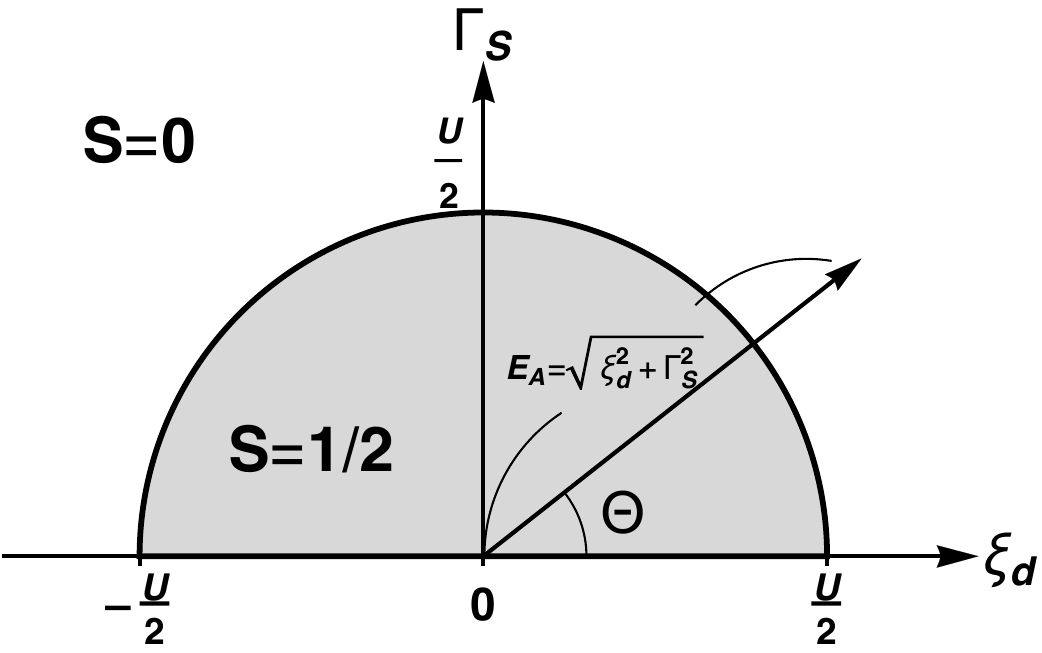}
\caption{Parameter space of $H_\mathrm{eff}^{}$ 
at zero magnetic field $b=0$,  
defined in Eqs.\ \eqref{eq:Heff_single} and \eqref{eq:SingleDot_new}. 
The semicircle represents the line along which 
the energy of the 
Andreev level  
 $E_A^{} \equiv \sqrt{\xi_d^2+\Gamma_S^2}$ 
coincides with one-half of the Coulomb interaction $U/2$,  
where $\xi_d^{} \equiv \varepsilon_d^{} +U/2$.   
In the atomic limit  $\Gamma_N ^{} =0$, 
the ground state is a magnetic spin doublet  inside 
the semicircle, which eventually 
is screened by 
conduction electrons to form the Kondo singlet 
when the tunnel coupling $\Gamma_N^{}$ is switched on,  
whereas the ground state 
is a spin singlet due to the Cooper paring outside the circle. 
 $\Theta$ is the Bogoliubov-rotation angle,  
which parametrizes the contributions of the Andreev scattering 
on the transport coefficients.
}
\label{fig:SingleDotPhase}
\end{figure}

The purpose of this paper is to provide a comprehensive view of the Andreev transport 
through a strongly-correlated quantum state, 
the characteristics of which vary 
due to the interplay between the Kondo, Zeeman, and Cooper-pair correlations.
To this end,  we calculate the transport coefficients,  
using the Fermi-liquid theory
\cite{NozieresFermiLiquid,YamadaYosida2,YamadaYosida4,ShibaKorringa,Yoshimori}   
in conjunction with Wilson's numerical renormalization group (NRG).
Specifically, 
we consider 
 a three-terminal quantum dot 
connected to two normal and one superconducting leads,
 as illustrated in Fig.\ \ref{fig:NRGOneDot},  
in the large SC gap limit.\cite{YoichTanaka_2007}
We first of all  derive the optical theorem for the CAR at zero temperature, 
using the Fermi-liquid theory that describes the interacting Bogoliubov particles 
moving throughout the entire system.
It elucidates the fact that 
the nonlocal conductance is determined by the transmission rate of Cooper pairs 
$\mathcal{T}_{\mathrm{CP}}^{}  =  
\frac{1}{4} \sin^2 \Theta\, 
\sin^2 \bigl(\delta_{\uparrow}+ \delta_{\downarrow})$ 
and that of the Bogoliubov particles  
$\mathcal{T}_{\mathrm{BG}}^{}
= \frac{1}{2}\sum_{\sigma} \sin^2 \delta_{\sigma}^{}$.  
Here,  $\Theta \equiv \cot^{-1} (\xi_d^{} /\Gamma_S^{})$ 
is the angular coordinate 
in the  $\xi_d^{} \equiv \varepsilon_d^{} +U/2$ vs $\Gamma_S^{}$ plane   
shown in Fig.\ \ref{fig:SingleDotPhase},       
and  $\varepsilon_d^{}$ and  $U$ are
the discrete level and  the Coulomb interaction of electrons in the QD, respectively, 
and $\Gamma_S^{}$ is 
the coupling strength   
between the QD and SC lead. 
In this case, 
the phase shift $\delta_{\sigma}^{}$ of  
the interacting Bogoliubov particles does not depend
on the angle $\Theta$ but varies along the radial coordinate 
 $E_A^{} = \sqrt{\xi_d^2 + \Gamma_S^2}$.

We also calculate the transmission rate 
$\mathcal{T}_{\mathrm{CP}}^{}$ and 
$\mathcal{T}_{\mathrm{BG}}^{}$ with the NRG 
in a wide range of the parameter space.
It is demonstrated that, at zero magnetic field, 
 the CAR contributions are significantly enhanced  
near the crossover region between the Kondo regime and 
the SC-proximity-dominated regime.  
Specifically, it takes place in a crescent-shaped region 
spreading over the range of  
  $ U/2 \lesssim E_A^{} \lesssim U/2 + \Gamma_N^{}$ in the radial direction  
and $\pi/4 \lesssim \Theta \lesssim 3\pi/4$ in  the angular direction: 
 $\Gamma_N^{}$ is the resonance width 
 due to the tunneling between the QD and normal leads. 
When a magnetic field is applied,
another crossover occurs between the Zeeman-dominated regime 
and the SC-proximity-dominated regime 
when the spin-polarized Andreev level crosses the Fermi level.  
We find that the CAR-dominated transport taking place in the crescent region   
is less sensitive to magnetic fields,     
and it emerges as a flat valley structure in the magnetic-field dependence    
 of the nonlocal conductance. 
This  parameter region provides an optimal condition 
 for observing  the Cooper-pair tunneling, i.e., a sweet spot,   
especially in the direction of  $\Theta \simeq \pi/2$  
where the Cooper pairs are most entangled 
and become equal-weight linear combinations of an electron and a hole.

This paper is organized as follows.
In Sec.\ \ref{Formulation}, 
we introduce  
an Anderson impurity model for quantum dots 
 connected to SC and normal leads,    
and rewrite the Hamiltonian and the Green's function 
in terms of interacting Bogoliubov particles.   
Then, the optical theorem and the formula for the nonlocal conductance 
are derived at zero temperature using the Fermi-liquid description 
for the interacting Bogoliubov particles 
 in Sec.\ \ref{sec:multi_terminal_formulation}.   
We investigate the CAR contributions to the nonlocal conductance, 
using the NRG,  at zero and finite magnetic fields in Secs.\ \ref{CARZeroMag} 
 and \ref{CARMag}, respectively.  
Summary and discussion are  given in Sec.\ \ref{summary}.

 \section{Fermi-liquid description for interacting Bogoliubov particles}

\label{Formulation}

In this section, we show how the contributions of the CAR 
to the nonlocal conductance of the multi-terminal network 
can be described in the context of 
the Fermi-liquid theory  for the interacting Bogoliubov particles 
at zero temperature.\cite{YoichTanaka_2007}

\subsection{Anderson impurity model for the CAR}

We start with an Anderson impurity model 
for a single quantum dot (QD) connected to
two normal (N) and one superconducting (SC) leads,
 as shown in Fig.\ \ref{fig:NRGOneDot}:  
\begin{align}
    H \,  = & \ \, 
H_\mathrm{dot}^{}
\,+\, H_{\text{N}}^{}  \,+\,H_{\text{TN}}^{} 
\,+\,H_{\text{S}}^{} +H_{\text{TS}}^{}\,,
\label{eq:total_H_single}
\\
H_\mathrm{dot}^{}
 = &  \  
\xi_d^{}  \bigl( n_d^{} -1 \bigr) 
- b \, \bigl(
n_{d,\uparrow}^{} - n_{d,\downarrow}^{} 
\bigr) 
 \,+\frac{U}{2}\bigl( n_d^{} -1 \bigr)^2 ,
\label{eq:H_dot}
\\
  H_{\text{N}}^{} =& \ \sum_{\nu=L,R}\sum_{\sigma} 
\int_{-D}^{D}  \! d\varepsilon \,\varepsilon\,
 c^\dagger_{\varepsilon,\nu,\sigma}c^{ }_{\varepsilon,\nu,\sigma},
\label{eq:H_N}
\\
    H_{\text{TN}}^{} = & \ \sum_{\nu=L,R} 
v_{\nu}^{}
\sum_{\sigma} 
\int_{-D}^{D}  \! d\varepsilon \,\sqrt{\rho_c^{}}\,
\Bigl( c^\dagger_{\varepsilon,\nu,\sigma}d^{}_\sigma 
+ \mathrm{H.c.}
\Bigr),
\label{eq:H_TN}
\\
H_{\text{S}} = & \  
\sum_{\sigma} \!
\int_{-D_S^{}}^{D_S^{}}  \!\! d\varepsilon \,\varepsilon\,
s^\dagger_{\varepsilon,\sigma} s^{}_{\varepsilon,\sigma} 
\nonumber \\
 & \quad \   
  + 
\int_{-D_S^{}}^{D_S^{}}  \!\! d\varepsilon 
\left( \Delta_S^{} \,s^\dagger_{\varepsilon,\uparrow} 
s^\dagger_{\varepsilon,\downarrow} 
+  \mathrm{H.c.} \right),
\\
H_{\text{TS}} = &  \ 
 v_{\text{S}}^{}
 \sum_{\sigma} \! 
 \int_{-D_S^{}}^{D_S^{}}  \!\!  d\varepsilon \,\sqrt{\rho_S^{}}\, 
 \Bigl( s^\dagger_{\varepsilon,\sigma}d^{}_\sigma 
 + \mathrm{H.c.}
 \Bigr).
\end{align}
Here, 
$H_\mathrm{dot}^{}$ describes the QD part:
$\xi_d^{} \equiv \varepsilon_d^{} +{U}/{2}$,  
with $\varepsilon_d^{}$ 
the discrete energy level,   $U$  the Coulomb interaction, 
and  $b$ ($\equiv \mu_B^{} B)$  the Zeeman energy due to 
 the magnetic field $B$ applied to the QD, 
with $\mu_B^{}$ the Bohr magneton. 
 $d^\dagger_\sigma$ is 
the creation operator for an electron with spin $\sigma$, 
and  $n_d^{} \equiv n_{d,\uparrow}^{}+n_{d,\downarrow}^{}$ 
is the number operator with 
$n_{d,\sigma}^{}\equiv d^\dagger_\sigma d^{ }_\sigma$. 
A constant energy shift, which does not affect the physics,  
is included in Eq.\ \eqref{eq:H_dot} 
 in order to describe clearly that the system 
has the electron-hole symmetry at $\xi_d^{}=0$.

  $H_{\text{N}}$ describes 
the conduction electrons in the normal leads, 
the density of states of which is assumed to be flat 
  $\rho_c^{}=1/(2D)$, with $D$ 
the half-width of the bands. 
 $c^\dagger_{\varepsilon,\nu,\sigma}$ is 
the creation operator 
for conduction electrons 
with spin $\sigma$ and energy $\varepsilon$. 
The operators for conduction electrons satisfy the 
 following anti-commutation relation that is 
 normalized by the Dirac delta function:  
$
\{ c^{}_{\varepsilon,\nu, \sigma}, 
c^{\dagger}_{\varepsilon', \nu',\sigma'}
\} = \delta_{\nu\nu'} \,\delta_{\sigma\sigma'}   
\delta(\varepsilon-\varepsilon')$.  
$H_{\text{TN}}^{}$ describes 
the tunnel couplings between the QD and the normal leads.
The level broadening of the discrete energy level 
 in the QD is given by $\Gamma_N \equiv \Gamma_L + \Gamma_R$, 
with $\Gamma_{\nu} \equiv \pi \rho_c^{} v_{\nu}^2$  
the contributions of the two normal leads  
on the left $\nu =L$ and right $\nu=R$.

 $H_{\text{S}}^{}$ and $H_{\text{TS}}^{}$ 
describe the contributions of the superconducting lead 
with an $s$-wave SC gap  
 $\Delta_{S}^{} \equiv \left| \Delta_{S}^{}\right|\,e^{i\phi_S^{}}$:  
$s^\dagger_{\varepsilon,\sigma}$ is 
the creation operator for electrons in the SC lead, 
with $D_S^{}$ the half-band width and $\rho_S^{} =1/(2D_S^{})$.
One of the key parameters for the SC proximity effects 
is $\Gamma_S \equiv \pi \rho_S^{} v_{\text{S}}^2$,   
i.e., the coupling strength between the QD and the SC lead.

In this paper, 
we study the crossed Andreev reflection occurring at low energies,
 much lower than 
the SC energy gap.
To this end, we consider the large gap limit  $\left| \Delta_{S}^{}\right| \to \infty$, 
which is taken at $\left| \Delta_{S}^{}\right|  \ll D_S^{}$ 
keeping $\Gamma_S^{}$ constant.\cite{YoichTanaka_2007}
In this case, the superconducting proximity effects can be described 
by the pair potential  penetrating into the QD:
\begin{align}
\Delta_d^{}  \equiv   \Gamma_{S}^{}\,e^{i\phi_S^{}} \,.
\end{align}

The Coulomb interaction $U$ induces the correlation effects 
for electrons in the QD and the symmetrized linear combination of 
the conduction bands defined in Eq.\  \eqref{eq:even_lead}, 
which can be described by an effective Hamiltonian $H_\mathrm{eff}^{}$  
given in Eq.\ \eqref{eq:Heff_single} (see  Appendix \ref{AppendixBogo}).
Furthermore, carrying out the Bogoliubov rotation   
defined in Eq.\ \eqref{eq:Bogoliubov_tans_single}, 
it can be transformed further into a system of interacting Bogoliubov particles 
described by a standard Anderson model:   
\begin{align}
H_\mathrm{eff}^{} 
=& \   E^{}_A
\left(\sum_\sigma \gamma^\dagger_{d,\sigma}\gamma^{}_{d,\sigma}-1\right)
- b\, \Bigl(
\gamma^\dagger_{d,\uparrow}\gamma^{}_{d,\uparrow}
-\gamma^\dagger_{d,\downarrow}\gamma^{}_{d,\downarrow}
\Bigr)
\nonumber \\
& + \frac{U}{2}\left(\sum_\sigma \gamma^\dagger_{d,\sigma}
\gamma^{}_{d,\sigma} - 1 \right)^2 
 + 
\sum_{\sigma} 
\int_{-D}^{D}  \! d\varepsilon \,\varepsilon\  
 \gamma^\dagger_{\varepsilon,\sigma}
 \gamma^{ }_{\varepsilon,\sigma} 
\nonumber \\
&
+ v_N^{}
\sum_{\sigma} 
\int_{-D}^{D}  \! d\varepsilon \,
\sqrt{\rho_c^{}}\,
\left(
\gamma^\dagger_{\varepsilon,\sigma}
\gamma^{ }_{d,\sigma}
+ \mathrm{H.c.} 
\right)\, , 
\label{eq:SingleDot_new}
\\
N_{\gamma}^{} 
=& \ 
\sum_\sigma  
\gamma^{\dagger}_{d,\sigma}\gamma^{}_{d,\sigma} 
+ 
\sum_\sigma  
\int_{-D}^{D}  \! d\varepsilon \,
\gamma^{\dagger}_{\varepsilon,\sigma}\gamma^{}_{\varepsilon,\sigma} 
\,.  
\label{eq:Bogolon_total_charge}
\end{align}
Here,   $E_A \equiv \sqrt{\xi_d^2 + \Gamma_S^2}$ is 
the effective impurity level, and 
 $v_N^{} \equiv \sqrt{v_L^2+v_R^2}$.   
The operators 
 $\gamma^{ }_{d,\sigma}$ 
and  $\gamma^{ }_{\varepsilon,\sigma}$ 
describe the Bogoliubov particles in the dot and 
the symmetrized part of the conduction band, respectively. 
The effective Hamiltonian conserves 
the total number of the Bogoliubov particles 
$N_{\gamma}^{}$, 
reflecting the $U(1)$ symmetry along the principal axis 
in the Nambu pseudo-spin space.

Figure \ref{fig:SingleDotPhase} 
illustrates the parameter space  of 
  $H_\mathrm{eff}^{}$  
at zero magnetic field  $b=0$. 
For finite $\Gamma_N^{}$, 
the Kondo screening due to the normal conduction electrons occurs 
inside the semicircle region,  
at which  the impurity level is occupied 
by a single Bogoliubov particle: 
   $Q\simeq 1.0$ with  
 \begin{align}
  Q \, & \equiv \, Q_\uparrow + Q_\downarrow , 
\qquad 
   Q_\sigma \,  \equiv \,  
\left \langle 
\gamma^\dagger_{d,\sigma}\gamma^{}_{d,\sigma}
\right\rangle .
\label{eq:def_Q}
 \end{align} 
Bogoliubov particles 
show also the valence-fluctuation behavior near $E_A^{} \simeq U/2$,
at which the crossover between the Kondo singlet and 
 the superconducting singlet occurs. 
The Bogoliubov rotation angle corresponds to   
 $\Theta =  \cot^{-1}(\xi_d^{}/\Gamma_S^{})$  
shown in Fig.\ \ref{fig:SingleDotPhase}.
In particular, 
the crossed Andreev scattering  is enhanced 
in the angular range of $\pi/4  \lesssim \Theta \lesssim 3\pi/4$ 
outside the semicircle  $E_A^{} \gtrsim U/2$, 
as discussed later in Secs.\ \ref{CARZeroMag}  and \ref{CARMag}.

\subsection{Renormalized Bogoliubov quasiparticles}

\label{subsec:def_renormalized_parameters}

In this work, 
we calculate the nonlocal conductance 
for the current flowing into the drain electrode,
using the retarded Green's function for electrons in the QD:  
\begin{align}
 &   
\!\!\!
\bm{G}^r_{dd} (\omega) 
\equiv  -i\int_0^{\infty} \! dt \, e^{i\left(\omega + i0^+ \right)t} 
\nonumber \\
 & \qquad \qquad \ \ 
\times 
 \begin{pmatrix}
\left\langle\left\{d^{}_{\uparrow}(t),\,d^\dagger_{\uparrow} \right\} 
\right\rangle & 
\left\langle\left\{d^{}_{\uparrow}(t),\,d^{}_{\downarrow} \right\} 
\right\rangle\\
\left\langle\left\{d^\dagger_{\downarrow}(t),\,d^\dagger_{\uparrow} \right\} 
\right\rangle & 
\left\langle\left\{d^\dagger_{\downarrow}(t),\,d^{}_{\downarrow} \right\} 
\right\rangle
\rule{0cm}{0.6cm}
\end{pmatrix}.
\label{eq:impurity_Nambu_Green's_function}
\end{align}
Here, $\langle \cdots \rangle$ denotes the thermal average 
at equilibrium. 
This matrix Green's function can be 
diagonalized with 
the Bogoliubov transformation:  
\begin{align}
\bm{\mathcal{U}}^\dagger\,  
\bm{G}^r_{dd}(\omega)\ 
\bm{\mathcal{U}}^{}  
\, &= \,
 \begin{pmatrix}
G^r_{\gamma,\uparrow}(\omega) & 0 \\
0 & - G^a_{\gamma,\downarrow}(-\omega)
\rule{0cm}{0.4cm}
\end{pmatrix} .
\label{eq:GreenBogoliubovTrans}
\end{align}
We will choose the Josephson phase of the pair potential  
to be $\phi_S^{} =0$ in the following, 
so that 
$\bm{\mathcal{U}}^{}$ is 
determined solely by a pseudo-spinor rotation 
with the angle $\Theta/2$: 
\begin{align}
\bm{\mathcal{U}}
\, &=\, 
\begin{pmatrix} 
\cos \frac{\Theta}{2} & 
\, -
\sin \frac{\Theta}{2} \\
\sin \frac{\Theta}{2} & 
\rule{0cm}{0.5cm} 
\quad   
\cos \frac{\Theta}{2}
\end{pmatrix}
.
\label{eq:Bofoliubov_trans_I}
\end{align}
The matrix elements of $\bm{\mathcal{U}}$ determine 
the behaviors of transport coefficients as  
the superconducting coherence factors,  
\begin{align}
& \cos \frac{\Theta}{2} =  
\sqrt{\frac{1}{2}\left(1+\frac{\xi_{d}}{E^{}_A}\right)} , 
\quad 
\sin \frac{\Theta}{2} =  
\sqrt{\frac{1}{2}\left(1-\frac{\xi_{d}}{E^{}_A}\right)} .
\nonumber 
\rule{0cm}{1cm}
\end{align}
The diagonal elements 
$G^r_{\gamma,\sigma}$ and  $G^a_{\gamma,\sigma}$ 
on the right-hand side of Eq.\ \eqref{eq:GreenBogoliubovTrans} are 
the retarded and advanced Green's functions 
for the interacting Bogoliubov particles, 
described by  $H_\mathrm{eff}^{}$.   
These diagonal elements can be expressed in the form, 
using  Eq.\ \eqref{eq:Bogoliubov_tans_single}, 
 \begin{align}
 G^r_{\gamma,\sigma}(\omega)
\equiv& \,   -i\int_0^{\infty} \!  dt \, e^{i\left(\omega + i0^+\right)t} 
\left\langle\left\{\gamma^{}_{d,\sigma}(t),\, 
\gamma^\dagger_{d,\sigma} \right\} \right\rangle  
\nonumber 
\\
=&  \ \frac{1}{\omega 
\,- E_{A,\sigma}^{}  
-\Sigma^U_{\gamma,\sigma}(\omega) +i\Gamma_{N}}, 
 \end{align}
and   $ G^a_{\gamma,\sigma}(\omega) 
= \left\{G^r_{\gamma,\sigma}(\omega)\right\}^*$.
Here, 
$E_{A,\sigma}^{} \equiv E_A^{}-\sigma\, b$, and 
 $\Sigma^U_{\gamma,\sigma}(\omega)$ represents 
the self-energy corrections due to the Coulomb  interaction term, 
$({U}/{2})\left(n_d -1 \right)^2$,  
defined in Eq.\ \eqref{eq:H_dot}.
The unperturbed part of the denominator describes 
the Andreev resonance level 
with 
 the width $\Gamma_{N}$  
 situated at $\omega=E_{A,\sigma}^{}$.

At low energies, 
effects of the electron correlations on the transport properties  
can be deduced from 
the behavior of  the self-energy 
near $\omega \simeq 0$ at zero temperature $T=0$:
 \begin{align}
 G^r_{\gamma,\sigma}(\omega)  
\,\simeq \, \frac{Z_\sigma}{\omega -\widetilde{E}_{A,\sigma}^{} 
+ i\widetilde{\Gamma}_{N,\sigma} } \, .
 \end{align}
The asymptotic form of the Green's function defines 
 a renormalized resonance level of quasiparticles in the Fermi liquid,   
the position $\widetilde{E}_{A,\sigma}^{}$ 
and  the width  $\widetilde{\Gamma}_{N,\sigma}$ 
of which are given 
by\cite{NozieresFermiLiquid,YamadaYosida2,YamadaYosida4,ShibaKorringa,Yoshimori} 
\begin{align}
\widetilde{\Gamma}_{N,\sigma} = & \ Z_{\sigma}\,\Gamma_N,  
\qquad 
 \frac{1}{Z_{\sigma}}\,=\,  
1-\left. \frac{\partial \Sigma^U_{\gamma,\sigma}(\omega)}{\partial \omega} 
\right|_{\omega=0}, 
\label{eq:z_def}
\\
\widetilde{E}_{A,\sigma}^{} \,=& \ Z_{\sigma}
\Bigl[\, 
E_{A,\sigma}^{}
 +\Sigma^U_{\gamma,\sigma}(0) \, \Bigr] 
\,.
\label{eq:EAren}
\end{align}
Furthermore, 
the phase shift $\delta_\sigma$ of the interacting Bogoliubov particles is 
defined 
by $G^r_{\gamma,\sigma}(0) 
= -\left| G^r_{\gamma,\sigma}(0)\right| e ^{i \delta_{\sigma}}$, 
i.e.,  
 \begin{align}
\delta_{\sigma} \,=& \  
\frac{\pi}{2}-\tan^{-1}\left(\frac{\widetilde{E}_{A,\sigma}}
{\widetilde{\Gamma}_{N,\sigma}} \right) \,,
\label{eq:phase_shift_def}
 \end{align}
plays a primary role in the ground-state properties.
These renormalized parameters  can be calculated, 
for instance, 
using the NRG approach described in the next section.

The Friedel sum rule also holds for the interacting Bogoliubov particles,  
and thus  the average number of the Bogoliubov particles  
in the QD is determined by the phase shift, 
 \begin{align}
  Q_\sigma & \xrightarrow{\,T \to 0 \,} \frac{\delta_\sigma}{\pi}.
\label{eq:Friedel_Q}
\end{align}
The phase shift 
varies in the range of  $0\leq \delta_{\sigma}^{} \leq \pi/2$  
along the radial coordinate  $E_A^{}$ in the  $\xi_d^{}$ vs $\Gamma_S^{}$ plane 
but is independent of the angle  $\Theta$.

The ground-state properties, such as 
the occupation number of electrons $\left\langle n_{d} \right \rangle$ 
 and the pair correlation function 
$\bigl \langle d^\dagger_{\uparrow}\,d^\dagger_{\downarrow}
 +d_{\downarrow}\,d_{\uparrow} \bigr \rangle$, 
can be deduced from $Q$: 
\begin{align}
   \left\langle n_{d}^{} \right \rangle - 1 
\ =& \, \left(Q-1\right)\,\cos \Theta \, , 
 \label{eq:SingleDotM}
\\
   \left \langle d^\dagger_{\uparrow}\,d^\dagger_{\downarrow}
 +d_{\downarrow}\,d_{\uparrow} \right \rangle
\ =&  \, \left(Q-1\right) \,\sin \Theta \,.
 \label{eq:SingleDotK}
\end{align}
These two averages correspond to the projection of a vector 
of magnitude $Q-1$ directed along the principal axis 
onto the $z$ axis and the $x$ axis of the Nambu space, respectively.
Furthermore, 
a local magnetization $m_d^{}$ is induced in the quantum dot at finite magnetic fields, 
\begin{align}
  m_d^{} \, \equiv \ &  \  \left \langle n_{d,\uparrow}^{} \right \rangle 
- \left \langle n_{d,\downarrow}^{} \right \rangle
\ = \, Q_\uparrow - Q_\downarrow \,.
 \label{eq:SingleDotm}
\end{align}
Therefore, the occupation number of electrons
with spin $\sigma$ is given by 
\begin{align}
 \left \langle n_{d,\sigma}^{} \right \rangle 
\, =\, 
 \frac{1+\cos\Theta}{2} \,\frac{\delta_{\sigma}^{}}{\pi}
\,+\,\frac{1-\cos\Theta}{2} \,
\left(1-
 \frac{\delta_{\overline{\sigma}}^{}}{\pi}
\right) \,,
 \label{eq:nd_in_BG_phase_shift}
\end{align}
where $\overline{\sigma}$ represents 
an opposite-spin component of $\sigma$.

\section{Linear-response theory for CAR}

\label{sec:multi_terminal_formulation}

\subsection{Cooper-pair transmission in a local Fermi liquid}

We consider the linear-response current $I_{R}^{}$ 
 flowing from 
the QD to the normal lead on the right,  
induced by bias voltages $V_L$ and $V_R$ 
 applied to the left and right leads, respectively.  
It  can be expressed in the following form at $T=0$  
(see Appendix \ref{sec:conductance_derivation}):
\begin{align}
I_{R}^{} \, = & \ \, I_{R}^{\mathrm{ET}} \,+\,  I_{R}^{\mathrm{CP}},
\label{eq:current_R_ET_CP}
\\
I_{R}^{\mathrm{ET}}
=  & \      
\frac{2e^2}{h} 
 \mathcal{T}_{\mathrm{ET}}^{} \   
\frac{4\Gamma_R^{}\Gamma_L^{}}{\Gamma_N^{2}}\,
\bigl(V_L-V_R\bigr)  ,  
\rule{0cm}{0.6cm}
\nonumber
\\
I_{R}^{\mathrm{CP}}
= & \       
- \frac{2e^2}{h}   \mathcal{T}_{\mathrm{CP}}^{}
\left[
\frac{4\Gamma_R^{}\Gamma_L^{}}{\Gamma_N^{2}}\, \bigl(V_L+V_R\bigr) 
+\frac{4\Gamma_R^{2}}{\Gamma_N^{2}} \,  2V_R
\right].
\nonumber
\end{align}
Correspondingly, the current  $I_{L}^{}$  
flowing from the left normal lead towards the QD 
takes the form, 
$I_{L}^{} 
 =  I_{L}^{\mathrm{ET}} +  I_{L}^{\mathrm{CP}}$, 
with 
$I_{L}^{\mathrm{ET}} =I_{R}^{\mathrm{ET}}$ 
and 
\begin{align}
I_{L}^{\mathrm{CP}}
= & \       
 \frac{2e^2}{h}   \mathcal{T}_{\mathrm{CP}}^{}
\left[
\frac{4\Gamma_R^{}\Gamma_L^{}}{\Gamma_N^{2}}\, \bigl(V_L+V_R\bigr) 
+\frac{4\Gamma_L^{2}}{\Gamma_N^{2}} \,  2V_L
\right].
 \label{eq:current_L_CP}
\end{align}
The two components of the current 
 $I_{\nu}^{\mathrm{ET}}$ and 
$I_{\nu}^{\mathrm{CP}}$  
represent the contribution of  the single-electron tunneling 
and that of the Cooper-pair tunneling, respectively.
The transmission probabilities  
$\mathcal{T}_{\mathrm{ET}}^{}$ and 
 $\mathcal{T}_{\mathrm{CP}}^{}$ 
are determined by 
the equilibrium Green's functions at the Fermi level $\omega=0$, 
and can be expressed in terms of the phase shifts and the Bogoliubov angle 
(see Appendix \ref{sec:conductance_derivation}):  
\begin{align}
\!\!\!   
\mathcal{T}_{\mathrm{ET}}^{} \equiv & \ 
\frac{\Gamma_N^{2}}{2} 
\left[\ 
\Bigl | \bigl\{\bm{G}_{dd}^{r}(0)\bigl\}_{11}^{}\Bigr|^2 
+\Bigl | \bigl\{\bm{G}_{dd}^{r}(0)\bigl\}_{22}^{}\Bigr|^2
\ \right] 
\nonumber 
\\
 = & \   \frac{1}{2}
\sum_{\sigma} \sin^2 \delta_{\sigma}^{} 
- \frac{1}{4}\, \sin^2 \Theta\, 
\sin^2 \bigl(\delta_{\uparrow}^{}+ \delta_{\downarrow}^{})  
\,, 
 \label{eq:T_ET_single}
\\
\!\!\!   
\mathcal{T}_{\mathrm{CP}}^{}  \equiv & \ 
\frac{\Gamma_N^{2}}{2} 
\left[\  \Bigl | \bigl\{\bm{G}_{dd}^{r}(0)\bigl\}_{12}^{}\Bigr|^2 + \Bigl | \bigl\{\bm{G}_{dd}^{r}(0)\bigl\}_{21}^{}\Bigr|^2 \ \right]
\nonumber \\
  = & \,   
\frac{1}{4}\, \sin^2 \Theta\, 
\sin^2 \bigl(\delta_{\uparrow}+ \delta_{\downarrow}) \,. 
 \label{eq:T_CP_single}
\end{align}
These two are bounded in the range,  
  $0 \leq \mathcal{T}_{\mathrm{ET}}^{} \leq 1$ 
and $0 \leq \mathcal{T}_{\mathrm{CP}}^{} \leq 1/4$, 
and are  
related to each other through the optical theorem 
(see Appendix \ref{sec:optical_theorem}): 
\begin{align}
\mathcal{T}_{\mathrm{ET}}^{} + \mathcal{T}_{\mathrm{CP}}^{} 
\,=\, 
\mathcal{T}_{\mathrm{BG}}^{}\,, 
\qquad 
\mathcal{T}_{\mathrm{BG}}^{}
\,\equiv\, 
\frac{1}{2}\,\sum_{\sigma} \sin^2 \delta_{\sigma}^{}\,.
\label{eq:T_BG_single}
\end{align}
Here,  $\mathcal{T}_{\mathrm{BG}}^{}$ can be regarded as   
a transmission probability of the Bogoliubov particles.

The linear-response coefficients, 
given in the above for the large gap limit $|\Delta_S^{}|\to \infty$,   
are determined by $\delta_\sigma^{}$ and $\Theta$.   
Therefore, 
the Cooper-pair contributions, 
which vary 
depending on the parameter regions   
shown in Fig.\ \ref{fig:SingleDotPhase},   
can systematically be explored  
by using the polar coordinate  ($E_A^{}$, $\Theta$)  
since the phase shift $\delta_\sigma^{}$ through which 
the many-body effects enter is independent of the angle $\Theta$ 
that determines the superconducting coherence factor 
 $\sin^2 \Theta$  for the transmission probability 
$\mathcal{T}_{\mathrm{CP}}^{}$.

\subsubsection{Nonlocal conductance for $I_{R}^{}$ at $V_L\neq 0$ and $V_R=0$}

Equations \eqref{eq:current_R_ET_CP}--\eqref{eq:T_CP_single} 
provide a set of formulas  
that describe how the single-electron and the Cooper-pair tunneling parts,  
 $I_{R}^{\mathrm{ET}}$ and $ I_{R}^{\mathrm{CP}}$,
contribute to the total current  $I_{R}^{}$ 
for arbitrary bias voltages $V_L$ and $V_R$.  
We next consider the situation, at which 
 the right lead is grounded $V_R=0$ 
in order to clarify the contributions of the CAR 
to the nonlocal conductance $g_{\mathrm{RL}}^{}$ for the current $I_R^{}$,  
\begin{align}
g_{\mathrm{RL}}^{}
\, \equiv \,   
\frac{\partial I_R^{}}{\partial V_L} 
\,= & \  
2\, g_0^{}\,  
\bigl( 
\mathcal{T}_{\mathrm{ET}}^{} 
-\mathcal{T}_{\mathrm{CP}}^{}
\bigr) 
\nonumber \\
= & \
2\, g_0^{}\,  
\bigl( 
\mathcal{T}_{\mathrm{BG}}^{} 
-2\mathcal{T}_{\mathrm{CP}}^{}
\bigr),
\label{eq:nonlocalConSingleQD}
\end{align}
 where $g_0^{} = 
 \frac{e^2}{h}\, {4\Gamma_R^{}\Gamma_L^{}}/{\Gamma_N^{2}}$. 
 In the last line, the Bogoliubov angle  $\Theta$ enters 
  $g_{\mathrm{RL}}^{}$ solely through  $\mathcal{T}_{\mathrm{CP}}^{}$    
 since $\mathcal{T}_{\mathrm{BG}}^{}$ does not depend on it.
The contribution of Cooper-pair tunnelings in $g_{\mathrm{RL}}^{}$  
 is negative as it induces the current flowing  
from the right lead towards the QD at the center.

The CAR efficiency $\eta_\text{CAR}^{}$ 
is one of the useful parameters for measuring 
the CAR contribution to 
the nonlocal conductance  $g_{\mathrm{RL}}^{}$: 
\begin{align}
    \eta_\text{CAR}^{}\ 
 &\equiv\, 
\frac{|I_{R}^{\mathrm{CP}}|}{|I_{R}^{\mathrm{ET}}|+|I_{R}^{\mathrm{CP}}|} 
    =
\frac{\mathcal{T}_{\mathrm{CP}}^{}}{ \mathcal{T}_{\mathrm{BG}}^{} } 
\nonumber\\
\ &= \, 
\frac{\sin^2\left(\delta_\uparrow +\delta_\downarrow\right)}
{\,\sin^2 \delta_\uparrow +\sin^2 \delta_\downarrow}\,  
\frac{\sin^2\Theta}{2}
\,.
\label{eq:CARefficiencyDef}
\end{align}
Alternatively, 
the nonlocal conductance 
can also be expressed in terms of the efficiency:  
\begin{align}
g_{\mathrm{RL}}^{}
\,= & \  \,  
2 g_0^{}\,  \mathcal{T}_{\mathrm{BG}}^{}
\Bigl( 1\,-\,   2 \,\eta_\text{CAR}^{}\Bigr) \,.
\label{eq:nonlocalConSingleQD_in_efficiency}
\end{align}
Here, the $\Theta$ dependence of $g_{\mathrm{RL}}^{}$ arises from 
 the efficiency $\eta_\text{CAR}^{}$.
The efficiency $\eta_\text{CAR}^{}$ is enhanced 
by the coupling between  the QD and the SC lead.  
In the limit  $\Gamma_S^{}\to 0$ where the SC lead is disconnected,  
Eq.\ \eqref{eq:nonlocalConSingleQD_in_efficiency} 
reproduces the usual Landauer formula 
with the single-electron tunneling probability $\mathcal{T}_{\mathrm{BG}}^{}$.

Similarly, the local conductance for the current from the left lead $I_{L}^{}$  
 can also be expressed in the following form,
\begin{align}
& 
\frac{\partial I_L^{}}{\partial V_L} 
\,= \,  
2 g_0^{}\,  \mathcal{T}_{\mathrm{BG}}^{}
\left( 1\,+\, 
\frac{2\Gamma_{L}^{}}{\Gamma_{R}^{}}\,\eta_\text{CAR}^{}
\right) \,.
\label{eq:gLL_in_efficiency}
\end{align}
Here, the second term on the right-hand side 
represents the contribution of the direct Andreev reflection (DAR),  
inducing the current component 
$I_{L}^{\mathrm{DAR}} \equiv 4 g_0^{}\,  
\mathcal{T}_{\mathrm{CP}}^{}\,(\Gamma_{L}^{}/\Gamma_{R}^{})\, V_L^{}$ 
$ \propto  \Gamma_{L}^{2}/\Gamma_{N}^{2}$ 
 for $V_R^{}=0$. 
Therefore, the ratio of the DAR contribution to $I_{L}^{}$ is 
determined by $\delta$ and $\Theta$  through  
the efficiency $\eta_\text{CAR}^{}$,  
\begin{align}
&\frac{I_{L}^{\mathrm{DAR}}}{I_{L}^{}} 
\,=\, 
\frac{2\Gamma_{L}^{}\,\eta_\text{CAR}^{}}
{\Gamma_{R}^{}\,+\, 2\Gamma_{L}^{}\eta_\text{CAR}^{}}
\;.
\end{align}

\subsubsection{Andreev transport for $V_L=V_R$}

Here we briefly discuss another setting, 
in which bias voltages are applied 
in a symmetrical way $V_L=V_R$ 
 ($\equiv V$).
In this case, 
the contribution of  single-electron process vanishes 
$I_R^\mathrm{ET} = I_L^\mathrm{ET} = 0$,  
and the Cooper-pair tunnelings determine   
both $I_R^{}$ and $I_L^{}$,  as 
\begin{align}
 I_R^{} 
\,  \xrightarrow{\,V_L=V_R=V\,}  &  \  
- \frac{4e^2}{h}
\mathcal{T}_{\mathrm{CP}}^{}
\left[\,
\frac{4\Gamma_R^{}\Gamma_L^{}}{\Gamma_N^{2}}\,
+ \,\frac{4\Gamma_R^{2}}{\Gamma_N^{2}}\,
\right]  V \,, 
\label{eq:IR_V_sym}
\\
 I_L^{} \,  \xrightarrow{\,V_L=V_R=V\,}  & \    
+ \frac{4e^2}{h}
\mathcal{T}_{\mathrm{CP}}^{}
\left[\,
\frac{4\Gamma_R^{}\Gamma_L^{}}{\Gamma_N^{2}}\,
+ \,\frac{4\Gamma_L^{2}}{\Gamma_N^{2}}\,
\right]  V \,. 
\label{eq:IL_V_sym}
\end{align}
For both $ I_R^{}$ and $ I_L^{}$,  
 the first and the second terms in the square brackets on the right-hand side 
represent the contributions of 
the crossed Andreev reflection 
and the direct Andreev 
reflection, 
respectively. 
These terms depend sensitively on 
the asymmetry of the tunnel couplings. 
For instance, 
the CAR dominates $I_R^{}$ for $\Gamma_R^{} \ll \Gamma_L^{}$,
 as the direct Andreev scattering occurring in the right lead is suppressed.

The current flowing into the SC lead through 
the QD is given by $I_L^{} -I_R^{}$. 
It reaches the maximum value $4e^2 V/h$ 
in the case at which  $\mathcal{T}_{\mathrm{CP}}^{}=\frac{1}{4}$ 
for symmetric junctions $\Gamma_L^{} = \Gamma_R^{}$ 
($\equiv\Gamma_N^{}/2$).    
Note that the behavior of 
this current $I_L^{} -I_R^{}$ into the SC lead 
is equivalent to the one flowing through an N-QD-SC junction, 
which was investigated in the previous work.\cite{YoichTanaka_2007}

\subsection{NRG approach to the CAR}

In the following two sections,   
we numerically investigate the contribution of  
the CAR over a wide range of the parameter space. 
To this end, we have calculated the phase shift $\delta_{\sigma}^{}$ and the other 
correlation functions of Bogoliubov quasiparticles, 
applying the NRG approach\cite{KWW1,KWW2,HewsonOguriMeyer} 
to the effective Hamiltonian $H_\mathrm{eff}^{}$ 
given in Eq.\ \eqref{eq:SingleDot_new},\cite{YoichTanaka_2007}  
choosing the discretization parameter to be $\Lambda=2.0$ 
and $\Gamma_{N}/D=1/(100\pi)$. 
We have also constructed the interpolating functions for 
the phase shift $\delta_{\sigma}^{}$  from a discrete set of the NRG data 
obtained along the radial-$E_A^{}$ direction  in the parameter space,
 described in Fig.\ \ref{fig:SingleDotPhase}. 
The dependence of the transport coefficients 
on the Bogoliubov-rotation angle  $\Theta$  of the polar coordinate  
has been determined by using the exact formulas presented in the above.

We will discuss the CAR contribution to the nonlocal transport 
at zero field in  Sec.\ \ref{CARZeroMag}, 
and then consider magnetic-field dependence in Sec.\ \ref{CARMag}.

\section{Crossed-Andreev transport at zero field $b=0$}
\label{CARZeroMag}

In this section, we show the NRG results 
for the nonlocal conductance and renormalized parameters   
calculated 
at zero magnetic field $b=0$,  
extending the previous results obtained for a 
two terminal N-QD-S system.\cite{YoichTanaka_2007}
Before going into the details, we describe some general features  
which can be deduced from the transport formulas presented above.

At zero magnetic field $b=0$, the phase shift becomes independent of spin component 
$\delta_\uparrow^{} =\delta_\downarrow^{}$ ($\equiv \delta$),
and thus the transport coefficients are determined 
by two angular parameters $\delta$ and $\Theta$.
The average occupation number of the 
Andreev level 
in this case is given by the phase shift $Q = 2\delta/\pi$.  
It decreases from the unitary limit value $Q=1$  
as $E_A^{}$ deviates from the origin,  $E_A^{}=0$,
of the parameter space illustrated in Fig.\ \ref{fig:SingleDotPhase}. 
In contrast,  the SC pair correlation function 
$\bigl \langle d^\dagger_{\uparrow}\,d^\dagger_{\downarrow}
 +d_{\downarrow}\,d_{\uparrow} \bigr \rangle$,  
 defined  in Eq.\ \eqref{eq:SingleDotK}, 
depends not only on the phase shift $\delta$ but also 
the coherence factor,  $\sin \Theta$, 
which takes a maximum at $\Theta = \pi/2$.

Similarly,  at zero magnetic field, the transmission probabilities 
defined in Eqs.\  \eqref{eq:T_ET_single} and \eqref{eq:T_CP_single}    
can be simplified, as 
\begin{align}
\mathcal{T}_{\mathrm{ET}}^{} 
 = &  \  \sin^2 \delta - \mathcal{T}_{\mathrm{CP}}^{}      
\,, %
\qquad 
\mathcal{T}_{\mathrm{CP}}^{} =      
\frac{1}{4} \sin^2 \Theta\, 
\sin^2  2\delta \,, 
\label{eq:T_ET_CP_b0}
\end{align}
and $\mathcal{T}_{\mathrm{BG}}^{} =\sin^2 \delta$.
Therefore, the Cooper-pairing part $\mathcal{T}_{\mathrm{CP}}^{}$ 
takes a maximum at $\Theta=\pi/2$ and $\delta =\pi/4$,
where the Andreev level for Bogliubov particles is  quarter-filling
  $Q_\sigma^{}=\frac{1}{4}$.
Correspondingly,  
the nonlocal conductance and the CAR efficiency 
defined in Eqs.\ 
\eqref{eq:nonlocalConSingleQD}--\eqref{eq:nonlocalConSingleQD_in_efficiency}   
can be expressed in the following forms, at $b=0$:   
\begin{align}
 g_{\mathrm{RL}}^{}
 \,  = & \  
2\, g_0^{} 
\sin^2 \delta 
\,  \Bigl( 1
-  2 \,\sin^2 \Theta\,\cos^2  \delta \Bigr)  , 
\label{eq:g_tot_b0}
 \\ 
    \eta_\text{CAR}^{} \,= & \  
\sin^2 \Theta\,\cos^2 \delta
\label{eq:eta_CAR_b0}
\,.
\end{align}
Thus, for the CAR to dominate the nonlocal conductance, 
taking a negative value $g_{\mathrm{RL}}^{} <0$,  
the Bogoliubov angle must be in the range $\pi/4<\Theta< 3\pi/4$,  
i.e., $2 \sin^2 \Theta >1$.

In particular,  Cooper pairs are most entangled at $\Theta =\pi/2$,
and  in this case the transport coefficients take the form, 
\begin{align}
\mathcal{T}_{\mathrm{ET}}^{} 
 \,\xrightarrow{\,\Theta = \frac{\pi}{2}\,} \,
 \, \sin^4 \delta , \qquad 
g_{\mathrm{RL}}^{}
 \xrightarrow{\,\Theta = \frac{\pi}{2}\,} \,
 -\,
2\, g_0^{}\,
\sin^2 \delta\, \cos 2\delta . 
\label{eq:nonlocalConSingleQD_b0_half-filled}
\end{align}
Hence, along the $\Gamma_S^{}$-axis in Fig.\ \ref{fig:SingleDotPhase}, 
the nonlocal conductance  $g_{\mathrm{RL}}^{}$  becomes negative 
for $0<\delta \lesssim \pi/4$,  
where the ground state of $H_\mathrm{eff}^{}$ 
is in the valence-fluctuation regime or the empty-orbital regime 
of the Bogoliubov particles. 
It takes a minimum 
of the depth  $g_{\mathrm{RL}}^{}/ g_0^{} = -1/4$  at $\delta =\pi/6$.  
As the phase shift approaches  $\delta  \simeq \pi/2$,  
the Kondo effect  dominates 
and the transmission probability of the Bogoliubov-particle 
shows a Kondo-ridge structure, at which   
$\mathcal{T}_{\mathrm{BG}}^{} \simeq 1.0$, 
 as we will demonstrate later.

\begin{figure}[b]
\includegraphics[width=0.49\linewidth]{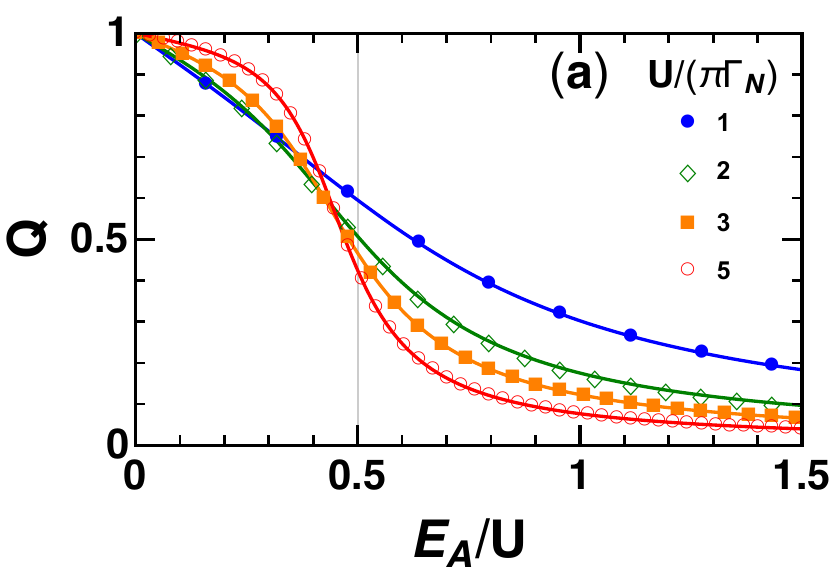}
\includegraphics[width=0.49\linewidth]{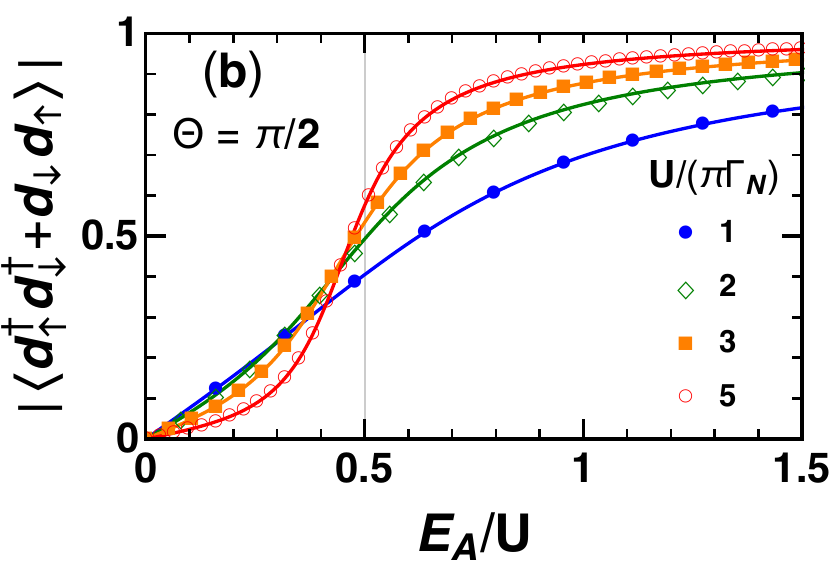}
 \\
\includegraphics[width=0.49\linewidth]{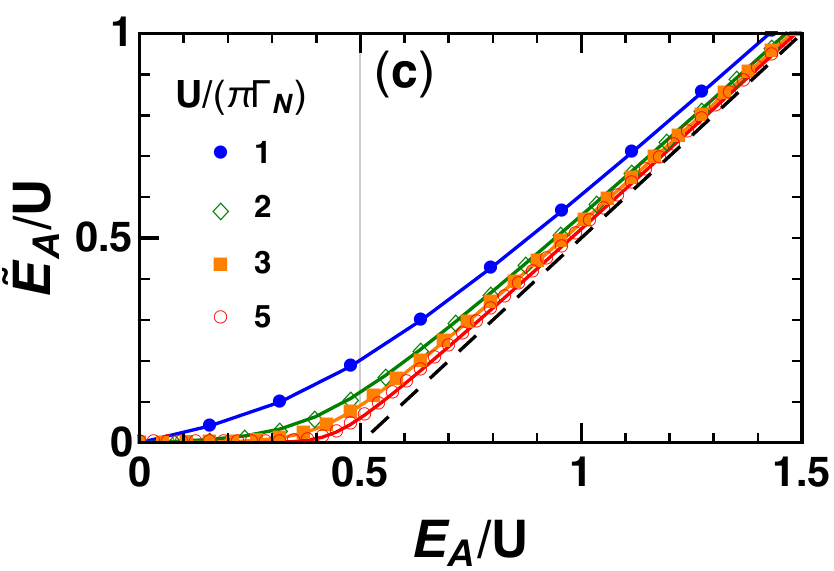}
\includegraphics[width=0.49\linewidth]{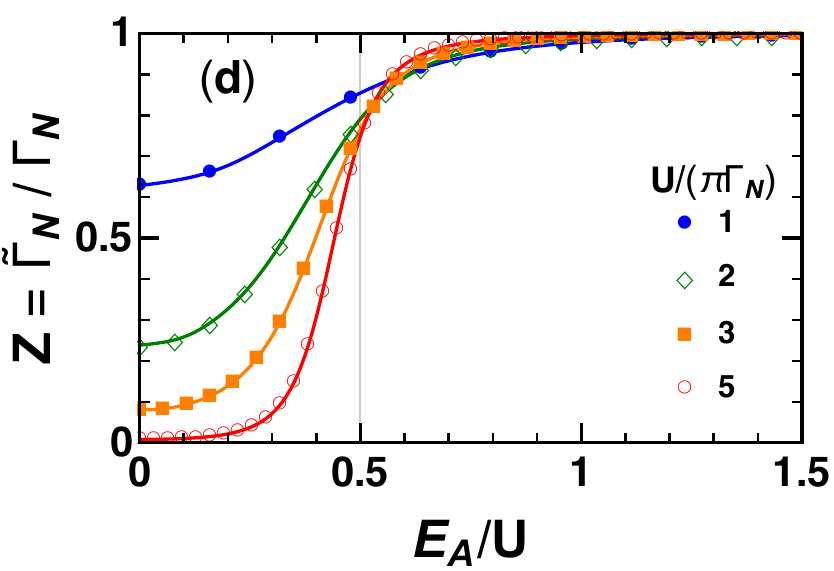}
\caption{Renormalized parameters 
plotted 
vs $E_A^{}/U$ for $U/(\pi \Gamma_{N})=1.0$, $2.0$, $3.0$, $5.0$ at $b=0$.
(a): Occupation number of Bogoliubov particles $Q$ ($=2\delta/\pi$).
 (b): Pair correlation  
$|\bigl \langle d^\dagger_\uparrow d^\dagger_\downarrow +  
        d^{}_\downarrow d^{}_\uparrow \bigr \rangle |$.  
        (c): Renormalized Andreev-resonance energy $\widetilde{E}_A$.
        (d): Renormalization factor 
       $Z = \widetilde{\Gamma}_{N}/\Gamma_N$, 
        which at $E_A^{}=0$ takes the  values  
        $0.629$,  $0.239$,  $0.080$,  $0.008$, 
        respectively, for the above four values of $U$.
        Note that   $\langle n_d^{} \rangle = 1.0$ and 
        $|\bigl \langle d^\dagger_\uparrow d^\dagger_\downarrow +  
        d^{}_\downarrow d^{}_\uparrow \bigr \rangle |=1-Q$, 
        at $\Theta = \pi/2$ 
       along the $\Gamma_S^{}$ axis of Fig.\ \ref{fig:SingleDotPhase}.  
The dashed line in (c) denotes the Hartree-Fock energy shift 
$E_{A}^\mathrm{HF} \xrightarrow{E_A^{} \gg U/2}  E_{A}^{} -U/2$, 
given in Eq.\ \eqref{eq:HF_energy_shift}.  
}
\label{fig:NRGOneDot_FLparametersxi0}
\end{figure}

\subsection{Ground state properties at $\Theta = \pi/2$}

\label{Electron-holeSymmetricCase}

We next consider how the ground state of $H_\mathrm{eff}^{}$ 
evolves as $E_A^{}$ varies 
along the radial direction in the $\xi_d^{}$ vs $\Gamma_S^{}$ plane, 
shown in Fig.\ \ref{fig:SingleDotPhase}.
 Note that the eigenstates and eigenvalues of the effective Hamiltonian  
defined in Eq.\ \eqref{eq:SingleDot_new} 
 do not depend on the angular coordinate $\Theta$.

Figure \ref{fig:NRGOneDot_FLparametersxi0}(a)    
shows the occupation number  $Q$ as a function of $E_A^{}$ 
for  $U/(\pi \Gamma_{N}) = 1.0$, $2.0$, $3.0$, and $5.0$.     
We see that $Q$ decreases as $E_A^{}$ increases, 
especially near  $E_A^{} \simeq U/2$,  
where the crossover from the Kondo regime 
to the valence-fluctuation regime of Bogoliubov particles occurs 
 for large interactions $U/(\pi \Gamma_N^{}) \gtrsim 2.0$.  
Figure \ref{fig:NRGOneDot_FLparametersxi0}(b) shows 
the magnitude of the pair correlation function for $\Theta =\pi/2$, 
where the absolute value  is given by 
$ |\bigl\langle d^\dagger_\uparrow d^\dagger_\downarrow 
+  d^{}_\downarrow d^{}_\uparrow \bigr\rangle| =1-Q$. 
 It increases significantly at  $E_A^{} \simeq U/2$, i.e., 
near the quarter-filling point $Q=0.5$ ($\delta =\pi/4$) of Bogoliubov particles,
and it saturates to the upper-bound value $1.0$ as $E_A^{}$ 
increases further towards the empty-orbital regime.

The Kondo behaviors of Bogoliubov particles are clearly seen for   
the renormalized Andreev level  $\widetilde{E}_A$ 
 and the wave-function renormalization factor 
$Z = \widetilde{\Gamma}_N/\Gamma_N$, plotted   
 in Figs.\ 
\ref{fig:NRGOneDot_FLparametersxi0}(c) and
\ref{fig:NRGOneDot_FLparametersxi0}(d), 
respectively.   
The renormalized level  is almost locked at the Fermi level 
$\widetilde{E}_A^{} \simeq 0.0$, 
 for large interactions $U/(\pi \Gamma_N^{})\gtrsim 2.0$, 
 over a wide Kondo-dominated region  $0\leq E_A^{} \lesssim U/2$, 
taking place the inside of the semicircle in Fig.\ \ref{fig:SingleDotPhase}. 
Correspondingly, the renormalization factor $Z$ is significantly suppressed in this region,  
and it indicates the fact that the Kondo energy scale $T^*$, 
\begin{align}
T^* \, \equiv \, \frac{Z}{4 \rho_d^{}} \,, 
\qquad \qquad 
\rho_d^{} \,=\, \frac{\sin^2 \delta}{\pi\Gamma_N^{}} \,,
\label{eq:T_star}
\end{align}
becomes much smaller than the bare tunneling energy scale  $\Gamma_N^{}$.

In contrast, at $E_A^{} \gtrsim U/2$, i.e.,  
 in the valence-fluctuation or  empty-orbital regime for Bogoliubov particles,  
the effects of electron correlations become less important:  
the renormalization factor  approaches $Z \simeq 1.0$  and the renormalized level 
$\widetilde{E}_{A,\sigma}^{}$ approaches the Hartree-Fock (HF) energy shift:    
\begin{align}
& E_{A,\sigma}^\mathrm{HF} 
\, \equiv \,  
E_{A}^{} -  \sigma b
\,+\, U\left( Q_{\overline{\sigma}}^{} -\frac{1}{2}\right) 
\label{eq:HF_energy_shift}
\\
&  \ \   \xrightarrow{\, E_{A}^{} \gg U/2, \, b=0\, } \,  E_{A}^{} 
\,-\, \frac{U}{2} \,, 
\nonumber 
\end{align}
since  $\,Q_{\overline{\sigma}}^{} \simeq  0.0$ 
at  $E_{A}^{} \gg U/2$ and $b=0$.

\begin{figure}[b]
\includegraphics[width=0.7\linewidth]{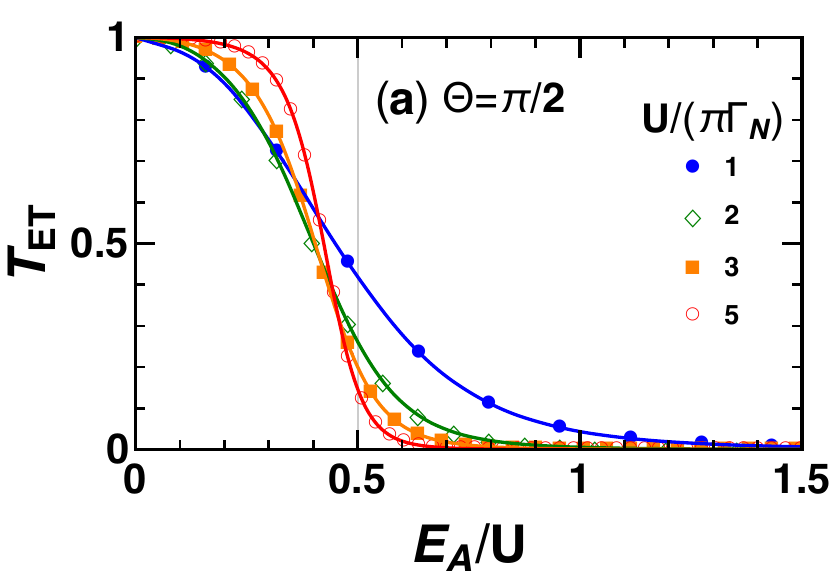}
\\
\includegraphics[width=0.7\linewidth]{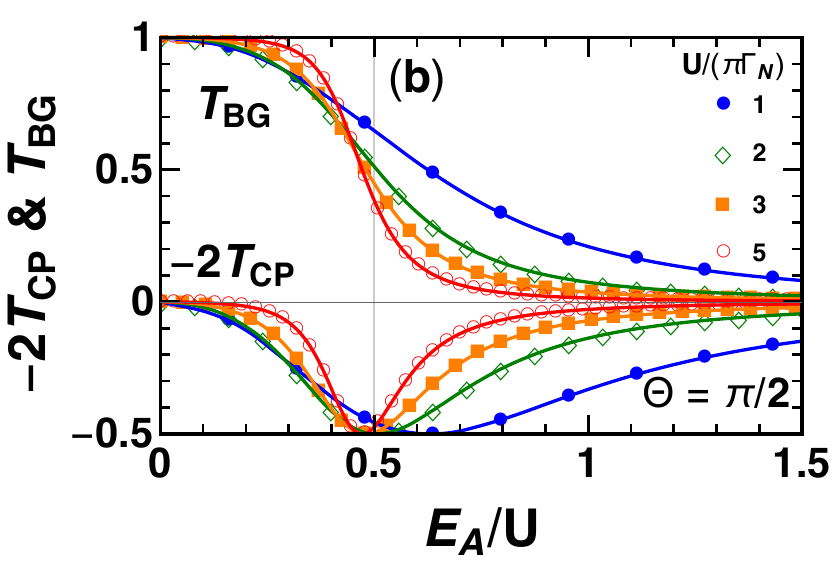}
\\
\includegraphics[width=0.7\linewidth]{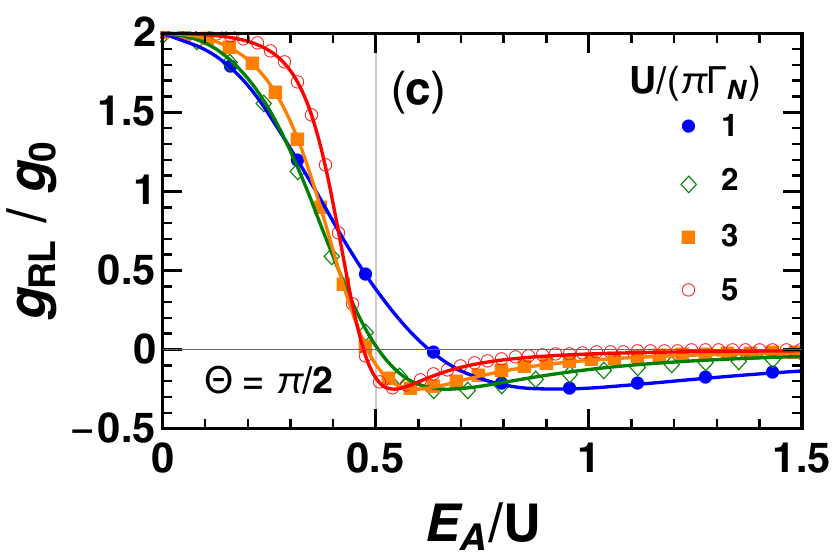}
\\
\includegraphics[width=0.7\linewidth]{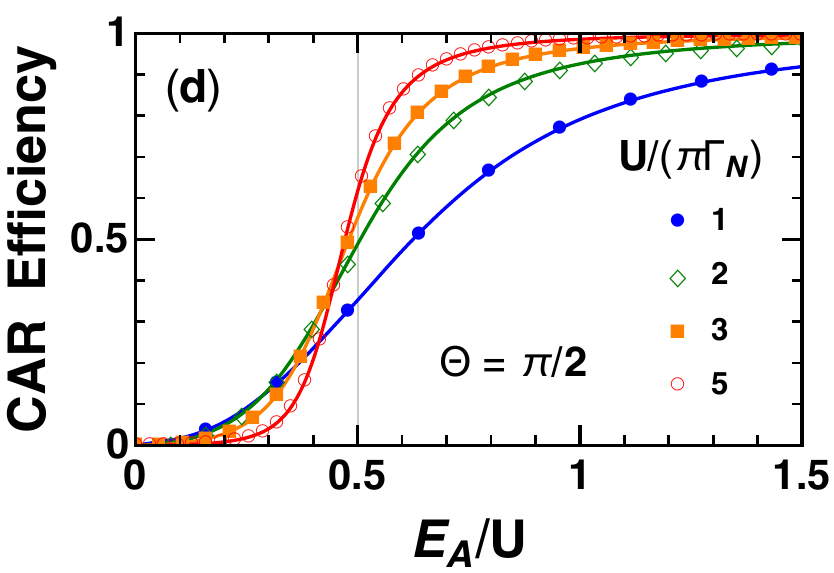}
\caption{Transport coefficients plotted vs $E_A/U$ 
    for  $U/(\pi \Gamma_{N})= 1.0$, $2.0$, $3.0$, $5.0$  
at $b=0$, 
    keeping the Bogoliubov angle fixed at $\Theta = \pi/2$ (i.e., $\xi_d=0$).  
    (a): Single-electron transmission $\mathcal{T}_{\mathrm{ET}}^{}$,
    defined in Eq.\  \eqref{eq:T_ET_single}. 
  (b): Cooper-pair contributions  
    $-2\mathcal{T}_{\mathrm{CP}}^{}$ ($< 0$)
    and the Bogoliubov-particle transmission 
$\mathcal{T}_{\mathrm{BG}}=\sin^2 \delta$, 
  defined in Eqs.\  \eqref{eq:T_CP_single} and \eqref{eq:T_BG_single}, 
     respectively.  
    (c): Nonlocal conductance 
 $g_{\mathrm{RL}}^{} / g_0^{} = 
 2(\mathcal{T}_{\mathrm{BG}} - 2\mathcal{T}_{\mathrm{CP}})$  
with $g_0^{} = \frac{e^2}{h}{4\Gamma_R^{}\Gamma_L^{}}/{\Gamma_N^{2}}$.
     (d): CAR efficiency 
    $\eta_{\text{CAR}}^{}$ 
     defined in Eq.\ \eqref{eq:CARefficiencyDef}.
     }
\label{fig:NRGOneDotXi0}
\end{figure}

\subsection{Transport properties at $\Theta = \pi/2$}

 We next discuss the transport properties. 
Specifically, in this subsection, 
we consider the case $\Theta =\pi/2$, where  $\xi_d^{}=0$    
and the occupation number of impurity electrons is fixed 
at  $\langle n_d^{}\rangle =1$, 
reflecting the electron-hole symmetry of  $H_\mathrm{eff}^{}$ 
defined in Eq.\ \eqref{eq:Heff_single}.
In this case, the Andreev level takes the value $E_A^{} = \Gamma_S^{}$,   
which is determined solely by the coupling strength between the QD and the SC lead  
and it breaks the particle-hole symmetry of the Bogoliubov particles 
even at  $\xi_d^{}=0$.

The transmission probability 
$\mathcal{T}_{\mathrm{ET}}^{}$ 
of the single-electron tunneling process  
is shown in Fig.\ \ref{fig:NRGOneDotXi0}(a). 
We see that the plateau of 
the unitary limit $\mathcal{T}_{\mathrm{ET}}^{} \simeq 1.0$ 
 evolves at $0\leq E_A^{} \lesssim U/2$, for large $U$. 
Since $\mathcal{T}_{\mathrm{ET}}^{} 
= \mathcal{T}_{\mathrm{BG}}^{} - 
\mathcal{T}_{\mathrm{CP}}^{}$ 
due to the optical theorem mentioned above,  
it is the Bogoliubov-particle part    
$\mathcal{T}_{\mathrm{BG}}^{} =\sin^2 \delta$ 
that shows the genuine  Kondo ridge, as  
demonstrated in Fig.\ 
\ref{fig:NRGOneDotXi0}(b). 
The single-particle contribution $\mathcal{T}_{\mathrm{ET}}^{}$ decreases  
outside of the Kondo regime  $E_A^{} \gtrsim U/2$, 
at which the occupation number $Q$ of Bogoliubov 
particles rapidly decreases and 
the SC pair correlation increases, as    
demonstrated in Figs.\ \ref{fig:NRGOneDot_FLparametersxi0}(a) 
and \ref{fig:NRGOneDot_FLparametersxi0}(b).

The Cooper-pair contribution 
 $\mathcal{T}_{\mathrm{CP}}^{}$  
 is also plotted in Fig.\ 
\ref{fig:NRGOneDotXi0}(b),  
choosing the Bogoliubov angle to be $\Theta=\pi/2$ and   
multiplying 
a  factor of $-2$ which emerges for the nonlocal conductance 
  $g_{\mathrm{RL}}^{} \propto 
\mathcal{T}_{\mathrm{BG}}^{}-2\mathcal{T}_{\mathrm{CP}}^{}$: 
the negative sign represents the fact that  
the crossed Andreev reflection induces 
the counterflow, flowing from the right lead towards the QD.  
In this case,  
Eq.\ \eqref{eq:T_ET_CP_b0}  can be rewritten further 
into a similar form to the current noise of normal electrons:  
 $\mathcal{T}_{\mathrm{CP}}^{} = \sin^2 \delta\, (1-\sin^2 \delta)$. 
\cite{BlanterButtiker,AO2022}
Thus, the contribution of $\mathcal{T}_{\mathrm{CP}}^{}$ 
 to the nonlocal conductance is maximized in the case at which  
the phase shift becomes $\delta = \pi/4$  
and it reaches the value $-2\mathcal{T}_{\mathrm{CP}}^{} = -\frac{1}{2}$. 
The corresponding dip emerges in Fig.\ \ref{fig:NRGOneDotXi0}(b)   
at the 
crossover region 
$E_A^{} \simeq U/2$, 
the width of which becomes of the order of $\Gamma_N^{}$.

Figure \ref{fig:NRGOneDotXi0}(c) shows the nonlocal conductance, 
which takes the form  $g_{\mathrm{RL}}^{}/g_0^{}  =  -2 \sin^2 \delta \cos 2\delta$ 
at $\Theta = \pi/2$, as mentioned.
It decreases from the unitary-limit value 
  $g_{\mathrm{RL}}^{}/g_0^{} = 2$ 
 as $E_A^{}$ deviates from $E_A^{}=0$,  
and  vanishes $g_{\mathrm{RL}}^{}=0$ at  $E_A^{} \simeq U/2$ 
 where the phase shift reaches $\delta = \pi/4$. 
The nonlocal conductance becomes 
negative  $g_{\mathrm{RL}}^{} <0$ at $E_A^{} \gtrsim U/2$ 
as the CAR contributions dominate in this region.
In particular, it has a dip of the depth $g_{\mathrm{RL}}^{}/g_0^{} = - 1/4$ 
at the point  where the phase shift takes the value $\delta = \pi/6$.

Similarly, the CAR efficiency 
takes a simplified form $\eta_{\text{CAR}}^{} = \cos^2\delta$ at $\Theta = \pi/2$,  
and 
the NRG results are plotted 
 in Fig.\ \ref{fig:NRGOneDotXi0}(d). 
The efficiency $\eta_{\text{CAR}}^{}$ increases with $E_A^{}$, 
and reaches $\eta_{\text{CAR}}^{} =0.5$ 
at $\delta=\pi/4$ where  $-2\mathcal{T}_{\mathrm{CP}}^{}$ has  
the dip of the depth $-1/2$ seen in Fig.\ \ref{fig:NRGOneDotXi0}(b).
The transient region of $\eta_{\text{CAR}}^{}$ varying from 0 to 1 
is estimated to be of the order of $\Gamma_N^{}$.  
Furthermore, at $E_A^{} \gg U/2$,  
the efficiency approaches the saturation value $\eta_{\text{CAR}}^{} = 1.0$ 
although the conductance $g_{\mathrm{RL}}^{}$ itself becomes very small.

\begin{figure}[b]
\includegraphics[width=0.7\linewidth]{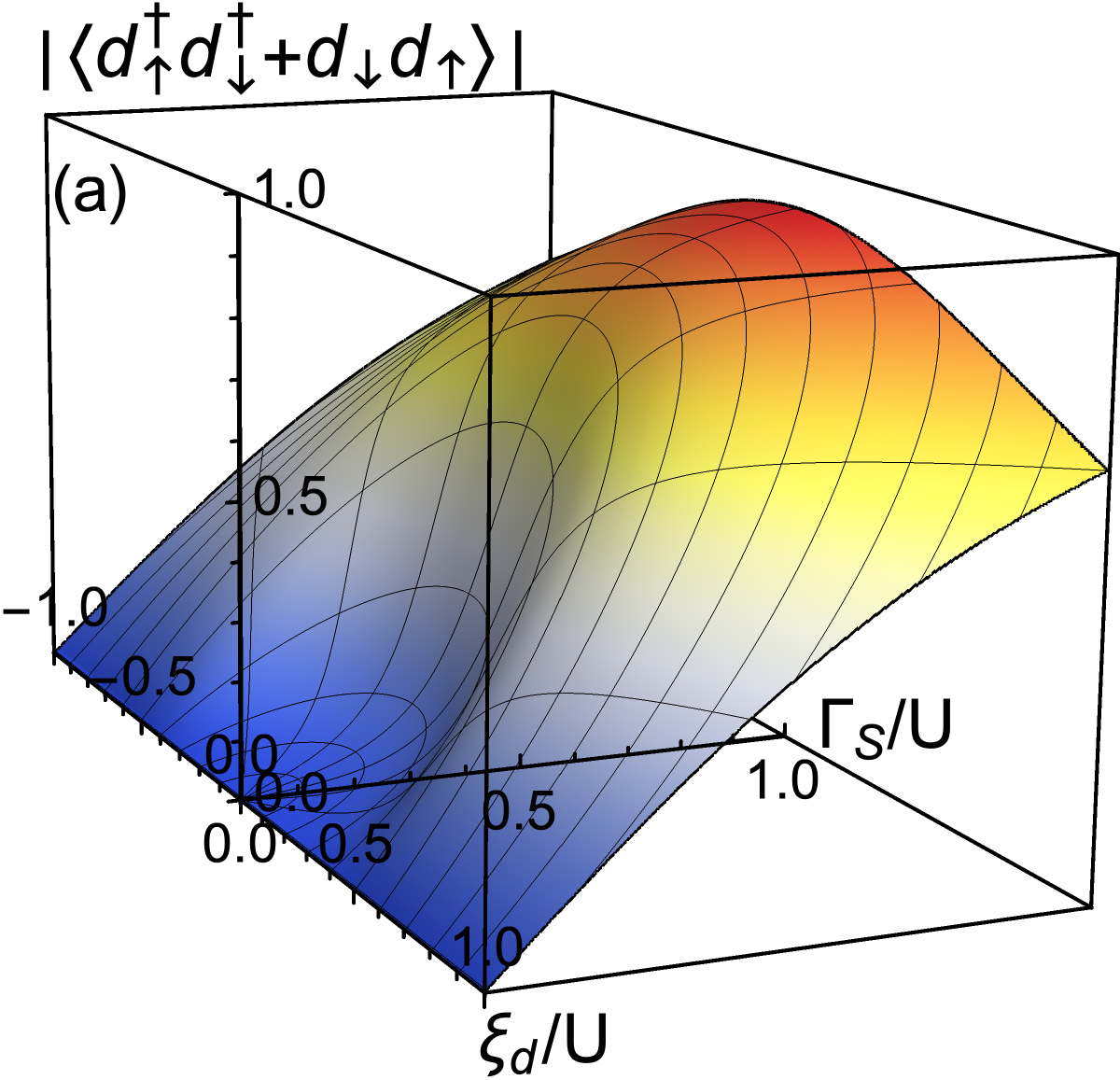}     \\
\includegraphics[width=0.9\linewidth]{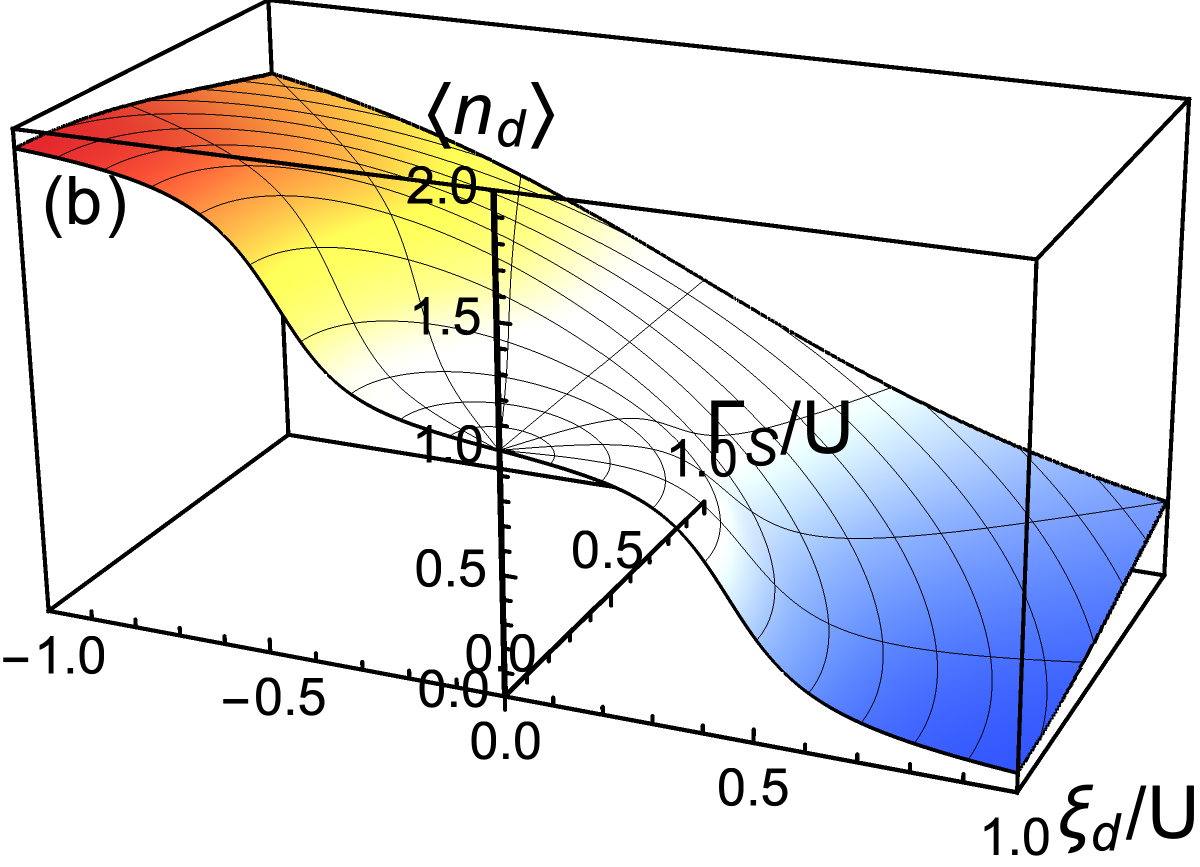}
\caption{Three-dimensional plots of 
(a)  $|\bigl\langle d^\dagger_\uparrow d^\dagger_\downarrow 
+  d^{}_\downarrow d^{}_\uparrow \bigr\rangle|$ and  (b) 
$\left\langle n_d^{}\right\rangle$,   
described 
as functions of  $\xi_d^{}$ and $\Gamma_S^{}$, 
choosing $U/(\pi\Gamma_N) = 5.0$. 
Mesh lines are drawn 
along the polar coordinate ($E_A^{}$, $\Theta$), 
with $E_A^{}=\sqrt{\xi_d^2+\Gamma_S^2}$ and 
$\Theta = \tan^{-1} (\Gamma_S^{}/\xi_d^{})$.
}
\label{fig:NRGOneDot_FLparametersxi}
\end{figure}

\subsection{The characteristics of CAR  along the \\ 
 polar coordinates $E_A^{}$ and $\Theta$ 
at $b=0$
}

\label{GateVoltage}

\begin{figure*}[t]
\includegraphics[width=0.32\linewidth]{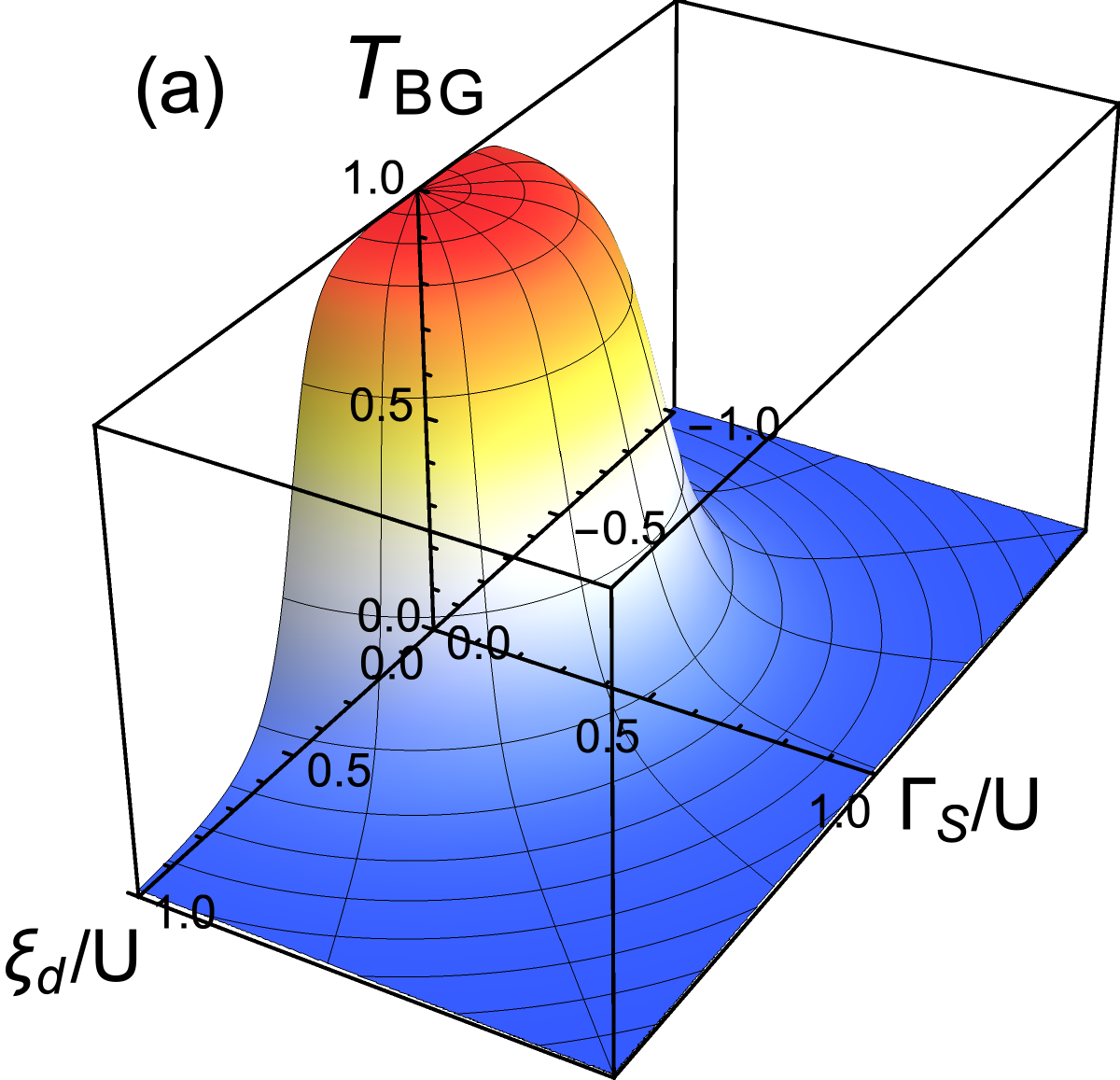}
\includegraphics[width=0.32\linewidth]{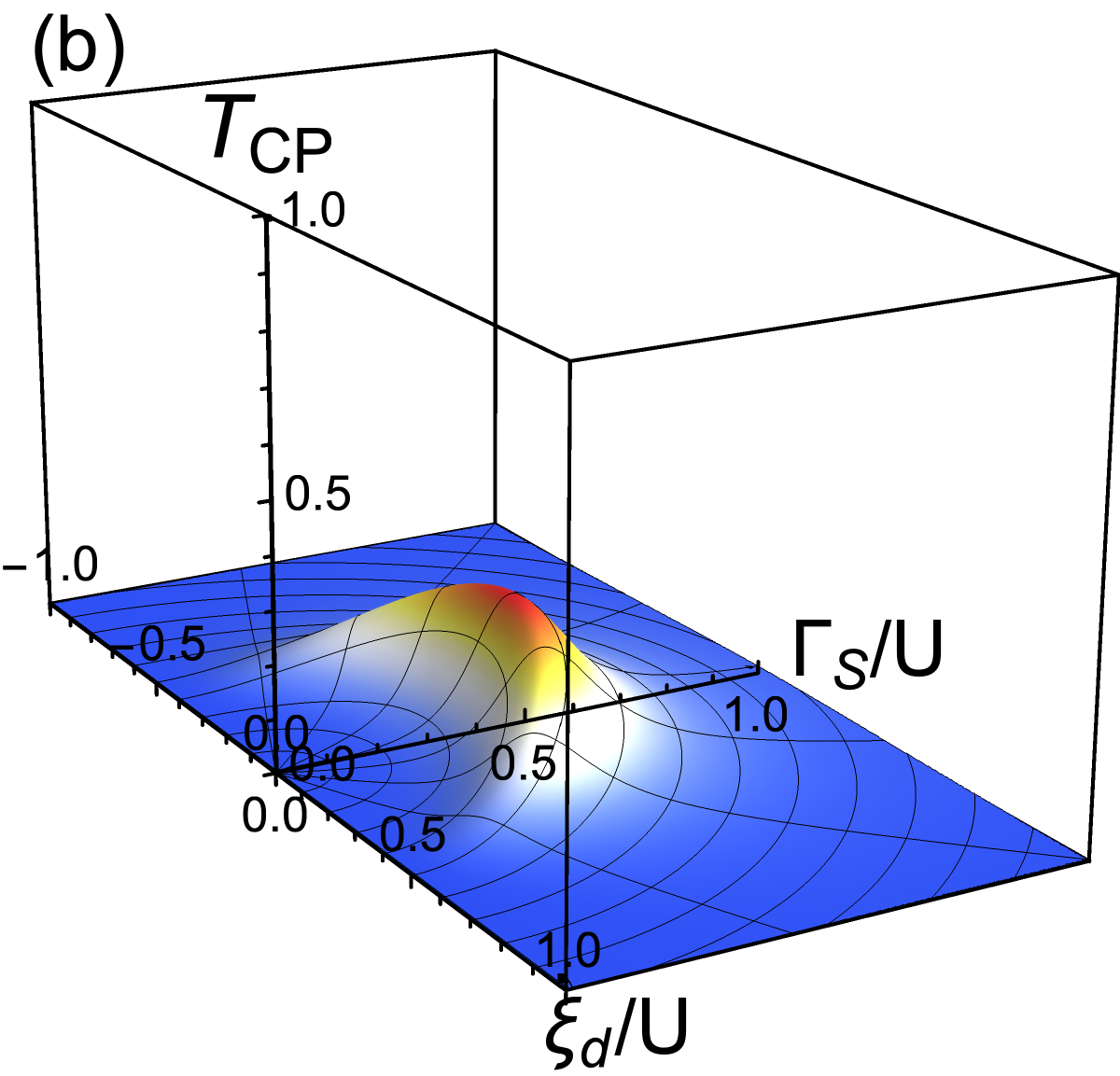}
\includegraphics[width=0.32\linewidth]{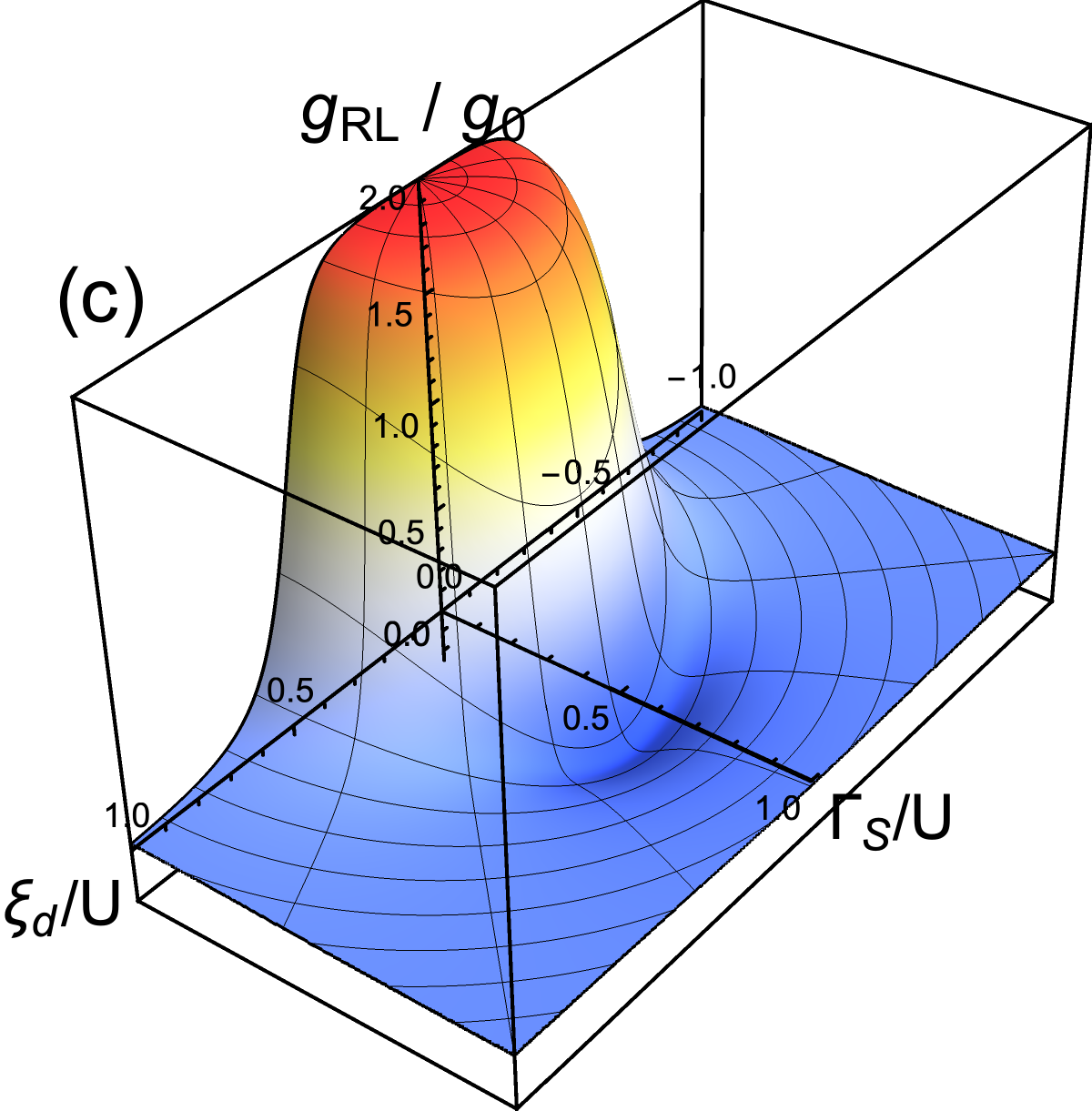}
\\
\includegraphics[width=0.4\linewidth]{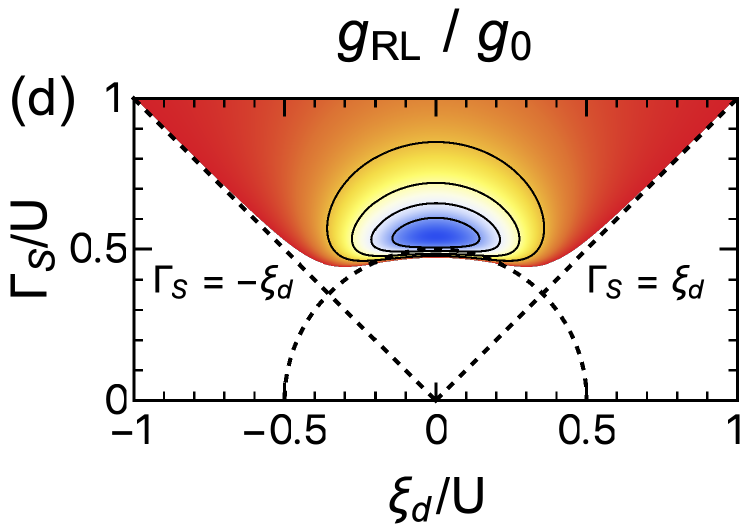}
\includegraphics[width=0.06\linewidth]{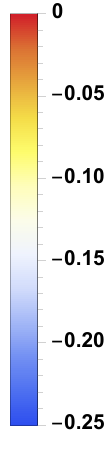}
\hspace{15pt}
\includegraphics[width=0.32\linewidth]{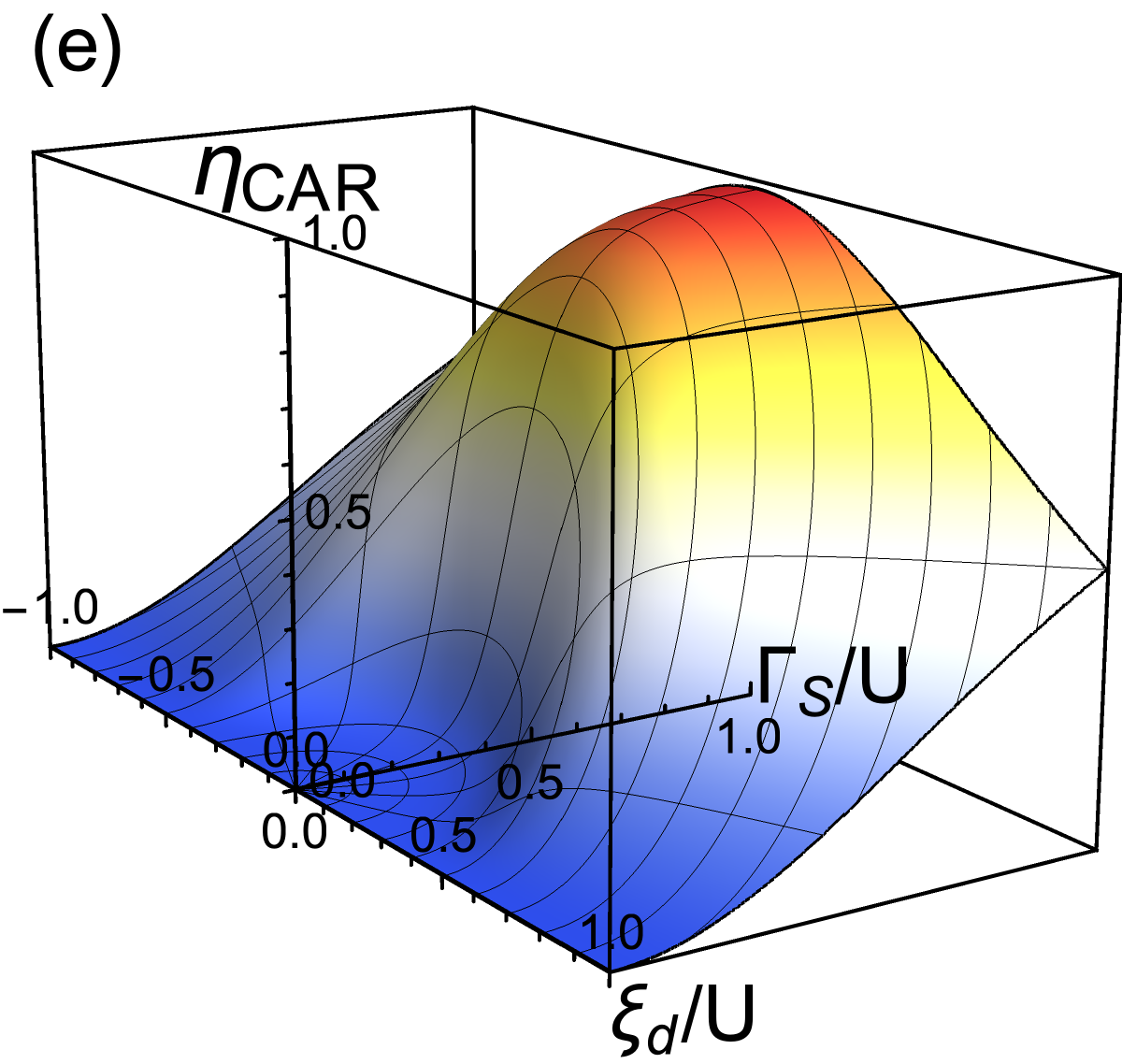}
\caption{Transport coefficients 
(a) $\mathcal{T}_{\mathrm{BG}}^{}$,  
(b)  $\mathcal{T}_{\mathrm{CP}}^{}$,   
(c) $g_{\mathrm{RL}}^{}$,  and 
(e) the  CAR efficiency $\eta_{\text{CAR}}^{}$ 
plotted as functions of $\xi_d^{}$ and $\Gamma_S^{}$,  
for $U/(\pi\Gamma_N) = 5.0$ and $b=0$.   
For these three-dimensional plots, mesh lines are drawn 
along the polar coordinate ($E_A^{}$, $\Theta$). 
Two-dimensional plot  (d) is a contour map 
for the region where the nonlocal conductance 
$g_{\mathrm{RL}}^{}$ becomes negative:  
contour lines are drawn 
with $0.05$ increments.
The CAR dominates $g_{\mathrm{RL}}^{}$ 
over the parameter region  $\Gamma_S^{} \gtrsim U/2$ and 
$\pi/4 < \Theta < 3\pi/4$, i.e.,  $\Gamma_S > |\xi_d|$.
}
\label{fig:NRGOneDotXi}
\end{figure*}

So far, we have discussed the transport properties at $\Theta = \pi/2$, 
along the vertical $\Gamma_S^{}$ axis 
in the $\xi_d^{}$ vs $\Gamma_S^{}$ plane.
As  $\xi_d^{}$  varies 
from the electron-hole symmetric point $\xi_d^{}=0$,     
the Bogoliubov angle  $\Theta$ deviates from $\pi/2$.  
Here we discuss how the ground-state and transport properties 
vary along the angular direction   
over the range  $0 \leq \Theta  \leq \pi$.

Figures \ref{fig:NRGOneDot_FLparametersxi} and \ref{fig:NRGOneDotXi}  
show the NRG results of the renormalized parameters 
and the transport coefficients 
as functions of $\xi_d^{}$ and $\Gamma_S^{}$  
for a relatively large Coulomb interaction $U/(\pi \Gamma_N^{})=5.0$. 
In these three-dimensional plots,  
mesh lines are drawn along the polar coordinates 
$(E_A^{}, \Theta)$. 
Note that the superconducting coherence factors, 
$\cos \Theta$ and $\sin \Theta$, vary in the angular direction:      
 Cooper pairs are strongly entangled at $\pi/2$  
and the SC proximity effect becomes weak as $\Theta$ deviates towards $0$ or $\pi$.  
In contrast, along the radial direction,  
the crossover  between the Kondo regime and valence fluctuation regime 
of the Bogoliubov particles occurs 
near the semicircle of radius $E_A^{} = U/2$, as mentioned.

\subsubsection{$\Theta$ dependence of 
$\,\bigl\langle d^\dagger_\uparrow d^\dagger_\downarrow 
+  d^{}_\downarrow d^{}_\uparrow \bigr\rangle$  
and  $\,\langle n_d^{}\rangle$}

Among the renormalized parameters plotted in
Fig.\ \ref{fig:NRGOneDot_FLparametersxi0}, 
the following three, 
 $Q$,  $Z$, and $\widetilde{E}_A^{}$  
do not depend on the Bogoliubov angle $\Theta$,  
and thus Figs.\ 
\ref{fig:NRGOneDot_FLparametersxi0}(a), 
\ref{fig:NRGOneDot_FLparametersxi0}(c), 
and \ref{fig:NRGOneDot_FLparametersxi0}(d) 
remain unchanged as angle $\Theta$ varies.
In contrast, the correlation functions which are defined 
with respect to electrons, such as 
$|\bigl\langle d^\dagger_\uparrow d^\dagger_\downarrow 
+  d^{}_\downarrow d^{}_\uparrow \bigr\rangle| 
= (1-Q) \sin \Theta$  
and 
 $\langle n_d^{}\rangle = 1+ (Q-1) \cos \Theta$, 
evolve with the Bogoliubov angle  $\Theta$.

We can see in Fig.\ \ref{fig:NRGOneDot_FLparametersxi} (a)  
that the pair correlation is suppressed due to the Kondo effect 
at $E_A^{} \lesssim U/2$, 
especially along the valley at $\Theta = \pi/2$, 
inside the semicircle shown in Fig.\ \ref{fig:SingleDotPhase}.   
The slope from the valley bottom towards 
the direction parallel to the $\xi_d^{}$-axis 
 is suppressed by the coherence factor $\sin \Theta$.  
Correspondingly, 
in Fig.\ \ref{fig:NRGOneDot_FLparametersxi}(b),    
the occupation number  $\langle n_d^{}\rangle$  
 of electrons clearly shows a plateau of a semicircle shape  
which spreads around  the origin $E_A^{}=0.0$ of 
 the $\xi_d^{}$ vs  $\Gamma_S^{}$ plane. 
Note that the occupation number is locked exactly 
at  $\langle n_d^{}\rangle =1.0$ 
along the $\Gamma_S^{}$ axis.
Outside the plateau  $E_A^{}  \gtrsim U/2$,  
the superconducting proximity effects dominate  
over the angular range of  $\pi/4 < \Theta <3\pi/4$,  
or equivalently at $\Gamma_S^{} > |\xi_d^{}|$.  
In particular, 
the ridge of the pair correlation  
develops at  $\Theta =\pi/2$,  
along the $\Gamma_S^{}$-axis  
 in Fig.\ 
\ref{fig:NRGOneDot_FLparametersxi}(a).

\subsubsection{$\Theta$ dependence of transport properties}

Figure \ref{fig:NRGOneDotXi}(a) 
shows the NRG results of transmission probability of Bogoliubov particles 
$\mathcal{T}_{\mathrm{BG}}^{} =\sin^2 \delta$ 
calculated for $U/(\pi \Gamma_N^{}) = 5.0$.  
It has an isotropic structure independent of $\Theta$. 
In particular, the semi-cylindrical elevation 
of the height $\mathcal{T}_{\mathrm{BG}}^{} \simeq 1.0$ 
at $E_A \lesssim 0.5U$ corresponds 
to a rotating body of the  Kondo ridge 
shown in Fig.\ \ref{fig:NRGOneDotXi0} (b). 
On the slopes of 
this semicylindrical hill at $E_A \simeq 0.5U$,    
 it spreads over 
the valence fluctuation region of the Bogoliubov particles,  
at which the transmission probability $\mathcal{T}_{\mathrm{BG}}^{}$ 
rapidly  decreases.

Figure \ref{fig:NRGOneDotXi}(b) shows 
 the transmission probability of Cooper pairs  
 $\mathcal{T}_{\mathrm{CP}}^{}=(1/4)\sin^2 \Theta \,
\sin^2 2\delta$.   
It is enhanced along the ridge of a crescent shape 
that is spreading over the angular range
of $\pi/4 < \Theta < 3\pi/4$ (at which  $\Gamma_S>|\xi_d^{}|$) 
on the arc of radius $E_A \simeq U/2$, 
where the crossover between the Kondo-singlet and 
 the superconducting-singlet states takes place. 
The ridge height of $\mathcal{T}_{\mathrm{CP}}^{}$ 
decreases from the maximum value $0.25$ as $\Theta$ deviates 
from $\Theta = \pi/2$, showing the $\sin^2 \Theta$ dependence. 
The width of the crescent region in the radial 
direction is of the order of $\Gamma_N^{}$ 
($\simeq 0.06U$ in Fig.\ \ref{fig:NRGOneDotXi}(b) ).

The nonlocal conductance $g_{\mathrm{RL}}^{}/g_0^{}=
 2(\mathcal{T}_{\mathrm{BG}}^{}-2\mathcal{T}_{\mathrm{CP}}^{})$ 
is shown in Fig.\ \ref{fig:NRGOneDotXi}(c).  
It also 
 features 
 a flat-topped semicylindrical elevation at $E_A \lesssim U/2$, 
which is mainly due to the contributions of  the Bogoliubov-particle part  
 $\mathcal{T}_{\mathrm{BG}}^{}$ 
seen in Fig.\ \ref{fig:NRGOneDotXi}(a).   
The nonlocal conductance $g_{\mathrm{RL}}^{}$ becomes 
negative at the foot of the hill, 
specifically at $E_A \gtrsim U/2$ 
 along the arc of the range $\pi/4 < \Theta < 3 \pi /4$,  
where the CAR dominates the transport. 
In order to see more precisely 
the profile of the negative-conductance region, 
a contour plot of $g_{\mathrm{RL}}^{}$ 
is shown in Fig.\ \ref{fig:NRGOneDotXi}(d).  
The dip in the profile becomes deepest at $\Theta=\pi/2$ 
and $E_A^{}/U \simeq 0.55$, 
as seen also  
in Fig.\ \ref{fig:NRGOneDotXi0}(c).
The behavior of $g_{\mathrm{RL}}^{}$ along the $\Theta$ direction 
is determined  
by the coherence factor $\sin^2 \Theta$ 
of the Cooper-pair part $\mathcal{T}_{\mathrm{CP}}^{}$
in Eq.\ \eqref{eq:T_ET_CP_b0}.
It suppresses the CAR contributions to the nonlocal conductance 
as $\Theta$ deviates from $\pi/2$. 
The crescent-shaped dip emerged for $g_{\mathrm{RL}}^{}$   
reflecting the corresponding one 
seen in Fig.\ \ref{fig:NRGOneDotXi}(b) for
 $\mathcal{T}_{\mathrm{CP}}^{}$,  
and the dip spreads from $E_A^{}\simeq U/2$
to $E_A^{} \sim U/2+\Gamma_N^{}$  
in the direction of the $\Gamma_S^{}$-axis.  
These results suggest that the crescent dip region 
will be a plausible target 
to probe the CAR contributions in experiments.

The NRG result of the CAR efficiency  at $b=0$, 
 $\eta_{\text{CAR}}^{}= \sin^2 \Theta\,\cos^2 \delta$,  
is shown in Fig.\ \ref{fig:NRGOneDotXi}(e). 
We can see that 
its behavior is similar to that of   
the pair correlation 
described in Fig.\ \ref{fig:NRGOneDot_FLparametersxi}(a): 
the ridge of $\eta_{\text{CAR}}^{}$  
evolves at $E_A^{}\gtrsim U/2$ in the direction of $\Theta =\pi/2$ 
along the $\Gamma_S^{}$ axis. 
In the valley region at $E_A^{}\lesssim U/2$, however, 
the slope of $\eta_{\text{CAR}}^{}$ in 
the direction 
parallel to the $\xi_d^{}$ axis becomes steeper   
as it is determined by the coherence factor $\sin^2 \Theta$, 
whereas that for the pair correlation function is $\sin \Theta$. 
There are also some quantitative 
differences 
between 
the profiles of the CAR efficiency and the pair correlation function 
in the radial direction:  it is 
because $\eta_{\text{CAR}}^{} \propto \cos^2 \delta$,   
whereas $|\bigl\langle d^\dagger_\uparrow d^\dagger_\downarrow 
+  d^{}_\downarrow d^{}_\uparrow \bigr\rangle| 
\propto 1- 2\delta/\pi$.

\begin{figure}[b]
\begin{tabular}{c}
  \includegraphics[width=0.95\linewidth]{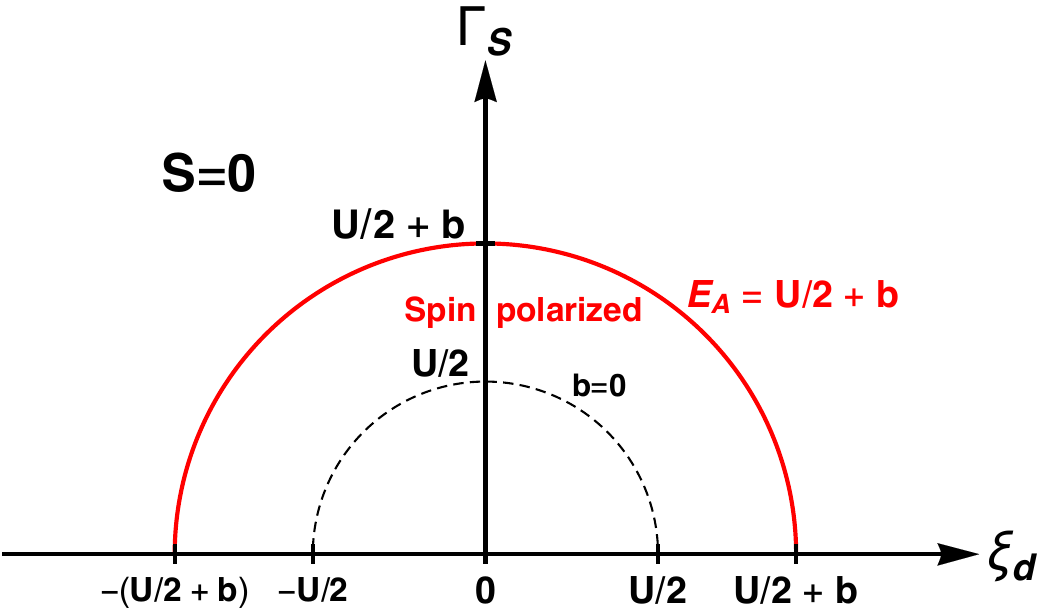}
\end{tabular}
\caption{Parameter space of $H_\mathrm{eff}^{}$ at finite magnetic fields.  
Near the semicircle 
of radius  $E_A^{}=U/2+b$  
with $E_A^{}=\sqrt{\xi_d^2+\Gamma_S^2}$,
the occupation number 
of the Bogoliubov particles 
in the 
Andreev level 
varies rapidly from $Q\simeq 1.0$ to 
$Q \simeq 0.0$ for large $U$.
Specifically, in the atomic limit $\Gamma_N ^{} \to 0$, 
 the ground state is  spin polarized  
$Q_{\uparrow}^{} \to 1.0$ inside the semicircle at finite fields $b\neq 0$.
The Andreev scattering can dominate the transport   
in the range of $\pi/4 < \Theta < 3\pi/4$ 
outside the semicircle which evolves with $b$.  
}
\label{fig:SingleDotPhaseMag}
\end{figure}

\section{Crossed-Andreev transport at finite magnetic fields $b \neq 0$}
\label{CARMag}

Both the Kondo effect and the  superconducting proximity effect  
are sensitive to a magnetic field.
Here we study precisely how the CAR contributions vary  
at finite magnetic fields.

Figure \ref{fig:SingleDotPhaseMag} 
shows   
the parameter space of the effective Hamiltonian $H_\mathrm{eff}^{}$ 
at finite magnetic fields ($b > 0$).
In the atomic limit  $\Gamma_N ^{} \to 0$, 
the phase boundary evolves with $b$,
and  the ground state of  the isolated QD is spin polarized 
inside the semicircle of radius  $E_A^{}=U/2+b$ 
where the 
Andreev level 
is occupied by a single 
Bogoliubov particle with majority spin: 
$Q_{\uparrow} \to 1.0$.
In contrast, outside the semicircle,    
the Andreev level  is empty $Q = 0$ 
and the ground state is unpolarized. 
The transition is caused by the level crossing between 
the energy level $E_{A,\uparrow} \equiv E_A^{} -b$ 
of the singly occupied majority-spin state 
and the spin-singlet empty state of energy $U/2$, 
and thus it 
takes place 
at the circumference 
of the semicircle  $E_{A,\uparrow}^{ }=U/2$.

The level crossing becomes a gradual crossover,   
the width of which is of the order of  $\Gamma_N^{}$,  
when  normal leads are connected. 
The CAR contribution to the nonlocal conductance 
is enhanced also at finite $b$ near the crossover region:  
specifically along the circumference of radius $E_A^{}\simeq U/2+b$  
over the angular range of $\pi/4 \lesssim \Theta \lesssim 3\pi/4$.  
We will consider magnetic-field dependence of the CAR contribution 
precisely in this section.

\subsection{
Ground-state and transport properties at $\Theta = \pi/2$}
\label{MagneticFieldDependence}

At finite magnetic fields, 
the renormalized parameters become dependent on 
spin components $\sigma$ 
and vary as Zeeman splitting increases. 
Our discussions in the following are based on 
the transport formulas for the ground state  
given in Eqs.\ \eqref{eq:T_ET_single}--\eqref{eq:CARefficiencyDef}.    
The NRG calculations have been carried out for a strong interaction 
 $U/(\pi \Gamma_N)=5.0$  in order to clarify
how the  electron correlations affect the crossed Andreev reflection 
in the multiterminal network.

\begin{figure}[b]
\begin{tabular}{cc}
\includegraphics[width=0.49\linewidth]{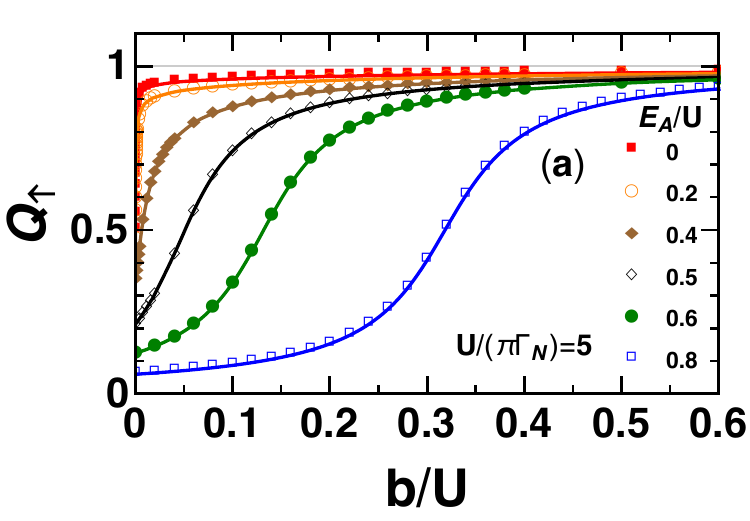}&
\includegraphics[width=0.49\linewidth]{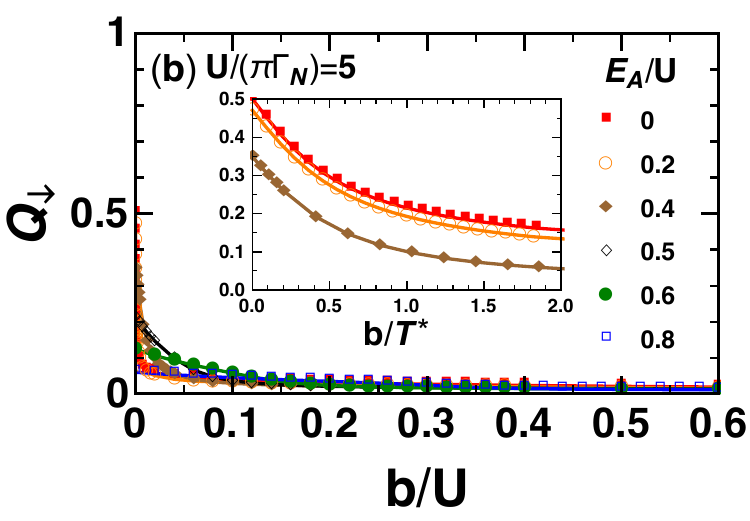}
\\
\includegraphics[width=0.49\linewidth]{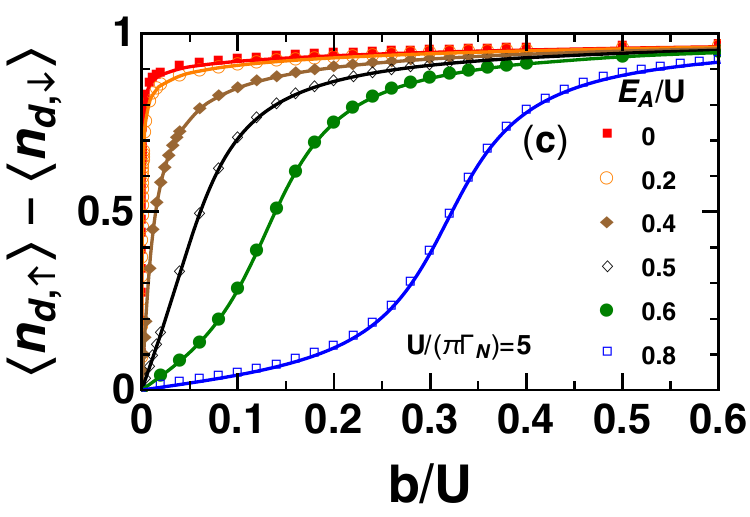}&
\includegraphics[width=0.49\linewidth]{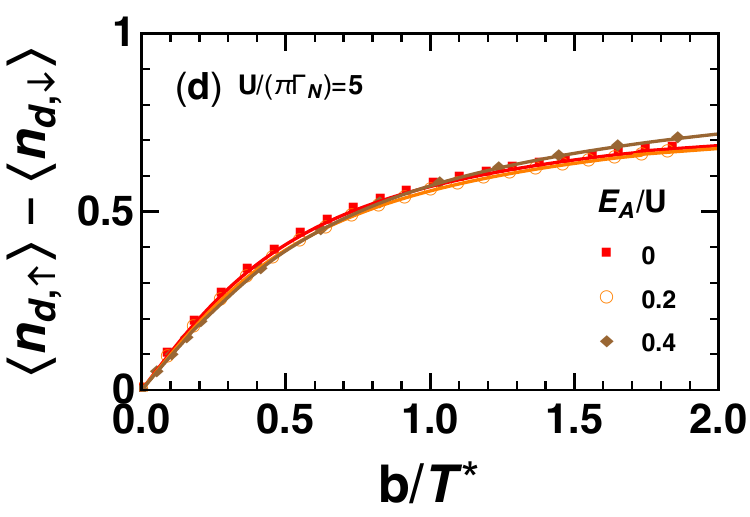}
\\
\includegraphics[width=0.49\linewidth]{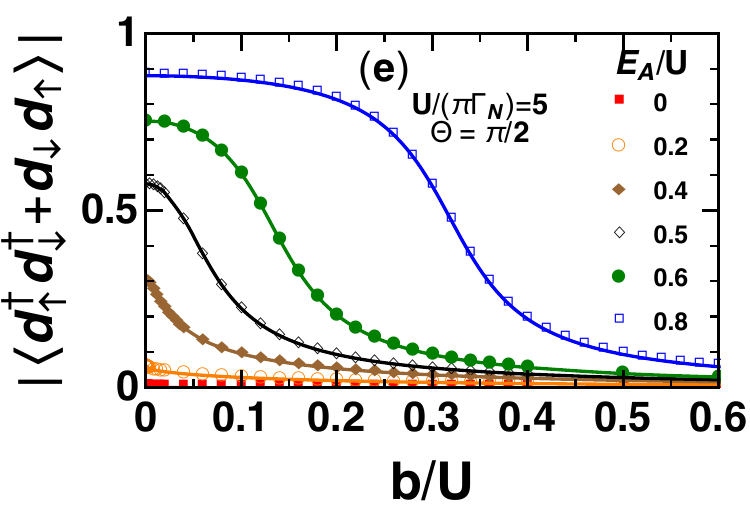}&
\includegraphics[width=0.49\linewidth]{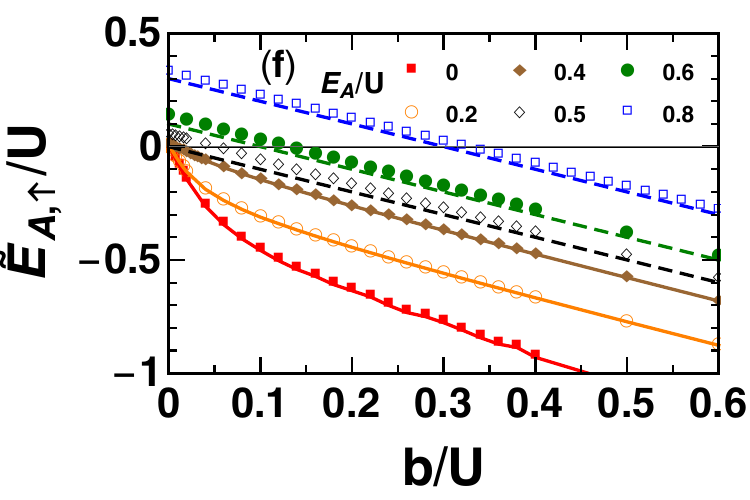}
\end{tabular}
\caption{Magnetic-field dependence of 
the 
renormalized parameters 
calculated at different Andreev-level positions 
$E_A/U = 0.0$, $0.2$, $0.4$, $0.5$, $0.6$,
and $0.8$, for a fixed interaction $U/(\pi\Gamma_N) = 5.0$. 
 (a), (b) Occupation number of Bogoliubov particles $Q_\uparrow^{}$ 
and $Q_\downarrow^{}$.
 Inset in (b) is an enlarged plot of $Q_\downarrow$ vs $b/T^*$,  
with  $T^*$  the characteristic energy scale of 
the Kondo regime defined 
 at $b=0$ in Eq.\ \eqref{eq:T_star}.
For $E_A^{}=0.0$, $0.2U$, and  $0.4U$,  
it takes the value 
 $T^*/(\pi\Gamma_N^{}) =0.002$, $0.005$, and $0.097$, respectively  
 (or $T^*/U=0.0004$, $0.001$, and  $0.019$).
The values of $Z = \widetilde{\Gamma}_N^{}/\Gamma_N^{}$ 
 at  these three points of $E_A^{}$ are given by  $Z=0.008$, $0.02$, $0.31$, 
respectively. 
(c) 
Magnetization 
$m_d^{} =\langle n_{d\uparrow}^{} \rangle - \langle n_{d\downarrow}^{}\rangle$, 
which does not depend on the Bogoliubov angle $\Theta$.
(d) Enlarged view of  $m_d^{}$ for $E_A^{}=0.0$, $0.2U$, and  $0.4U$ 
plotted vs  $b/T^*$ in the Kondo regime. 
(e) Pair correlation 
$|\bigl\langle d^\dagger_\uparrow d^\dagger_\downarrow +  d^{}_\downarrow d^{}_\uparrow \bigr\rangle |$ at  $\Theta=\pi/2$, 
which in this case is given by $1-Q$ and varies with $b$ and $E_A^{}$,
in contrast to the electron filling $\langle n_d^{}\rangle \equiv 1.0$ 
that remains unchanged along the $\Gamma_S^{}$ axis at $\xi_d^{}=0$.
(f) Renormalized Andreev levels $\widetilde{E}_{A,\uparrow}$.
The dashed straight  lines represent the Hartree-Fock (HF) result 
 $E_A^\mathrm{HF} \xrightarrow{\, E_{A}^{} \gg U/2\,}
E_A^{} - U/2 - b$. 
} 
\label{fig:NRGOneDot_FLparametersMag}
\end{figure}

\subsubsection{$b$ dependence of renormalized parameters}

Figures 
\ref{fig:NRGOneDot_FLparametersMag}(a)--\ref{fig:NRGOneDot_FLparametersMag}(f) 
show the magnetic-field dependence of the renormalized parameters,  
calculated 
for several different positions of the Andreev level:   
 $E_A^{}/U = 0.0$, $0.2$, $0.4$, $0.5$, $0.6$, and $0.8$.  
The results commonly reflect the Fermi-liquid properties 
of the Bogoliubov particles, which evolve as  $E_A^{}$ and $b$ vary.

For $0\leq E_A^{} \lesssim U/2$, 
the renormalized parameters exhibit 
a universal  $b/T^*$ scaling behavior at small magnetic fields,  
with $T^*$  the Kondo energy scale  defined  
at zero field  $b=0$ by Eq.\ \eqref{eq:T_star}. 
The magnitude of $T^*$ depends sensitively on the interaction strength 
and the level position $E_A^{}$, 
and, for instance, for $U/(\pi \Gamma_{N})=5.0$, 
 it is given by $T^*/(\pi\Gamma_N^{}) =0.002$, 
$0.005$, and $0.097$ for $E_A/U = 0.0$, $0.2$, and $0.4$,  respectively. 
At the magnetic field of order at $b \sim T^*$,  
the Kondo resonance of Bogoliubov particles 
splits into two as the Zeeman effect dominates.  
In contrast, in the  parameter region of  $E_A^{} \gtrsim U/2$      
where the Andreev level deviates further from the Fermi level, 
the renormalization effects due to the electron correlations 
are suppressed,  
and the low-energy states depend significantly 
on whether $U/2 \lesssim E_A^{} \lesssim U/2 +b$ 
or $ U/2 +b \lesssim E_A^{}$. 
The magnetization $m_d^{}$ of 
 quantum dot is almost fully polarized 
at  $U/2 \lesssim E_A^{} \lesssim U/2 +b$,   
where the Zeeman effect dominates.
In contrast, the superconducting proximity effect dominates 
outside the semicircle of radius  $E_A^{}\gtrsim U/2 +b$
in the angular range of  $\pi/4 <\Theta < 3\pi/4$. 
 We will discuss these of variations 
of the ground state properties 
 in more detail in the following.

Figures 
\ref{fig:NRGOneDot_FLparametersMag}(a)--\ref{fig:NRGOneDot_FLparametersMag}(d) 
describe  the occupation number  
 $Q_\sigma^{}$ and the magnetization 
 $m_d^{} \equiv \langle n_{d\uparrow}^{}\rangle  - 
\langle n_{d\downarrow}^{}\rangle$ 
as functions of magnetic fields. 
Note that two of them,  
Fig.\ \ref{fig:NRGOneDot_FLparametersMag}(d) 
and the inset presented for $Q_{\downarrow}^{}$ 
in Fig.\ \ref{fig:NRGOneDot_FLparametersMag}(b),  
are plotted vs  $b/T^*$ for small magnetic fields.  
We see  in Fig.\ \ref{fig:NRGOneDot_FLparametersMag}(d) 
that the magnetization $m_d^{}$ 
for $E_A^{}/U \lesssim 0.4$ exhibits the universal Kondo scaling behavior 
at $b \lesssim T^*$.  
It can also be recognized that the three curves 
for $Q_{\downarrow}$ shown in the inset 
in Fig.\ \ref{fig:NRGOneDot_FLparametersMag}(b)  
will collapse into a single universal curve 
if we introduce the offset values along the vertical axis  
which is determined at $b=0$ for each $E_A^{}$: 
note that the occupation number takes the value 
 $Q_{\sigma}^{} = 0.5$ at $E_A^{}=b=0$.

However,  
as seen in Figs.\ \ref{fig:NRGOneDot_FLparametersMag}(a) 
and \ref{fig:NRGOneDot_FLparametersMag}(c), 
the Zeeman effect dominates at  strong magnetic fields. 
Note that the magnetization can also be written as 
 $m_d^{} = Q_{\uparrow}^{} - Q_{\downarrow}^{}$ 
and does not depend on the Bogoliubov angle $\Theta$.   
The Kondo behavior disappears for $E_A^{}/U \gtrsim 0.5$, 
at which the Bogoliubov particles are in the valence fluctuation 
or empty orbital regime at $b=0$.
In this region of $E_A^{}/U$,  
the occupation number  $Q_{\uparrow}^{}$ 
 of the majority-spin component 
and the magnetization $m_d^{}$ 
show a steep increase at magnetic fields of $b \simeq E_{A}^{} -U/2$
 where the level crossing occurs.
As magnetic field increases further  $b \gtrsim E_A-U/2$,
the magnetization 
rapidly approaches the saturation value $m_d^{} \to 1.0$.

Figure \ref{fig:NRGOneDot_FLparametersMag}(e) 
shows the magnetic-field dependence of  the SC pair correlation function 
$| \bigl\langle d^\dagger_\uparrow d^\dagger_\downarrow 
+  d^{}_\downarrow d^{}_\uparrow \bigr\rangle |$  
which becomes equal to $1-Q$ in the direction of $\Theta = \pi/2$.   
While the pair correlation increases with $E_A^{}$,  
it decreases as $b$ increases.  
We can see that the SC proximity effect dominates  
at small fields  $b \lesssim E_A^{}-U/2$ 
in the parameter region of $E_A^{} \gtrsim U/2$, 
i.e, the outside of the semicircle of radius $E_A^{} \gtrsim U/2+b$ 
shown in Fig.\ \ref{fig:SingleDotPhaseMag}.
In this region, the SC pair correlation function can 
take the value  of the order of 
10\% 
of the maximum possible value  
$| \bigl\langle d^\dagger_\uparrow d^\dagger_\downarrow 
+  d^{}_\downarrow d^{}_\uparrow \bigr\rangle | = 1$, 
as seen in Fig.\ref{fig:NRGOneDot_FLparametersMag}(e) 
 at magnetic fields of the  order of $b \sim 0.1U$.
However, 
as magnetic field approaches $b \simeq E_A^{} -U/2$,   
the crossover to the Zeeman-dominated spin-polarized regime occurs, 
and the pair correlation  rapidly decreases. 
The sum of the phase shifts 
takes the value $\delta_\uparrow + \delta_\downarrow \simeq \pi/2$ 
 in the crossover region. 
Therefore, the Andreev scattering is most enhanced at this point 
since the factor $\sin^2 (\delta_\uparrow +\delta_\downarrow)$ 
that determines $\mathcal{T}^{}_{\mathrm{CP}}$  
 takes the maximum value.

Figure \ref{fig:NRGOneDot_FLparametersMag}(f) 
shows the results for the majority-spin component of the 
renormalized Andreev level  $\widetilde{E}_{A,\uparrow}^{}$  
which includes the Zeeman energy and the many-body corrections   
 defined in Eq.\ \eqref{eq:EAren}.
For  $E_A^{}\lesssim U/2$, 
the slope of  $\widetilde{E}_{A,\uparrow}^{}$ 
is steep at small magnetic fields $b \simeq 0$. 
This is because the spin susceptibility, $\chi_s^{}= m_d^{}/b$,  
is enhanced in this region by the Kondo effect as seen 
in Fig.\ \ref{fig:NRGOneDot_FLparametersMag}(c).
The slope becomes gradual, however, as $b$ increases 
and the crossover to the Zeeman-dominated regime occurs at $b\sim T^*$.
In contrast, for $E_A^{} \gtrsim U/2$, 
the renormalized level $\widetilde{E}_{A,\uparrow}^{}$ 
at small magnetic fields $b \simeq 0$  
shifts away from the Fermi level, 
and the occupation number of the Bogoliubov particles 
$Q$ decreases 
as $E_A^{}$ increases. 
However, as magnetic field increases further, 
the renormalized level 
 $\widetilde{E}_{A,\uparrow}^{}$ crosses the Fermi level  
at  $b \simeq E_A^{} -U/2$,  
and  the occupation number of the majority spin component  $Q_\uparrow^{}$ 
increases abruptly at the crossing point. 
The dashed straight lines in Fig.\ \ref{fig:NRGOneDot_FLparametersMag}(f) 
represent the Hartree-Fock energy shift 
$E_{A,\uparrow}^\mathrm{HF}$, 
which asymptotically takes the following form at  $E_{A}^{} \gg U/2$:  
\begin{align}
& E_{A,\uparrow}^\mathrm{HF} 
\, \equiv \,  
E_{A}^{} - b
\,+\, U\left( Q_{\downarrow}^{} -\frac{1}{2}\right) 
\nonumber 
\\
&  \ \   \xrightarrow{\, E_{A}^{} \gg U/2\,} \,  
E_{A}^{} - \frac{U}{2} - b  \,.
\label{eq:HF_energy_shift_mag}
\end{align}
The NRG results 
for 
 $\widetilde{E}_{A,\uparrow}^{}$ 
and the Hartree-Fock energy shifts show 
a close agreement for  $E_A^{}\gtrsim U/2$. 
This is caused by the fact that 
the renormalization factor approaches  $Z_\sigma^{}\simeq 1.0$  
and 
it becomes less important  at the crossover region 
between the Zeeman-dominated spin-polarized regime 
and the SC-dominated regime.

\begin{figure}[b]
 \begin{tabular}{cc}
\includegraphics[width=0.5\linewidth]{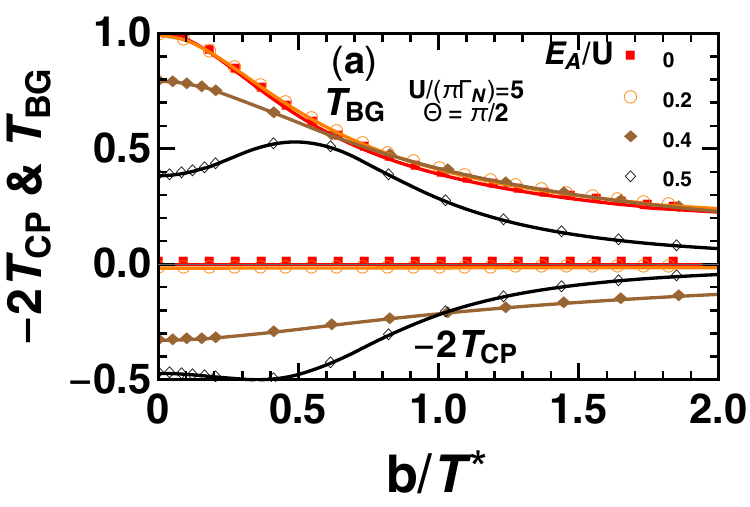}&
\includegraphics[width=0.5\linewidth]{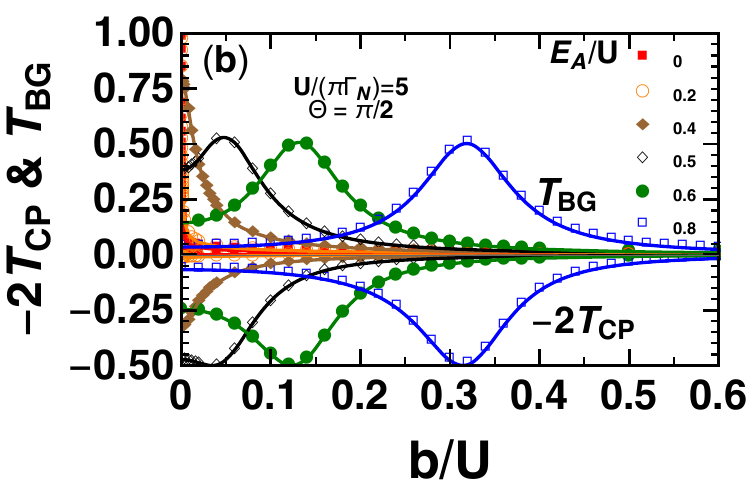}\\
\includegraphics[width=0.5\linewidth]{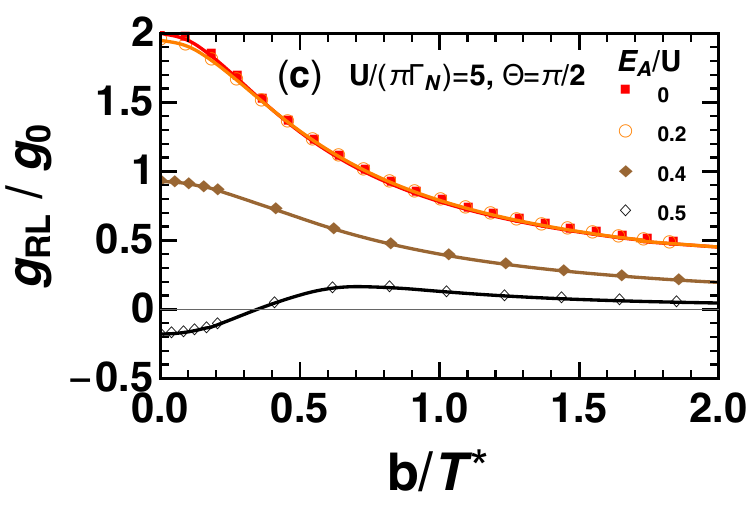}&
\includegraphics[width=0.5\linewidth]{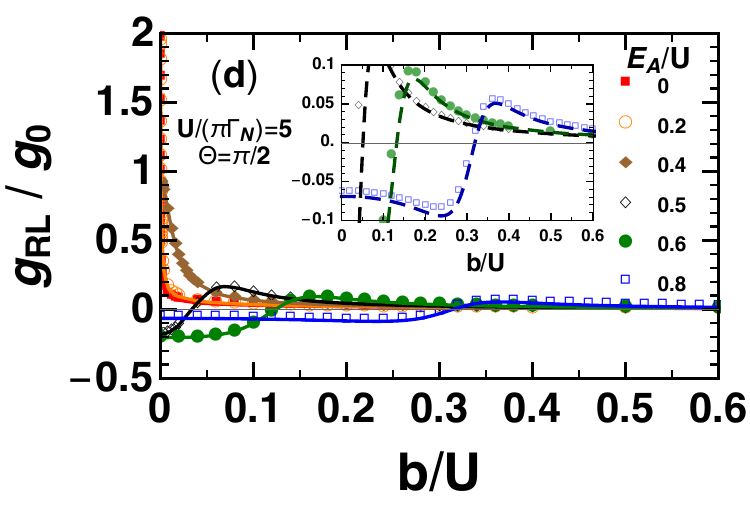}\\
\includegraphics[width=0.5\linewidth]{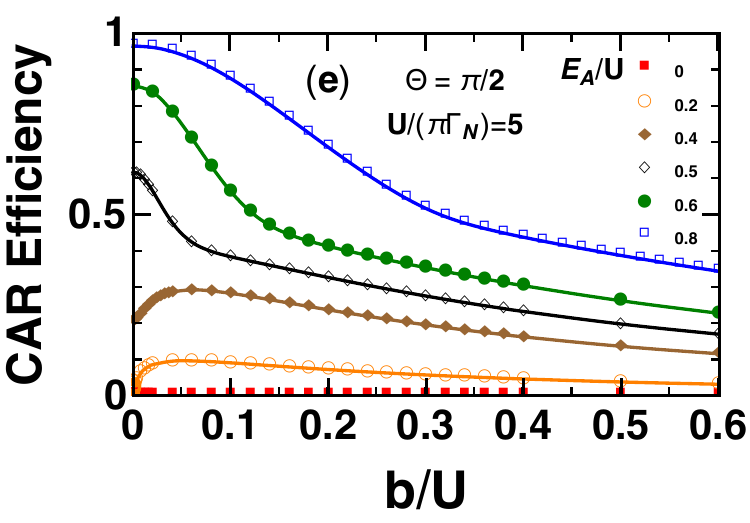}&
\includegraphics[width=0.5\linewidth]{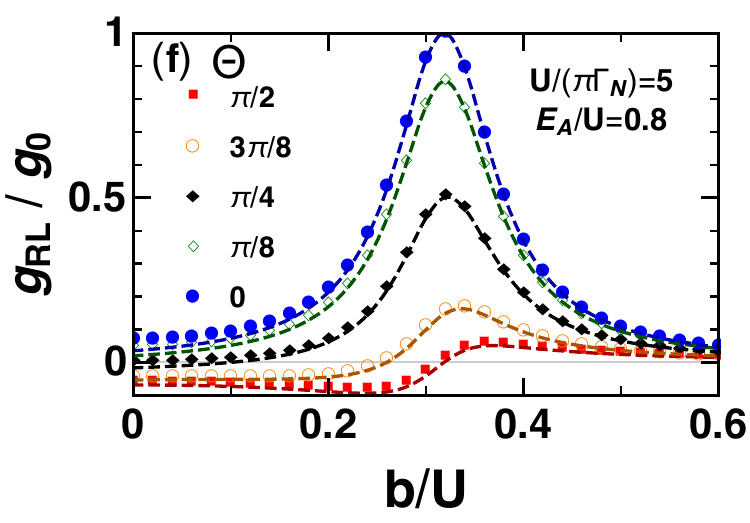}
\end{tabular}
\caption{Magnetic-field dependence of  
the transport coefficients at $\Theta = \pi/2$ 
for different positions 
of  $E_A/U = 0$, $0.2$, $0.4$, $0.5$, $0.6$,  and 
$0.8$, for a strong interaction  $U/(\pi\Gamma_N) = 5.0$.
Top panel describes  $\mathcal{T}_{\mathrm{BG}}^{} 
=\sum_\sigma \sin^2 {\delta_{\sigma}^{}}/2$,  
and $\Theta$ dependent part $-2\mathcal{T}_{\mathrm{CP}}^{}$ 
for (a) small $E_A^{}\lesssim 0.5U$  
and  (b) large  $E_A^{}\gtrsim 0.5U$.  
(c), (d) Nonlocal conductance $g_{\mathrm{RL}}^{} / g_0^{}$.
Inset in (d) is an enlarged view in the region around 
$g_{\mathrm{RL}}^{}/g_0 \approx 0.0$:  
the dashed lines 
represent the perturbation results obtained 
with Eq.\ \eqref{eq:gtot_expended_to_first_order}. 
The characteristic energy is given by 
 $T^*/U=0.0004$, $0.001$,   $0.019$, and 0.097   
for $E_A^{}/U=0.0$, $0.2$,  $0.4$, and $0.5$,  respectively. 
(e) CAR efficiency $\eta_{\text{CAR}}^{}$.
(f) $g_{\mathrm{RL}}^{} / g_0^{}$ at a 
fixed Andreev level position $E_A^{}=0.8U$ 
for several different angles 
$\Theta = 0$, $\pi/8$, $\pi/4$, $3\pi/8$, $\pi/2$: 
the dashed lines here also represent the perturbation results.
}
\label{fig:NRGOneDotMag}
\end{figure}

\subsubsection{$b$ dependence of transport properties at $\Theta =\pi/2$}

We next discuss the magnetic-field dependence of 
transport coefficients in the direction of $\Theta = \pi/2$, i.e., 
along the $\Gamma_S^{}$ axis  ($\xi_d = 0$).
The NRG results are shown in Fig.\ \ref{fig:NRGOneDotMag}:    
the magnetic field $b$ is scaled by $T^*$ 
in two of the panels \ref{fig:NRGOneDotMag}(a) and \ref{fig:NRGOneDotMag}(c),   
whereas the other panels are plotted vs $b/U$. 

We see in Fig.\ \ref{fig:NRGOneDotMag}(a) 
that the transmission probabilities  
of Bogoliubov particles $\mathcal{T}_{\mathrm{BG}}^{}$ 
 for $E_A^{} \lesssim 0.2U$ collapse into a single curve at 
small magnetic fields $b \lesssim T^*$,
showing a universal $b/T^*$  Kondo scaling behavior, 
whereas the probability of the Cooper-pairs  
 $\mathcal{T}_{\mathrm{CP}}^{}$  
is suppressed in this region. 
Correspondingly, the nonlocal conductance $g_{\mathrm{RL}}^{}$   
exhibits the scaling behavior for $E_A^{} \lesssim 0.2U$, 
as shown in Fig.\ \ref{fig:NRGOneDotMag}(c). 
The results for $\mathcal{T}_{\mathrm{BG}}^{}$ 
and $g_{\mathrm{RL}}^{}$ at $E_A^{} = 0.4U$ 
still show a similar monotonic decrease  
although they did not collapse into the universal curves.
Therefore, 
the CAR efficiencies $\eta_{\mathrm{CAR}}
=\mathcal{T}_{\mathrm{CP}}^{}/\mathcal{T}_{\mathrm{BG}}^{}$ 
 for $E_A^{}/U = 0.2$ and $0.4$, 
described in Fig.\ \ref{fig:NRGOneDotMag}(e),  
increase clearly with $b$ at small magnetic fields near $b \simeq 0$.  
It shows that the Zeeman splitting suppresses the Kondo correlations 
and assists the contributions of the Cooper-pair tunneling.

In contrast, 
when  the Andreev level situates further away 
from the Fermi level  $E_A^{} \gtrsim  U/2$,   
the ground state evolves from 
the superconductivity-dominated regime  
to the Zeeman dominated spin-polarized regime, 
as magnetic field increases.  
In particular, at  $b \simeq E_A^{} -U/2$, i.e., 
the crossover region between these two regimes, 
the transmission probability of the Bogoliubov particles 
$\mathcal{T}_{\mathrm{BG}}^{} = 
\left( \sin^2\delta_\uparrow +\sin^2\delta_\downarrow\right)/2$ 
has a peak, which emerges in Fig.\ \ref{fig:NRGOneDotMag}(b),  
as the phase shifts take the value    
$\delta_\uparrow \simeq \pi/2$ and $\delta_\downarrow \simeq 0$. 
Similarly, the Cooper-pair contribution 
$\mathcal{T}_{\mathrm{CP}}^{}= (1/4)\sin^2 \Theta\, 
\sin^2 \bigl(\delta_{\uparrow}+ \delta_{\downarrow}) $ 
reaches the maximum value $1/4$ 
at  $b \simeq E_A^{} -U/2$ in the crossover region.  
This is consistent with the previous work that examined an N-QD-SC system 
with the modified second-order perturbation theory. 
\cite{YAMADA2007265} 
It revealed the fact that the Andreev scattering 
is significantly enhanced under the condition that 
the renormalized parameters satisfy 
$\prod_{\sigma}(\widetilde{E}_{A,\sigma}
/\widetilde{\Gamma}_{N,\sigma})
=1$: 
  this can be rewritten into the form 
$\cot \delta_{\uparrow}^{}\cot \delta_{\downarrow}^{}=1$ 
and is fulfilled  
at $\delta_{\uparrow}^{}+\delta_{\downarrow}^{}=\pi/2$.

We can see in Fig.\ \ref{fig:NRGOneDotMag}(d)  that, for $E_A^{}  \gtrsim  U/2$, 
 the nonlocal conductance 
$g_{\mathrm{RL}}^{}=\mathcal{T}_{\mathrm{BG}}^{} 
- 2 \mathcal{T}_{\mathrm{CP}}^{}$ 
becomes negative in the SC-dominated regime  
$b \lesssim  E_A^{} -U/2$, 
whereas $g_{\mathrm{RL}}^{}$ takes a positive value  
in the Zeeman-dominated regime 
$b \gtrsim  E_A^{} -U/2$.  
In particular,   for $E_A^{}\gtrsim 0.6U$, 
the nonlocal conductance 
forms 
a flat valley structure at $0 \leq b \lesssim  E_A^{} -U/2$  
the bottom of which is negative and is less sensitive to $b$. 
This is caused by the fact that 
the peak of $\mathcal{T}_{\mathrm{BG}}^{}$
and the dip  of $-2\mathcal{T}_{\mathrm{CP}}^{}$ 
move almost synchronously, in Fig.\ \ref{fig:NRGOneDotMag}(b),  
as $E_A^{}$ increases from $0.5U$.  
For observing the flat valley structure of $g_{\mathrm{RL}}^{}$, 
the depth of which should not be too shallow, 
and thus $E_A^{}-U/2$ should be of the order of  $\Gamma_N^{}$, 
 or should not be too much  larger than $\Gamma_N^{}$.  
Note that, 
in this magnetic-field region $0 \leq b \lesssim  E_A^{} -U/2$,
the occupation number of the Bogoliubov particles with the minority spin
 is almost empty 
 $Q_\downarrow^{} \simeq 0.0$ and 
the transport coefficients 
are determined by the majority-spin component $Q_\uparrow^{}$.
In order to verify this quantitatively, 
we expand the nonlocal conductance into the following form, 
keeping the terms up to the first order with respect to $\delta_{\downarrow}^{}$,
\begin{align}
g_\mathrm{RL}^{}
\, \approx & \ 
\, g_0^{}
\Bigl[\, 
\cos^2\Theta \, \sin^2 \delta_{\uparrow}^{} 
-  
\delta_{\downarrow}^{}
\sin^2\Theta \,
\sin 2\delta_{\uparrow}^{} 
\,\Bigr]
 \, + \,O\! \left(\delta_{\downarrow}^{2} \right) 
\nonumber \\ 
\xrightarrow{\,\Theta = \frac{\pi}{2}\,} & \  
g_0^{}\,\left[\, 
-  \delta_{\downarrow}^{} \,
\sin 2\delta_{\uparrow}^{}  \, + \,O\! \left(\delta_{\downarrow}^{2} \right) 
\,\right]
\,. 
\label{eq:gtot_expended_to_first_order}
\end{align}
The dashed lines plotted in the inset of Fig.\ \ref{fig:NRGOneDotMag}(d) 
are the results evaluated with Eq.\ \eqref{eq:gtot_expended_to_first_order}, 
using the NRG results for $\delta_{\sigma}^{}$. 
These results show close agreement with the full NRG calculations  
of $g_\mathrm{RL}^{}$ plotted with the symbols, i.e.,  
for $E_A^{}/U = 0.6$ ($\bullet$) and $0.8$ ($\square$).

So far, we have considered behaviors 
along the angular direction of $\Theta = \pi/2$.
Figure \ref{fig:NRGOneDotMag}(f) compares 
the magnetic-field dependence of  $g_\mathrm{RL}^{}$
 for several different angles $\Theta$,  
keeping the Andreev-level position at $E_A^{} =0.8U$. 
The dashed lines, 
which also show nice agreement with the full NRG results (symbols)
of $g_\mathrm{RL}^{}$ in this figure,   
 are the perturbation results obtained from Eq.\ \eqref{eq:gtot_expended_to_first_order}.  
We find that the flat structure with negative $g_\mathrm{RL}^{}$ 
remains for $\Theta =3 \pi/8$, where the angle largely deviates from $\pi/2$.
As $\Theta$ derives further, however, 
$g_\mathrm{RL}^{}$ becomes positive  
at $0<\Theta < \pi/4$, or  $3\pi/4 < \Theta <\pi$. 
Note that the $\Theta$ dependence  
enters the nonlocal conductance through the coherence factor $\sin^2 \Theta$ 
in $\mathcal{T}_{\mathrm{CP}}^{}$, and thus  
 $g_\mathrm{RL}^{}$ 
is symmetrical with respect to the $\Gamma_S^{}$-axis 
in parameter space shown in Fig.\ \ref{fig:SingleDotPhaseMag}.

In the SC-dominated regime $0\leq  b \lesssim E_A^{} -U/2$,  
both components of the phase shift approach zero 
as $E_A^{}$ increases keeping $b$ unchanged, i.e., 
$\delta_\uparrow^{} \simeq 0$ and $ \delta_\downarrow^{} \simeq 0$ 
as seen in Figs.\ \ref{fig:NRGOneDot_FLparametersMag}(a) and 
\ref{fig:NRGOneDot_FLparametersMag}(b). 
The CAR efficiency $\eta_{\mathrm{CAR}}^{}$ is enhanced in this region 
although it decreases as $b$ increases,   
as seen in Fig.\ \ref{fig:NRGOneDotMag}(e)  for $E_A/U=0.6$ and $0.8$.
In particular, for $E_A^{} \gg U$, the efficiency approaches 
saturation value $\eta_{\mathrm{CAR}} \to 1.0$ 
at small magnetic fields near $b \simeq 0.0$.

\subsection{The characteristics of CAR  along the \\ 
  polar coordinates $E_A^{}$ and $\Theta$ 
at $b \neq 0$}

\label{MagneticFieldGateDependence}

\begin{figure}[b]
 \begin{tabular}{ccc}
\includegraphics[width=0.45\linewidth]{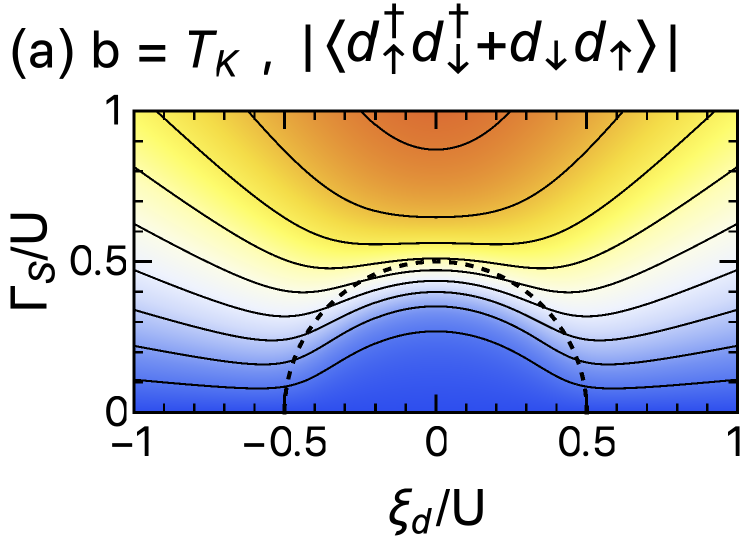}&
\includegraphics[width=0.45\linewidth]{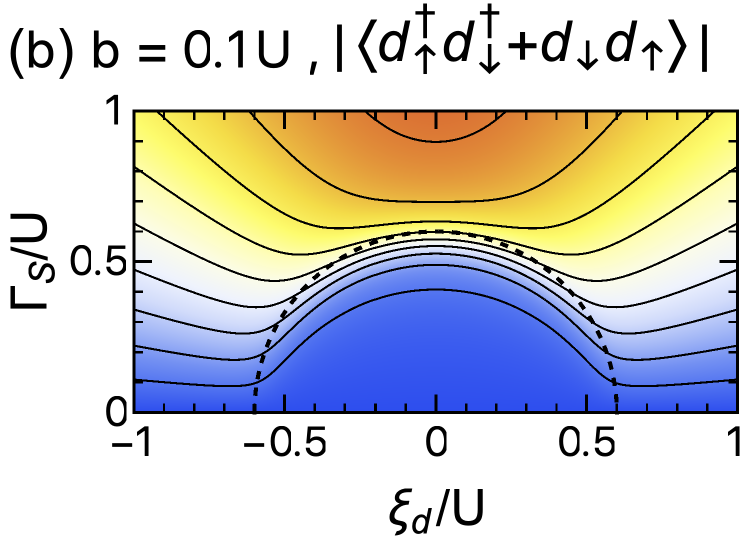}&
\includegraphics[width=0.05\linewidth]{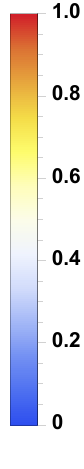}\\
\end{tabular}
\caption{Contour maps of  
 $|\bigl\langle d^\dagger_\uparrow d^\dagger_\downarrow 
+  d^{}_\downarrow d^{}_\uparrow \bigr\rangle |$ 
at finite magnetic fields: (a) $b = T_K$ and (b) $b = 0.1U$,  
for $U/(\pi\Gamma_N) = 5.0$. 
Here, $T_K=0.0004U$ is defined as the value of $T^\ast$ at $E_A=b=0$.
The contours are drawn with 0.1 increments, 
and the dashed line represents the semicircle of radius $E_A = U/2 + b$.
}
\label{fig:NRGOneDot_FLparametersMagGate}
\end{figure}

\begin{figure*}[t]
\begin{tabular}{ccc}
\includegraphics[width=0.22\linewidth]{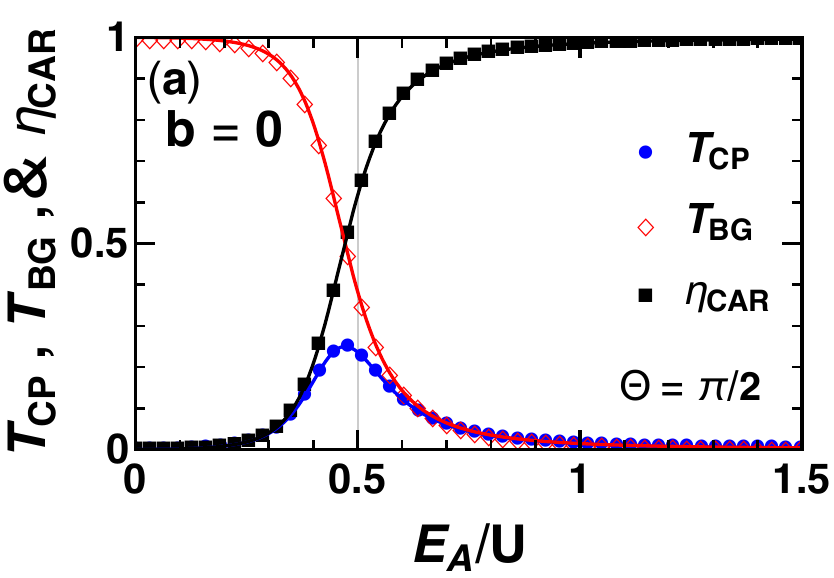}&
\includegraphics[width=0.22\linewidth]{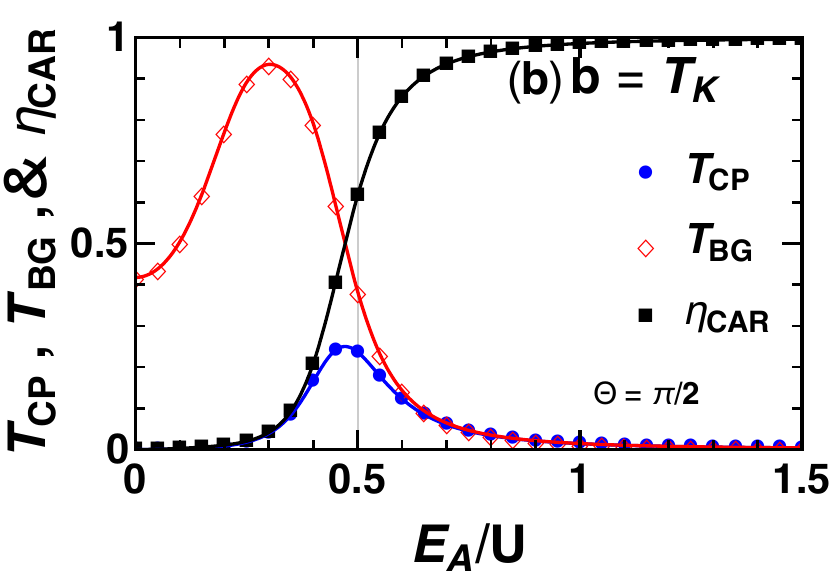}&
 \includegraphics[width=0.22\linewidth]{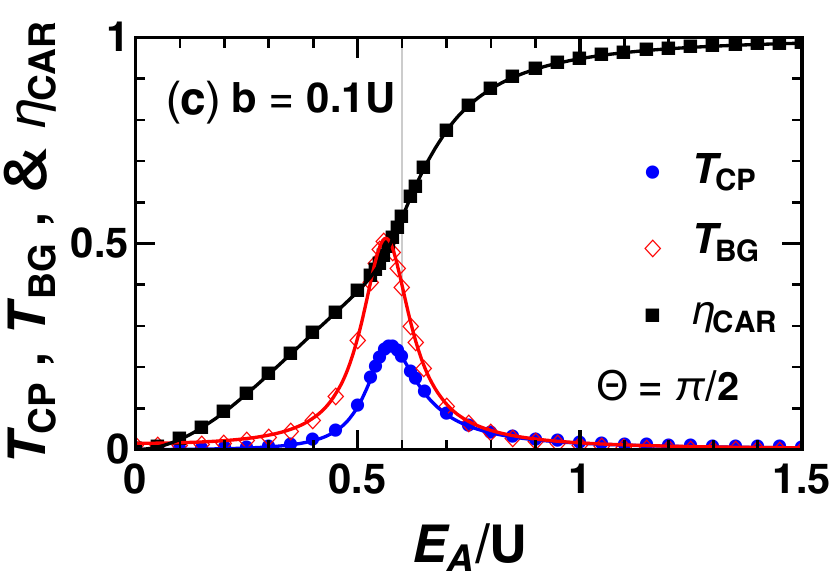}
 \end{tabular}
\begin{tabular}{cc}
 \begin{tabular}{ccc}
\includegraphics[width=0.27\linewidth]{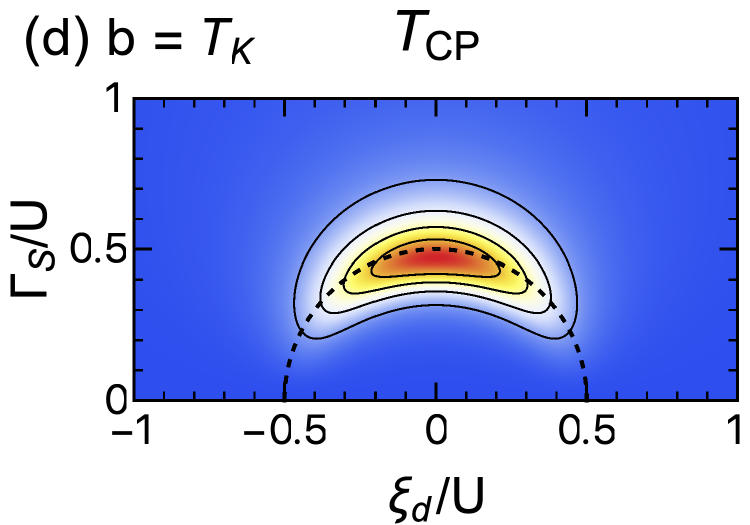}&
\includegraphics[width=0.27\linewidth]{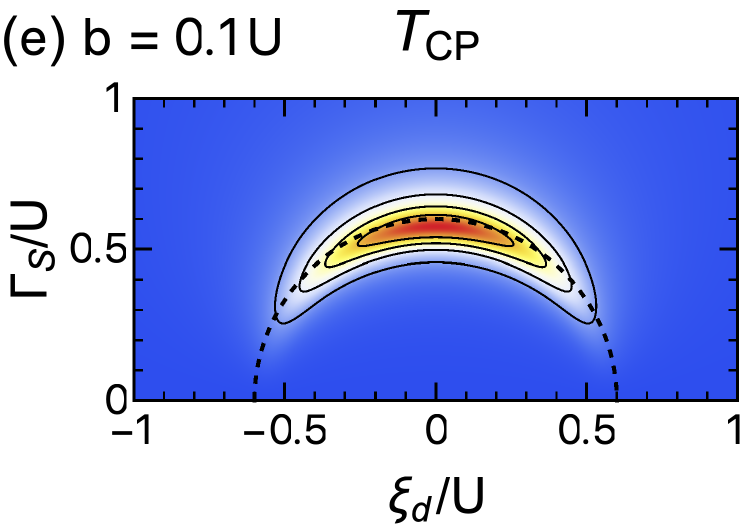}&
\includegraphics[width=0.03\linewidth]{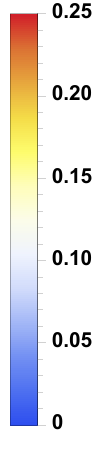}
 \\
\includegraphics[width=0.27\linewidth]{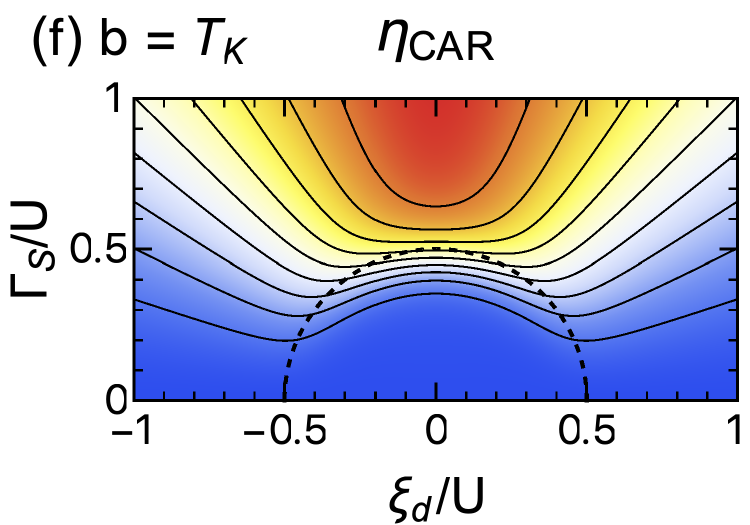}&
\includegraphics[width=0.27\linewidth]{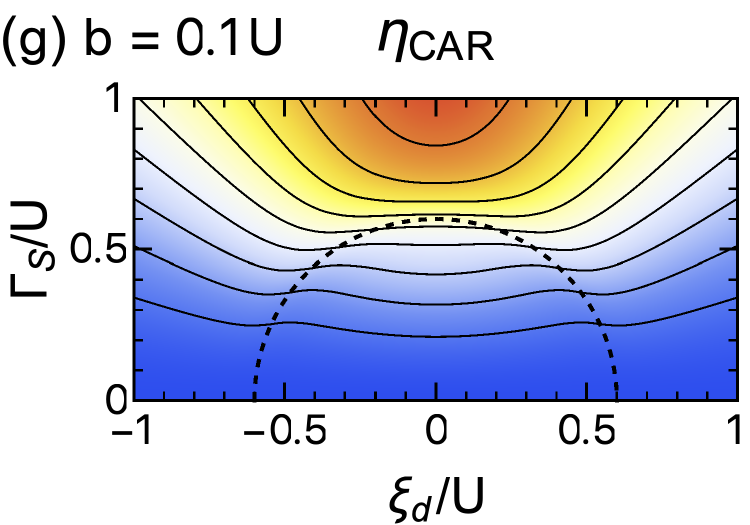}&
\includegraphics[width=0.03\linewidth]{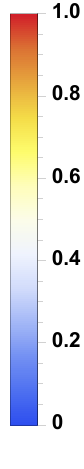}
\\
\includegraphics[width=0.27\linewidth]{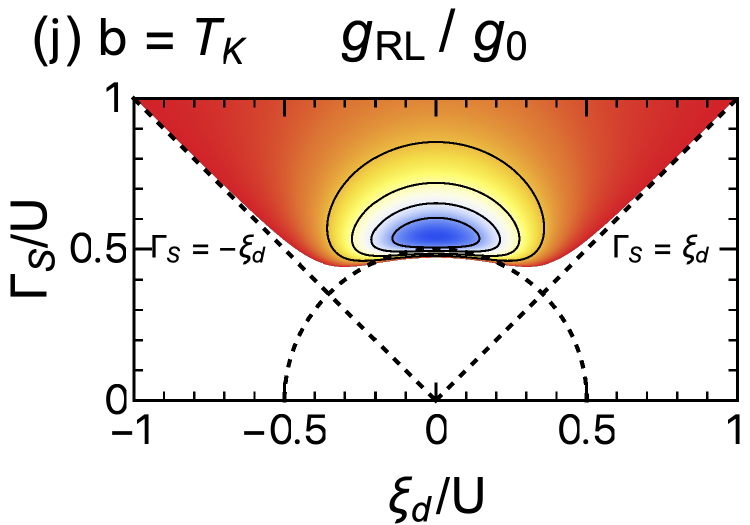}&
 \includegraphics[width=0.27\linewidth]{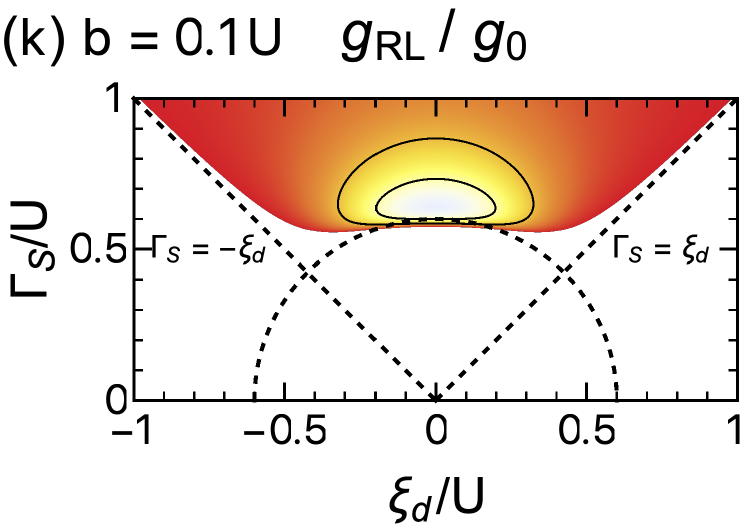}&
 \includegraphics[width=0.04\linewidth]{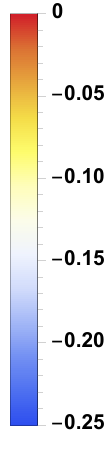}
\end{tabular}&
\begin{tabular}{c}
\includegraphics[width=0.3\linewidth]{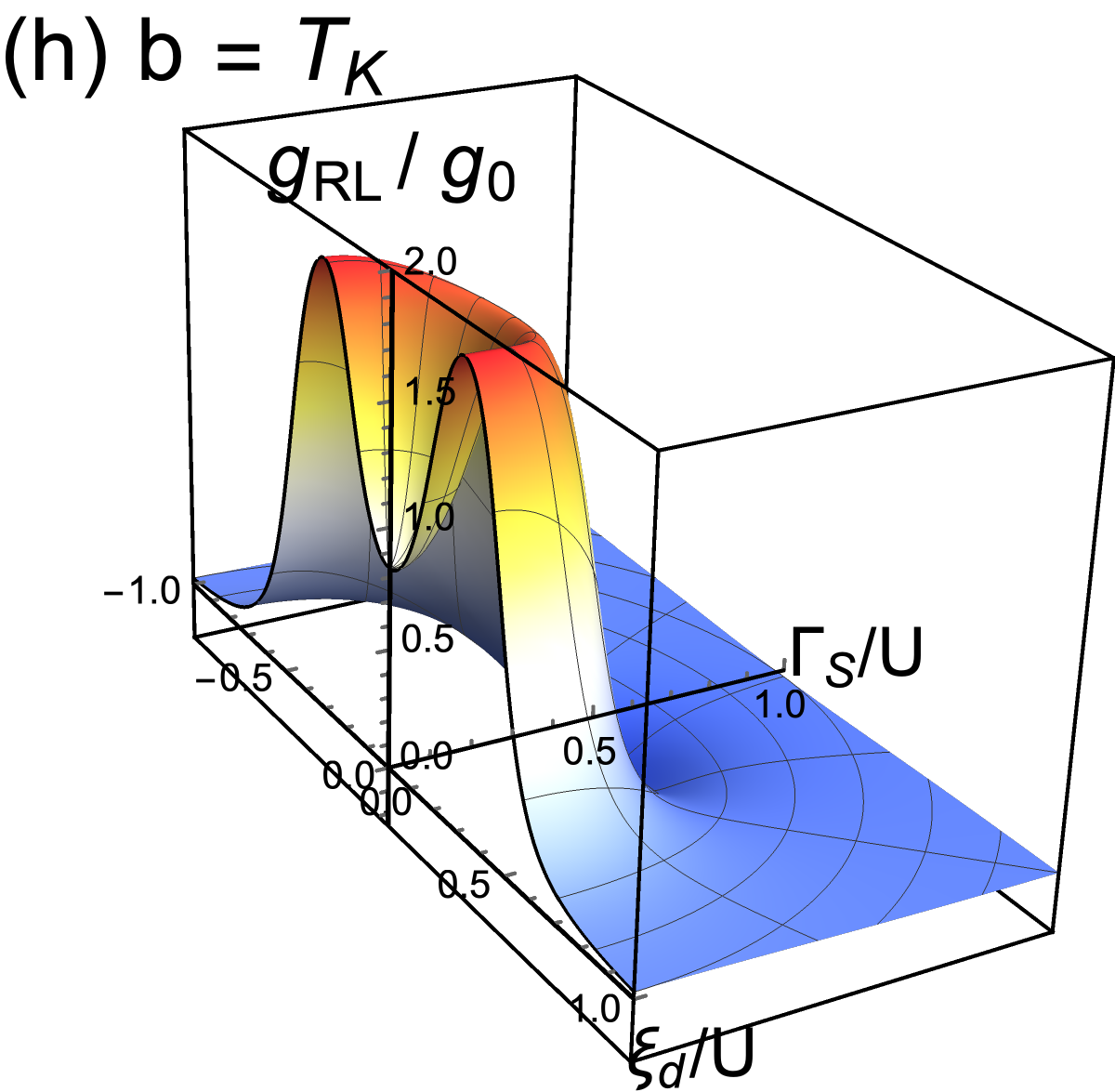}\\
\includegraphics[width=0.3\linewidth]{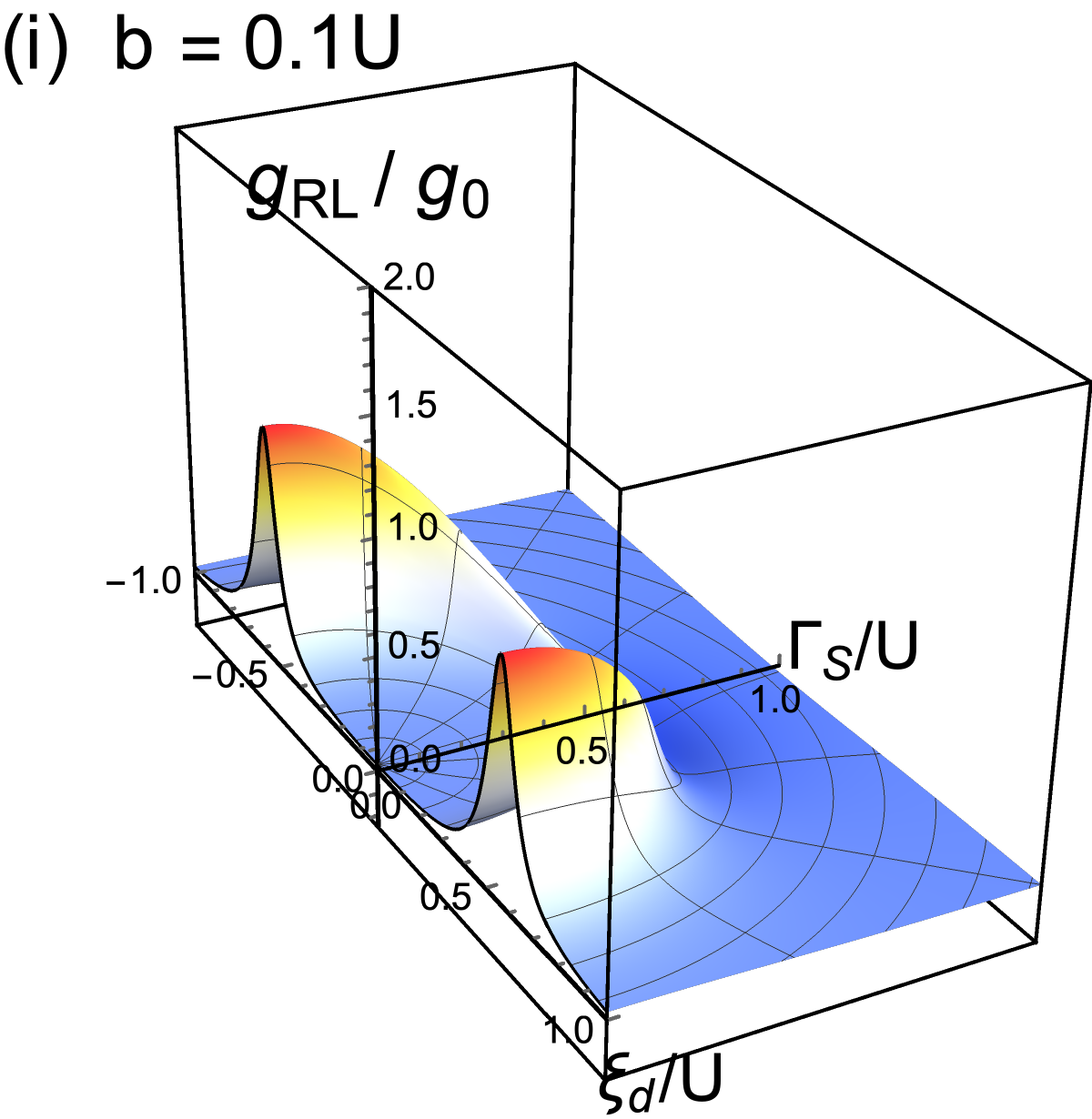}	
\end{tabular}
\end{tabular}
\caption{Bogoliubov-angle dependence of 
transport coefficients in a strong interaction case $U/(\pi\Gamma_N) = 5.0$ 
at small $b = T_K$ and large $b = 0.1U$ fields,   
with  $T_K/U=0.0004$ defined as the value of $T^\ast$ at $E_A=b=0$.
In the  top panels for (a) $b=0$,  (b) $b=T_K^{}$, and (c) $b=0.1U$, 
the coefficients  
$\mathcal{T}_{\mathrm{BG}}^{}$,  $\mathcal{T}_{\mathrm{CP}}^{}$, 
and CAR efficiency $\eta_{\mathrm{CAR}}^{}$ at $\Theta = \pi/2$ 
are plotted vs  $E_A^{}/U$.   
(d), (e) Contour maps of $\mathcal{T}_{\mathrm{CP}}^{}$, 
drawn with 0.05 increments.
(f), (g) Contour maps of $\eta_{\mathrm{CAR}}^{}$,
drawn with 0.1 increments.
(h)--(k) Nonlocal conductance  $g_{\mathrm{RL}}^{}/g_0^{}$.
In particular, (j) and (k) are the contour maps 
for negative conductance region $g_{\mathrm{RL}}^{}<0$, 
for which contours are drawn 
with $0.05$ increments. 
The dashed semicircle of radius $E_A = b + U/2$ 
in (d)--(g) and (j)--(k) corresponds to the one in Fig.\ \ref{fig:SingleDotPhaseMag}.
}
\label{fig:NRGOneDotMagGate}
\end{figure*}

In this subsection, we consider the $\Theta$ dependence 
of the transport properties at finite magnetic fields in more detail.  
Specifically, 
in order to clarify in which situations the CAR contribution 
can dominate the nonlocal conductance by overcoming 
the disturbance of the SC proximity effects 
by the Coulomb interaction and magnetic field,
we explore   a wide range of the parameter space, 
i.e., the  $\xi_d^{}$ vs $\Gamma_S^{}$ plane.      
Our discussion here is based on the NRG results in  
Figs.\ \ref{fig:NRGOneDot_FLparametersMagGate} 
and \ref{fig:NRGOneDotMagGate}, 
 obtained for a relatively large interaction  $U/(\pi \Gamma_N) = 5.0$:  
the Kondo temperature in this case is given by 
 $T_K/U = 0.0004$, 
which is defined as the value of 
the characteristic energy scale  $T^\ast$ 
at  $E_A^{}=b=0$.  
These results 
can be compared with {those} for zero field  
presented in Figs.\ \ref{fig:NRGOneDot_FLparametersxi} 
and \ref{fig:NRGOneDotXi}.

\subsubsection{$\Theta$ dependence 
of $\bigl\langle d^\dagger_\uparrow d^\dagger_\downarrow 
+  d^{}_\downarrow d^{}_\uparrow \bigr\rangle $ at $b \neq 0$}

Figure\ \ref{fig:NRGOneDot_FLparametersMagGate} shows 
the pair correlation 
$|\bigl\langle d^\dagger_\uparrow d^\dagger_\downarrow 
+ d^{}_\downarrow d^{}_\uparrow \bigr\rangle | = (1-Q) \sin \Theta$  
for two different magnetic-field strengths: (a) $b = T_K$ and (b) $b = 0.1U$.  
Here, the occupation number $Q$ of Bogoliubov particles 
does not depend on $\Theta$ but varies with $b$ and $E_A^{}$, 
as mentioned and shown in Figs.\ 
\ref{fig:NRGOneDot_FLparametersMag}(a) and 
\ref{fig:NRGOneDot_FLparametersMag}(b).

The pair correlation function for a small magnetic field $b = T_K$, 
shown in Fig.\ \ref{fig:NRGOneDot_FLparametersMagGate}(a),  
is enhanced  
outside the semicircle of radius $E_A^{} \gtrsim U/2 + b$   
in the angular range of  $\pi/4 < \Theta < 3\pi/4$, 
especially along the $\Gamma_S^{}$-axis ($\Theta = \pi/2$),  
where  the SC proximity effects dominate. 
In contrast, it is suppressed by the Kondo effect  
inside the semicircle $E_A^{} \lesssim U/2 + b$.
Note that $b$ is much smaller than $U$ in this case
 ($T_K/U=0.0004$). 

Figure \ref{fig:NRGOneDot_FLparametersMagGate}(b) shows  
the pair correlation function for a large magnetic field $b= 0.1U$. 
Although it is qualitatively similar 
to Fig.\ \ref{fig:NRGOneDot_FLparametersMagGate}(a),  
we can see that  the slope just inside of the circumference 
becomes steeper than that for $b= T_K$. 
The radius of the dashed semicircle at the crossover region 
in this case is $E_A \simeq U/2 +b$ ($= 0.6U$), 
and thus the expansion 
of the circumference due to $b$ becomes  
visible in Fig.\ \ref{fig:NRGOneDot_FLparametersMagGate}(b).

\subsubsection{$\Theta$  dependence of transport properties at $b \neq 0$}

The top panels of Fig.\ \ref{fig:NRGOneDotMagGate}
show  $\mathcal{T}_{\mathrm{BG}}^{}$,  
 $\mathcal{T}^{}_{\mathrm{CP}}$,  
and $\eta_{\mathrm{CAR}}^{}$ as functions of $E_A^{}$  
for three different magnetic fields:  (a) $b=0$, (b) $b=T_K$, and (c) $b=0.1U$, 
taking the Bogoliubov angle to be $\Theta = \pi/2$, i.e., 
the direction in which the SC proximity effect is most enhanced. 
We can see that, as $b$ increases, 
 the peak of the Cooper-pair tunneling part 
$\mathcal{T}^{}_{\mathrm{CP}}$, 
emerging at $E_A^{} \simeq U/2 + b$,  
moves with the crossover region 
 towards the larger $E_A^{}$ side along the horizontal axis.   
The peak height is $1/4$ and 
the width becomes of the order of $\Gamma_N^{}$ 
 ($\simeq 0.06U$ in this case).   
The Bogoliubov-particle part  $\mathcal{T}_{\mathrm{BG}}^{}$ 
exhibits the flat Kondo plateau at zero field, 
plotted in Fig.\ \ref{fig:NRGOneDotMagGate}(a) for comparison. 
However, 
the Zeeman splitting dominates at magnetic fields of the order of $b \simeq T_K$  
and the top of the Kondo plateau 
caves in 
around $E_A^{}\simeq 0.0$, 
as seen in Fig.\ \ref{fig:NRGOneDotMagGate}(b).
As magnetic field increases further, 
the peak of $\mathcal{T}_{\mathrm{BG}}^{}$  
in Fig.\ \ref{fig:NRGOneDotMagGate}(c) 
becomes small and approaches  
the peak of $\mathcal{T}_{\mathrm{CP}}^{}$ 
that situates close to the crossover region.

The CAR efficiency 
 $\eta_{\mathrm{CAR}}^{}=\mathcal{T}_{\mathrm{CP}}^{}/\mathcal{T}_{\mathrm{BG}}^{}$ plotted in Fig.\ \ref{fig:NRGOneDotMagGate}(c)
 takes relatively large value $ 0.1 \lesssim \eta_{\mathrm{CAR}}^{} \lesssim 0.5$ 
even at $E_A \lesssim U/2 + b$.    
 Such a behavior is not seen for small $b$ 
in Figs.\ \ref{fig:NRGOneDotMagGate}(a) and \ref{fig:NRGOneDotMagGate}(b), 
and reflects the suppression of $\mathcal{T}_{\mathrm{BG}}^{}$ 
 caused by a large magnetic field $b=0.1U$. 
Outside the crossover region  $E_A \gtrsim U/2 + b$, 
however,  $\eta_{\mathrm{CAR}}^{}$  
shows a similar behavior in  Figs.\ 
\ref{fig:NRGOneDotMagGate}(a)--\ref{fig:NRGOneDotMagGate}(c): 
it approaches the saturation value 
 $\eta_{\mathrm{CAR}}^{}\to 1.0$ as $E_A^{}$ increases.

The Bogoliubov angle $\Theta$ enters the nonlocal conductance $g_{\mathrm{RL}}^{}$
through $\mathcal{T}_{\mathrm{CP}}^{}$  
since the Bogoliubov part $\mathcal{T}_{\mathrm{BG}}^{}$ is independent of it.  
Figures \ref{fig:NRGOneDotMagGate}(d) and  \ref{fig:NRGOneDotMagGate}(e) 
show the contour maps of $\mathcal{T}_{\mathrm{CP}}^{}$    
described on the $\xi_d^{}$ vs $\Gamma_S^{}$ plane, 
for magnetic fields of (d) $b = T_K$ and (e) $b = 0.1U$. 
The Cooper-pair tunneling part $\mathcal{T}_{\mathrm{CP}}^{}$ 
 is enhanced along in the crescent-shaped region  
on the arc of radius $E_A = U/2 + b$ 
in the angular range of $\pi/4 < \Theta < 3\pi/4$. 
The crescent region spreads over the direction of the $\Gamma_S^{}$-axis 
with the width of the order of $\Gamma_N^{}$ ($\simeq 0.06U$ in this case).  
As the magnetic field increases,  the crescent region moves 
upward along the $\Gamma_S^{}-$axis, 
together with the arc indicated as a dashed semicircle in 
Figs.\ \ref{fig:NRGOneDotMagGate}(d) and  \ref{fig:NRGOneDotMagGate}(e).  
This evolution of the crescent region 
causes the CAR-dominated flat structure of 
nonlocal conductance $g_{\mathrm{RL}}^{}$  
that emerged 
in the magnetic-field dependence   
in Figs.\ \ref{fig:NRGOneDotMag}(d) and \ref{fig:NRGOneDotMag}(f).    

Figures \ref{fig:NRGOneDotMagGate}(f) and \ref{fig:NRGOneDotMagGate}(g) show 
the contour maps of 
the CAR efficiency $\eta_{\mathrm{CAR}}^{}$ 
for magnetic fields of (f) $b =T_K$ and (g) $b = 0.1 U$.
Figure \ref{fig:NRGOneDotMagGate}(f) 
captures the typical profile of  $\eta_{\mathrm{CAR}}^{}$ 
at small fields of order $b \simeq T_K$:  
the CAR efficiency is enhanced  
in the SC-dominated regime $E_A \gtrsim U/2 + b$ and $\pi/4 < \Theta < 3\pi/4$, 
whereas  it decreases rapidly outside this region,
especially  just inside the semicircle,  $E_A \lesssim U/2 + b$, 
in the edge of the Zeeman-dominated spin-polarized regime. 
It reflects the steep slope along the direction of $\Theta =\pi/2$,
seen in  Fig.\ \ref{fig:NRGOneDotMagGate}(b),  
at the crossover region $E_A^{} \simeq U/2+b$. 
In contrast,  at large fields of order $b \simeq 0.1 U$, 
the corresponding slope of $\eta_{\mathrm{CAR}}^{}$ 
inside the semicircle shows a slow modest decline 
as seen in Fig.\ \ref{fig:NRGOneDotMagGate}(g). 
This modest decline of 
$\eta_{\mathrm{CAR}}^{}=
\mathcal{T}_{\mathrm{CP}}^{}/
\mathcal{T}_{\mathrm{BG}}^{}$ 
 is caused by the behavior of 
the transmission probability of the {Bogoliubov} particles 
in the denominator 
that is suppressed in {the} Zeeman-dominated regime 
for large magnetic fields, 
as seen in Fig.\ \ref{fig:NRGOneDotMagGate}(c).    

Figures \ref{fig:NRGOneDotMagGate}(h) and \ref{fig:NRGOneDotMagGate}(i)  
describe three-dimensional plots 
of the nonlocal conductance $g_{\mathrm{RL}}^{}/g_0^{}$
for magnetic fields of (h) $b= T_K$ and (i) $b = 0.1U$, respectively. 
These two examples show quite different behaviors 
inside the semicircle of radius  $E_A^{} \lesssim U/2 + b$. 
While the Kondo plateau of $g_{\mathrm{RL}}^{}$ starts to cave in around $E_A^{}=0$ 
for magnetic fields of the order of $T_K$, 
it is significantly suppressed at large magnetic fields of order $0.1U$, 
almost in the whole region 
inside the semicircle $E_A^{} \lesssim U/2+b$,   
except for the rim of the semicircle. 
However, in both cases,  
there spreads commonly 
a CAR-dominated region 
with negative nonlocal conductance,   
outside the semicircle 
 $E_A^{} \gtrsim U/2 + b$ in the direction of the $\Gamma_S^{}$-axis, 
which also emerges at zero magnetic field in Fig.\ \ref{fig:NRGOneDotXi}(c).

Figures \ref{fig:NRGOneDotMagGate}(j) and \ref{fig:NRGOneDotMagGate}(k) are 
the contour maps of the region,
at which the nonlinear conductance becomes negative  $g_{\mathrm{RL}}^{}<0.0$, 
for magnetic fields of (j) $b= T_K$ and (k) $b = 0.1U$. 
It spreads in the $\xi_d^{}$ vs $\Gamma_S^{}$ plane,  
over the region of  $E_A \gtrsim U/2 + b$ and $\pi/4 < \Theta < 3\pi/4$. 
These plots clearly show that the CAR contribution is enhanced, 
particularly at the crescent region just outside the circumference of the dashed semicircle.
The CAR dominates the nonlocal conductance in this region, 
and the dip structure of $g_{\mathrm{RL}}^{}$  
still remains for finite magnetic fields of order $0.1U$  
although the depth decreases as $b$ increases.
Furthermore, these results demonstrate how 
the flat structure can emerge in the magnetic-field-dependence 
of $g_{\mathrm{RL}}^{}$,  
seen in Figs.\ \ref{fig:NRGOneDotMag}(d) and \ref{fig:NRGOneDotMag}(f).         
For example, at the point $(E_A^{}=0.6U, \Theta=\pi/2)$ 
in the $\xi_d^{}$ vs $\Gamma_S^{}$ plane situates  
in the dip region of  $g_{\mathrm{RL}}^{}$ 
when the magnetic field $b$ varies from $0$ to the order $0.1U$.  

These results suggest that, in order to experimentally probe 
the CAR contributions in the nonlocal conductance $g_{\mathrm{RL}}^{}$,  
this crescent region will be a plausible target to be examined. 
The  CAR-dominated transport occurs 
in the parameter region $\Gamma_S^{}\gtrsim U/2 +b$,
where  the Cooper pairs can penetrate into quantum dots, 
 overcoming the repulsive interaction and magnetic field.  
Although we have chosen a rather strong interaction 
 $U/(\pi \Gamma_N^{})=5.0$ in this section,  
the sweet spot for the measurements, 
at which $g_{\mathrm{RL}}^{}$ exhibits a dip structure, 
emerges for any $U$,
 as demonstrated in Fig.\ \ref{fig:NRGOneDotXi0}(c) for $b=0$.

\subsection{Spin-polarized current between normal leads}
 \label{subsec:spin_current}

So far, we have mainly considered the charge transport. 
In particular, we have seen 
in Fig.\ \ref{fig:NRGOneDotMagGate}(i)  that 
 for a magnetic field of $b = 0.1U$,  
 the nonlocal conductance has a peak 
in the angular directions  $\Theta\simeq 0$ and  $\Theta\simeq \pi$,  
along the rim of the semicircle of radius $E_A^{} \simeq U/2 + b$.   
Here we discuss 
the resonant spin-polarized current 
which is significantly enhanced in this region 
where the crossover takes place   
between the Zeeman-dominated  regime and 
the SC-proximity-dominated regime.

The spin current 
$I_{R,\mathrm{spin}}^{} \equiv I_{R,\uparrow}^{} - I_{R,\downarrow}^{}$ 
flowing from 
the quantum dot to the right lead 
can be expressed in the following form,  
as shown in Appendix \ref{sec:spin_current_appendix}:
\begin{align}
I_{R,\mathrm{spin}}^{} \, 
     = & \  \frac{e^2}{h} \,
\frac{4\Gamma_L^{}\Gamma_R^{}}{\Gamma_N^2} 
\, \mathcal{T}_{\mathrm{spin}}^{}\,
\left(V_L - V_R \right) \, ,
\label{eq:Tspin_text}
\end{align}
where   
$\mathcal{T}_{\mathrm{spin}}^{} =
 \left( \sin^2\delta_\uparrow - \sin^2\delta_\downarrow\right) \cos\Theta$.  
The magnetic-field dependence of the spin current 
is determined by 
the difference $\sin^2\delta_\uparrow^{} - \sin^2\delta_\downarrow^{}$ 
between the transmission probability of the  $\uparrow$-spin  and 
that of the $\downarrow$-spin Bogoliubov particles.
Similarly,  the normalized current polarization is defined by
\cite{PhysRevLett.90.116602,PhysRevB.78.155303,PhysRevB.74.161301,
PhysRevB.91.155302}   
\begin{align}
    P_R \, &\equiv  \frac{I_{R,\uparrow}^{} - I_{R,\downarrow}^{}}
{I_{R,\uparrow}^{} + I_{R,\downarrow}^{}} 
   \xrightarrow{\,\Gamma_L^{} = \Gamma_R^{}\,}  
 \frac{\sin^2\delta_\uparrow^{} - \sin^2\delta_\downarrow^{}}
{\sin^2\delta_\uparrow^{} + \sin^2\delta_\downarrow^{}}\cos\Theta .
    \label{eq:PR_text}
\end{align}

\begin{figure}[b]
\includegraphics[width=0.7\linewidth]{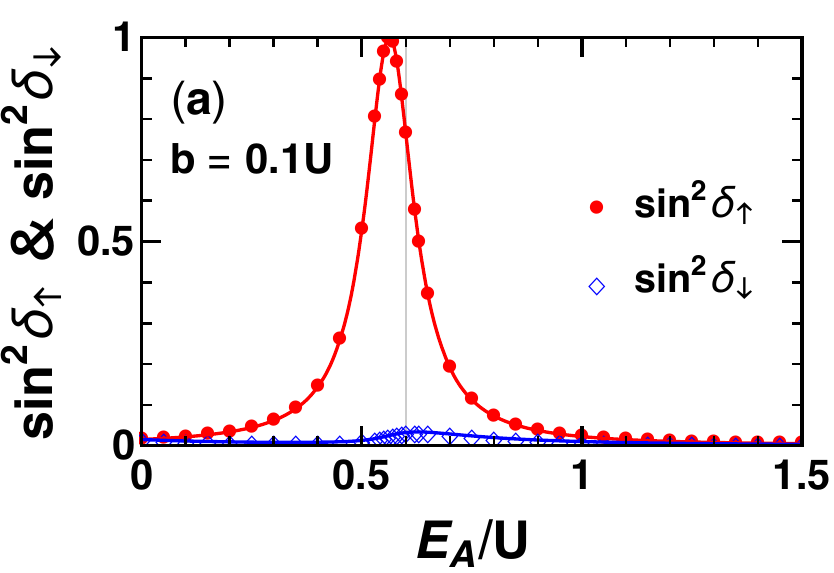}\\
\includegraphics[width=0.95\linewidth]{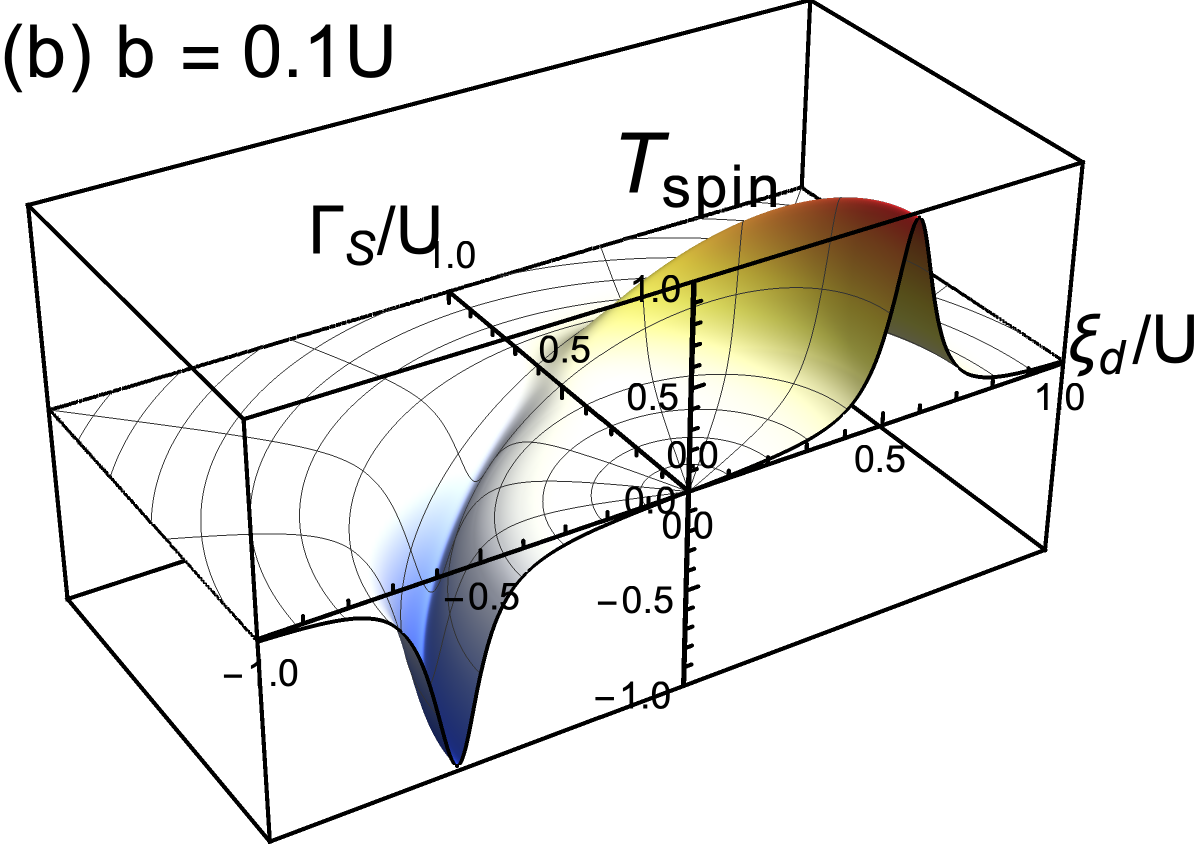}\\
\includegraphics[width=0.95\linewidth]{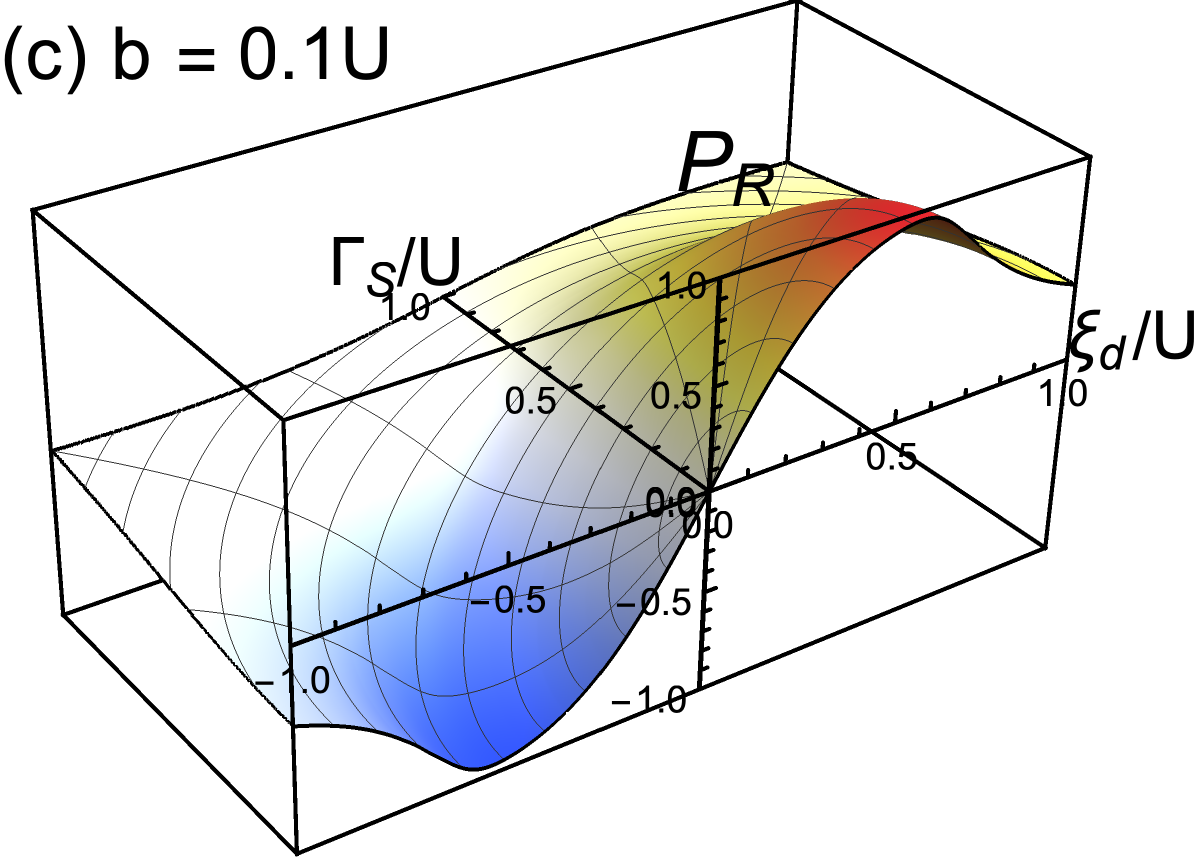}
\caption{Spin-dependent transport coefficients 
at finite magnetic field $b=0.1U$, for $U/(\pi\Gamma_N) = 5.0$.
Top panel (a) shows  
$\sin^2 \delta_\uparrow$ and $\sin^2 \delta_\downarrow$,  plotted vs $E_A^{}$.
Three-dimensional figures represent     
(b) spin transmission coefficient  $\mathcal{T}_{\mathrm{spin}}$ 
and (c) current polarization $P_R$, 
plotted as functions of $\xi_d^{}$ and $\Gamma_S^{}$,  
for $\Gamma_L = \Gamma_R$.     
}
\label{fig:NRGOneDot_SpinCurrent}
\end{figure}

Figure \ref{fig:NRGOneDot_SpinCurrent} shows the NRG result of  
the spin-resolved transport coefficients 
calculated 
at a magnetic field of $b=0.1U$,   
for a strong interaction  $U/(\pi \Gamma_N) =5.0$.
In this case,  the renormalized Andreev level for the 
majority spin $\widetilde{E}_{A,\uparrow}^{}$ crosses the Fermi level    
at  $E_{A}^{} \simeq U/2 +b$ since $\widetilde{E}_{A,\uparrow}^{}$ 
can be approximated by the Hartree-Fock energy shift,
defined in Eq.\ \eqref{eq:HF_energy_shift_mag}, in the  crossover region.

Figure  
\ref{fig:NRGOneDot_SpinCurrent}(a) 
shows that the resonant tunneling of the unitary limit 
 $\sin^2\delta_\uparrow^{}=1$
 occurs for the majority $\uparrow$-spin Bogoliubov particles,
whereas the minority one  $\sin^2\delta_\downarrow^{}$ 
 is very small and does not give any significant contribution to the current.
Note that the occupation number of electrons 
 $\langle n_{d,\sigma}^{}\rangle$ depends on 
the coherence factor  $\cos \Theta$, 
and is given by a linear combination 
of the phase shifts as shown in Eq.\ \eqref{eq:nd_in_BG_phase_shift}. 
Therefore,  the occupation number of $\downarrow$-spin electrons fluctuates 
significantly at the crossover region in the direction of $\Theta =\pi$ 
in such a way that $\langle n_{d,\downarrow}^{}\rangle\xrightarrow{\,\Theta =\pi\,}
1-\delta_{\uparrow}/\pi$,  
 whereas the $\uparrow$-spin electrons fluctuate in the direction of  $\Theta =0$ 
as $\langle n_{d,\uparrow}^{}\rangle \xrightarrow{\,\Theta =0\,} 
\delta_{\uparrow}/\pi$.

Figures  \ref{fig:NRGOneDot_SpinCurrent}(b) and \ref{fig:NRGOneDot_SpinCurrent}(c) 
clearly show that  $\mathcal{T}_{\mathrm{spin}}^{}$ and $P_R$ 
are enhanced at the level-crossing point  $E_A \simeq b +U/2$  
 near the $\xi_d^{}$ axis. 
The $\Theta$ dependencies 
of  $\mathcal{T}_{\mathrm{spin}}^{}$ and $P_R$  
are determined by the coherence factor  $\cos \Theta$, 
as shown in Eqs.\ \eqref{eq:Tspin_text} and  \eqref{eq:PR_text}. 
Therefore, these coefficients become most significant 
in the directions of $\Theta = 0$ and $\pi$,  
where the resonant tunneling occurs for  
the  $\uparrow$-spin and $\downarrow$-spin electron components, 
respectively.   
As the Bogoliubov angle $\Theta$ deviates away from the $\xi_d^{}$ axis, 
the spin polarization is suppressed, especially in  
the SC-proximity-dominated regime at $\pi/4 < \Theta < 3\pi/4$, 
and 
the spin current $I_{R,\mathrm{spin}}^{}$ 
vanishes at $\Theta = \pi/2$.

\section{Summary}
\label{summary}

We have studied the interplay
 between 
the crossed Andreev reflection,  Kondo effect, and Zeeman splitting, 
occurring in a multi-terminal quantum dot, 
consisting of two normal and one SC leads.

It has been shown 
that the linear-response currents 
flowing through quantum dot  at zero temperature $T=0$  
are determined by two angular variables, i.e., 
the phase shift $\delta_{\sigma}^{}$ of 
Bogoliubov particles and 
the Bogoliubov rotation angle $\Theta =\cot^{-1}(\xi_d^{}/\Gamma_S^{})$ 
 in the  the limit of large SC gap $|\Delta_S^{}| \to \infty$. 
In this limit, 
the phase shift can be deduced from an effective 
Anderson model for interacting Bogoliubov particles,   
which has a global U(1) symmetry along the principal axis 
in the Nambu pseudo-spin space. 
The Bogoliubov angle $\Theta$ 
enters the transport coefficients  
 through the SC coherence factors,  
and plays an essential role in the conductance,  
together with the position of the  Andreev level 
 $E_A^{}=\sqrt{\xi_d^2+\Gamma_S^2}$.

In the first half of the paper,  
we have described the role of the many-body optical theorem on the CAR,  
and have shown that the multi-terminal conductance 
at finite magnetic fields 
is determined by the transmission probability 
$\mathcal{T}_{\mathrm{BG}}^{} = 
\frac{1}{2}\sum_{\sigma} \sin^2 \delta_{\sigma}^{}$ of the Bogoliubov particles, 
which  does not depend on $\Theta$,  
and by  the Cooper-pair tunneling part 
$\mathcal{T}_{\mathrm{CP}}^{} =
\frac{1}{4} \sin^2 
(\delta_{\uparrow}+ \delta_{\downarrow}) \sin^2 \Theta$. 
In the second half,  
we have discussed the behaviors of  nonlocal conductance,
obtained by using the NRG approach  
in a wide range of the parameter space 
which consists of 
  $\xi_d^{}$,  $\Gamma_S^{}$,   $\Gamma_N^{}$, 
the Coulomb interaction  $U$, 
and the magnetic field $b$.

At zero magnetic field, the nonlocal conductance $g_\mathrm{RL}^{}$ 
becomes negative at $E_A^{} \gtrsim U/2$ and  $\pi/4 < \Theta < 3\pi/4$,   
where the CAR dominates.  
In particular,  the contribution of Cooper-pair tunnelings  
 $\mathcal{T}_{\mathrm{CP}}^{}$ 
is maximized at a crescent-shaped crossover region 
between the Kondo-dominated and the SC-dominated regimes, 
emerging  at $E_A^{} \simeq U/2$ 
in the angular direction of  $\Theta \simeq \pi/2$.     
The width of the crescent region along the  $\Gamma_S^{}$-axis 
is of the order of $\Gamma_N^{}$.  
The enhanced CAR occurring in this region is caused by  
 the valence fluctuation of the Bogoliubov particles,  
in the middle of which the occupation number takes the value  $Q=1/2$ 
and the phase shift due to the Cooper-pair tunneling reaches 
the unitary limit  $\delta_\uparrow + \delta_\downarrow = \pi/2$.

Magnetic fields lift the spin degeneracy of the 
Andreev resonance level.
In the strongly-correlated case where $Q \simeq 1.0$ with  
$E_A^{} \lesssim U/2$ and $U \gg \Gamma_N^{}$,  
the crossover between 
the Kondo regime and Zeeman-dominated regime occurs  
at a magnetic field  $b\sim T^*$ of the order 
of the Kondo energy scale $T^*$. 
In contrast,  at  $E_A^{} \gtrsim U/2$ in 
the valence-fluctuation region of the Bogoliubov particles, 
magnetic fields induce a crossover  
between the SC-proximity dominated regime and 
the Zeeman-dominated regime 
at $b \simeq E_A -U/2$,  
where the renormalized Andreev level $\widetilde{E}_{A,\uparrow}^{}$
for the majority spin  component ($\sigma=\uparrow$) crosses the Fermi level.
It induces the resonant tunneling  
of the Bogoliubov particles 
and the Cooper-pair tunneling,  
 the transmission probabilities 
of which are determined by the phase shifts  
  $\delta_{\uparrow}^{} \simeq \pi/2$ and
 $\delta_{\uparrow}^{} +\delta_{\downarrow}^{} \simeq \pi/2$,  
respectively.
 Note that  $\delta_{\downarrow} \simeq 0.0$ 
as the renormalized Andreev level for the minority-spin component 
becomes almost empty   $Q_{\downarrow}^{} \simeq 0.0$   in this region.   
It has also been demonstrated that the resonant spin current is enhanced 
in the angular direction of $\Theta \simeq 0$ or $\Theta \simeq \pi$   
when the Andreev level 
of the majority spin crosses the Fermi level.

The nonlocal conductance $g_\mathrm{RL}^{}$ becomes negative 
in the parameter region of $E_A^{} \gtrsim U/2 +b$ and $\pi/4 < \Theta < 3\pi/4$. 
In particular, the CAR contribution is maximized in the crescent-shaped region,
which moves in the $\xi_d^{}$ vs $\Gamma_S^{}$ plane,  
together with the semi-circular boundary of radius 
$E_A^{} \simeq U/2 +b$  as $b$ increases.   
The crescent region evolves with the magnetic field  
and yields  a flat valley structure which emerges 
in the $b$ dependence of $g_\mathrm{RL}^{}$,  
at $0\leq b \lesssim E_A^{} - U/2$.
These results suggest that, in order to experimentally probe 
the CAR contributions measuring the nonlocal conductance,  
the crescent parameter region will be a plausible target to be examined.

\begin{acknowledgments}

This work was supported by JSPS KAKENHI 
Grants No.\ JP18K03495 and No.\ JP23K03284,
and 
JST Moonshot R \& D-MILLENNIA  Program Grant No.\ JPMJMS2061.
Y.\ Teratani was supported by the Sasakawa Scientific Research Grant 
from the Japan Science Society Grant No.\ 2021-2009.

\end{acknowledgments}

\appendix

\section{Effective Hamiltonian for $|\Delta_S^{}| \to \infty$}

\label{AppendixBogo}

The Hamiltonian $H$ defined in Eq.\ \eqref{eq:total_H_single} 
can be separated into two independent parts since 
only the symmetrized linear combination  
$\alpha_{\varepsilon,\sigma}^{}$ of conduction electrons 
has a finite tunnel coupling to the QD, 
whereas the anti-symmetrized linear combination
$\beta_{\varepsilon,\sigma}^{}$ is decoupled 
from the rest of the system:  
\begin{align}
 \alpha_{\varepsilon, \sigma}^{}
\,\equiv& \ \frac{v_L^{}\, c_{\varepsilon,L,\sigma}^{}
+v_R^{} \,c_{\varepsilon,R,\sigma}^{}}{\sqrt{v_L^2+v_R^2}} 
\,,
\label{eq:even_lead}
\\
 \beta_{\varepsilon,\sigma}^{} 
\,\equiv& \ \frac{-v_R^{} \,c_{\varepsilon,L,\sigma}^{}
+v_L^{} \,c_{\varepsilon,R,\sigma}^{}}{\sqrt{v_L^2+v_R^2}} \,.
\label{eq:odd_lead}
\end{align}
Correspondingly, the conduction-electron part  and  the normal-tunneling part  
of the  Hamiltonian can be rewritten in the form
\begin{align}
H_{\text{N}}^{}\,=& \ 
\sum_{\sigma} 
\int_{-D}^{D}  \! d\varepsilon \,\varepsilon\,
\Bigl( \alpha^\dagger_{\varepsilon,\sigma} 
\alpha^{ }_{\varepsilon,\sigma}
+ \beta^\dagger_{\varepsilon,\sigma} 
\beta^{ }_{\varepsilon,\sigma} 
\Bigr) , 
\label{eq:H_N_eo}
\\
H_{\text{TN}}^{} 
= & \ 
v_N^{}
\sum_{\sigma} 
\int_{-D}^{D}  \! d\varepsilon \,
\sqrt{\rho_c^{}}\,
\Bigl(
\alpha^\dagger_{\varepsilon,\sigma} 
d_{\sigma}
+ \mathrm{H.c.} 
\Bigr) \,,
\label{eq:H_TN_eo}
\end{align}
where  $v_N^{} \equiv \sqrt{v_L^2+v_R^2}$.

Furthermore, in the large gap limit  
 $\left| \Delta_{S}^{}\right| \to \infty$ 
which is taken at $\left| \Delta_{S}^{}\right|  \ll D_S^{}$ 
 keeping $\rho_S$ constant,  
the superconducting proximity effects 
can be described by the pair potential
$\Delta_d^{}  \equiv   \Gamma_{S}^{}\,e^{i\phi_S^{}}$ 
penetrating into the QD.
\cite{YoshihideTanaka_2007, YoichTanaka_2007} 
 Therefore, at low energies, the subspace to which the QD belongs 
can be described by the following effective Hamiltonian:
\begin{align}
H_\mathrm{eff}^{}
\, \equiv& 
\ \ 
\bm{\psi}_{d}^{\dagger}\  
\bm{\mathcal{H}}_\mathrm{dot}^\mathrm{SC} \  
\bm{\psi}_{d}^{} 
\,+ \, \frac{U}{2}\left(n_d -1 \right)^2 
\nonumber
\\
& + 
v_N^{}
\int_{-D}^{D} \!\! d\varepsilon\,\sqrt{\rho_c^{}}
\left[\,
\bm{\psi}_{\alpha}^{\dagger}(\varepsilon)\,\bm{\psi}_{d}^{} 
+\bm{\psi}_{d}^{\dagger}\,\bm{\psi}_{\alpha}^{}(\varepsilon) 
\,\right]
\nonumber
\\
&+ 
\int_{-D}^{D} \!\! d\varepsilon\,\varepsilon\ 
\bm{\psi}_{\alpha}^{\dagger}(\varepsilon)\,
\bm{\psi}_{\alpha}^{}(\varepsilon) 
\ +\,b
\,.
\label{eq:Heff_single}
\end{align}
Here,  $\bm{\mathcal{H}}_\mathrm{dot}^\mathrm{SC}$  is 
the following matrix defined in the Nambu pseudo-spin space,
\begin{align}
\bm{\mathcal{H}}_\mathrm{dot}^\mathrm{SC} \, 
 \equiv &  \ 
 \begin{pmatrix}
 \xi_d^{} &  \Delta_{d}^{} \cr
 \Delta_{d}^* & - \xi_d^{}
\rule{0cm}{0.35cm}
 \end{pmatrix}
\, -\, b \,\bm{1} \,,
\end{align}
with $\bm{1}$  the $2\times 2$ unit matrix,  and  
\begin{align}
 \bm{\psi}_{d}^{} \equiv & \  
 \begin{pmatrix}
  d_{\uparrow}^{} \cr
  d_{\downarrow}^{\dagger} \rule{0cm}{0.5cm}\cr
 \end{pmatrix}
 , 
\qquad  
 \bm{\psi}_{\alpha}^{}(\varepsilon) \equiv 
 \begin{pmatrix}
  \alpha_{\varepsilon, \uparrow}^{} \cr
  -\alpha_{-\varepsilon,\downarrow}^{\dagger} 
\rule{0cm}{0.5cm}\cr
 \end{pmatrix}
 .
\rule{0cm}{0.8cm}
\end{align}

The effective Hamiltonian $H_\mathrm{eff}^{}$ 
has a global  U$(1)$ symmetry with respect to the principal axis 
along the three-dimensional vector  $\widehat{\bm{n}} \propto 
(\mathrm{Re}\,\Delta_{d}^{},\, -\mathrm{Im}\,\Delta_{d}^{},\,\xi_d)$ 
in the Nambu space. 
The conserved charge associated with this U$(1)$ symmetry corresponds to 
the total number of Bogoliubov particles, 
the operators for which are given by
\begin{align}
 \begin{pmatrix}
  \gamma_{d,\uparrow}^{\phantom{\dagger}} \cr
  \gamma_{d,\downarrow}^{\dagger} \rule{0cm}{0.6cm}\cr
 \end{pmatrix}
= & \ 
 \bm{\mathcal{U}}^\dagger
\bm{\psi}_{d}^{} , 
\quad 
 \begin{pmatrix}
  \gamma_{\varepsilon, \uparrow}^{\phantom{\dagger}} \cr
  -\gamma_{-\varepsilon,\downarrow}^{\dagger} 
\rule{0cm}{0.6cm}\cr
 \end{pmatrix}
 =  \, 
 \bm{\mathcal{U}}^\dagger
\bm{\psi}_{\alpha}^{}(\varepsilon) \,.
\label{eq:Bogoliubov_tans_single}
\end{align}
Here, $\bm{\mathcal{U}}$ is the unitary matrix which 
diagonalizes $\bm{\mathcal{H}}_\mathrm{dot}^\mathrm{SC}$: 
\begin{align}
&\!\! 
\bm{\mathcal{U}}^\dagger\,
\bm{\mathcal{H}}_\mathrm{dot}^\mathrm{SC} \  
\bm{\mathcal{U}}^{}  
\, =\, E^{}_A \bm{\tau}_3
-b \bm{1} ,
 \qquad
\bm{\tau}_3  =
 \begin{pmatrix}
  1 &  \  0\cr
  0 & -1
 \rule{0cm}{0.2cm}
  \end{pmatrix} 
,
 \label{eq:E_A_def}
\end{align}
with $E^{}_A \equiv \sqrt{\xi_{d}^2+\Gamma_S^{2}}$. 
For example, in the case where the Josephson phase  $\phi_S^{} =0$, 
the matrix  $\bm{\mathcal{U}}$ is 
determined by a single Bogoliubov angle  $\Theta$,  
 as shown in Eq.\ \eqref{eq:Bofoliubov_trans_I}.

\section{Derivation of linear nonlocal current}
\label{sec:conductance_derivation}

In this appendix, we provide a brief derivation of the nonlocal conductance 
defined in Eqs.\ \eqref{eq:current_R_ET_CP}--\eqref{eq:T_CP_single}

The current flowing from the quantum dot to the normal lead on the right is 
described by the operator, 
\begin{equation}
\widehat{I}_{R,\sigma} \,= \,  - i\,
e v_{R}^{}  \int_{-D}^{D} \!\! d\varepsilon\,
\sqrt{\rho_c^{}}\, 
\Bigl(
c^{\dagger}_{\varepsilon,R,\sigma} d^{}_{\sigma} 
-d^{\dagger}_{\sigma} c_{\varepsilon,R,\sigma}^{}\Bigr)  
\end{equation}
for spin $\sigma$ component. 
The steady-state average of the total current 
 $I_{R}\equiv\langle \widehat{I}_{R,\uparrow} \rangle 
+ \langle \widehat{I}_{R,\downarrow} \rangle$ with 
$I_{R,\sigma}\equiv\langle \widehat{I}_{R,\sigma} \rangle$ 
can be expressed in terms of the Green function in the Keldysh formalism, 
\cite{YoichTanaka_2007} 
\begin{align}
& \!\! 
I_R\, = \,
\frac{e}{h} \int_{-\infty}^{\infty} \! d\omega
\ 
(-i)
\Gamma_R^{} 
\,\mathrm{Tr} \Biggl[\, 
\bm{G}_{dd}^{r}
\bm{\Sigma}_\mathrm{tot}^\mathrm{K} 
\, \bm{G}_{dd}^{a}\,\bm{\tau}_3^{} 
\nonumber \\
& \qquad  \quad  
 -( \bm{1}-2\bm{f}_R)\,
\bm{G}_{dd}^{r}
\Bigr\{
\bm{\Sigma}_\mathrm{tot}^{-+} -\bm{\Sigma}_\mathrm{tot}^{+-} 
\Bigr\} 
\, \bm{G}_{dd}^{a} 
\,\bm{\tau}_3^{}\,\Biggr].
\label{eq:C_vertex}
\end{align}
Here, $\mathrm{Tr}$ denotes the trace of the $2\times 2$ 
matrices in the Nambu pseudo-spin space.
 $\bm{\Sigma}^{-+}_\mathrm{tot}$ and
$\bm{\Sigma}^{+-}_\mathrm{tot}$ are 
the lesser and greater self-energies, respectively,  
and $\bm{\Sigma}^\mathrm{K}_\mathrm{tot} \equiv 
-\bm{\Sigma}^{-+}_\mathrm{tot} 
-\bm{\Sigma}^{+-}_\mathrm{tot}$. 
The matrix  $\bm{f}_\nu^{}$  is defined as
\begin{align}
\bm{f}_{\nu}^{}(\omega) = 
 \left[\,
 \begin{matrix}
  f_{\nu}^{}(\omega) & \ 0 \cr
  0 \rule{0cm}{0.5cm}&\  \overline{f}_{\nu}^{}(\omega) \cr
 \end{matrix}
\,\right] 
, \qquad  \quad \nu=L,\,R. 
\end{align}
The bias voltage $eV_\nu^{}$ 
 is applied to the leads 
such that 
 $f_\nu(\omega) \equiv  f(\omega-eV_\nu^{})$ 
and  $\overline{f}_\nu(\omega) \equiv  f(\omega+eV_\nu^{})$,  
with  $f(\omega)=1/[e^{\omega/T}+1]$ 
the Fermi distribution function.

Each self-energy matrix can be separated into two parts, e.g.,  
\begin{equation}
\bm{\Sigma}^\mathrm{K}_\mathrm{tot}(\omega) \,=\,
\bm{\Sigma}^\mathrm{K}_{0}(\omega) 
+\bm{\Sigma}^\mathrm{K}_{U}(\omega)  \;.
\label{eq:self_enengy_2parts}
\end{equation}
Here, the first term on the right-hand side represents 
the tunnel contributions at $U=0$, 
\begin{align}
& \!\!\!\!
\bm{\Sigma}^\mathrm{K}_{0}(\omega)
    =   -2i \sum_{\nu=L,R} 
\Gamma_{\nu}^{}\Bigl[\, \bm{1} - 2\bm{f}_j^{}(\omega) \,\Bigr], 
\label{eq:SigmaK_U0}
\\
& \!\!\!\! 
\bm{\Sigma}_{0}^{-+}(\omega) 
-\bm{\Sigma}_{0}^{+-} (\omega)  \,=\, -2i  
\bigl( \Gamma_L^{}+\Gamma_R^{} \bigr) \, \bm{1}\,, 
\label{eq:SigmaR_U0}
\end{align}
with $\bm{1}$  the $2\times 2$ unit matrix in the pseudo-spin space. 
The second term on the right-hand side of Eq.\ \eqref{eq:self_enengy_2parts}, 
$\bm{\Sigma}^\mathrm{K}_{U}(\omega)$, 
represents the self-energy corrections 
due to  the Coulomb interaction $U$.  
This and  the corresponding terms of the lesser and greater self-energies,
 $\bm{\Sigma}^{-+}_{U}(\omega)$  and 
 $\bm{\Sigma}^{+-}_{U}(\omega)$ 
are also pure imaginary in the frequency domain,
 and represent the damping of quasiparticles due to the multiple collisions.
These imaginary parts of the interacting self-energies 
 vanish at $T=0$, $eV_{\nu}^{}=0$, and $\omega=0$, 
and thus they  do not contribute to the linear-response current 
at zero temperature. 
Furthermore,  
 the integrand on the right-hand side of Eq.\  \eqref{eq:C_vertex}  
vanishes  at $eV_{\nu}^{}=0$, since 
 $\left.
\bm{\Sigma}^\mathrm{K}_\mathrm{tot}(\omega)
\right|_{eV_\nu^{} = 0} =  
 (1-2f)\Bigr[
 \bm{\Sigma}_\mathrm{tot}^{-+} -\bm{\Sigma}_\mathrm{tot}^{+-} 
 \Bigr]_{eV_{\nu}^{}=0}$.

Consequently, at $T=0$, 
the linear-response current can be calculated using Eq.\  \eqref{eq:C_vertex}, 
by replacing $\bm{\Sigma}_\mathrm{tot}^{+-}$, 
$\bm{\Sigma}_\mathrm{tot}^{-+}$, 
and $\bm{\Sigma}^\mathrm{K}_\mathrm{tot}$ 
in the integrand with their noninteracting counterparts: 
\begin{align}
 I_{R}
= \,    
\frac{4e^2}{h} 
&\Biggl[\,
\left(
\Bigl | \bigl\{\bm{G}_{dd}^{r}(0)\bigl\}_{11}^{}\Bigr|^2 
+\Bigl | \bigl\{\bm{G}_{dd}^{r}(0)\bigl\}_{22}^{}\Bigr|^2 
\right)\nonumber\\
&\qquad \times \Gamma_R^{}\Gamma_L^{}
\,(V_L-V_R)\nonumber \\
& 
- \left(
\Bigl | \bigl\{\bm{G}_{dd}^{r}(0)\bigl\}_{12}^{}\Bigr|^2 
+ \Bigl | \bigl\{\bm{G}_{dd}^{r}(0)\bigl\}_{21}^{}\Bigr|^2
\right)
\nonumber\\
&\qquad \times 
\Bigl\{\Gamma_R^{}\Gamma_L^{}\,(V_L+V_R) +2\Gamma_R^{2}\,V_R\Bigr\}\,
\,\Biggr] .
\label{eq:current_R_magX}
\end{align}
Note that the anomalous Green's functions are 
related to  
each other through 
$\bigl\{\bm{G}_{dd}^{r}(\omega)\bigl\}_{21}^{}
=\bigl\{\bm{G}_{dd}^{a}(\omega)\bigl\}_{12}^{*}$.
Equation  \eqref{eq:current_R_magX} 
can be rewritten further  in terms of the phase shifts $\delta_{\sigma}$ 
and the Bogoliubov angle $\Theta$, by using Eq.\ \eqref{eq:GreenBogoliubovTrans}  
 to obtain Eqs.\ \eqref{eq:T_ET_single} and  \eqref{eq:T_CP_single}.

\section{Optical theorem for Andreev scattering}

 \label{sec:optical_theorem}

We provide a derivation of the optical theorem, 
which  emerges in the form  
\begin{align} 
\sum_{\sigma}\sin^2 \delta_{\sigma}^{} \,=\,
2 \bigl(\mathcal{T}_{\mathrm{ET}}^{} + \mathcal{T}_{\mathrm{CP}}^{} \bigr) \,.
\label{eq:opritcal_theorem_1}
\end{align} 

We start with the matrix  identity for
the impurity  Green's function in the Nambu form
\begin{align}
& \bm{G}_{dd}^{r}(\omega) -
\bm{G}_{dd}^{a}(\omega) 
\nonumber \\
& \quad 
=
\bm{G}_{dd}^{r}(\omega) 
\Bigl[
\left\{\bm{G}_{dd}^{a}(\omega)\right\}^{-1}
 - 
\left\{\bm{G}_{dd}^{r}(\omega)\right\}^{-1}
\Bigr]
\bm{G}_{dd}^{a}(\omega)
\nonumber \\
& \quad =   
\bm{G}_{dd}^{r}(\omega) 
\Bigl[
\bm{\Sigma}_\mathrm{tot}^{r}(\omega) - 
\bm{\Sigma}_\mathrm{tot}^{a}(\omega) 
\Bigr]
\bm{G}_{dd}^{a}(\omega) \,.
\end{align}
At $\omega=T=eV=0$,
it can be rewritten further in the form 
\begin{align}
- \frac{\Gamma_N^{}}{2i}
\Bigl[\,
\bm{G}_{dd}^{r}(0)   - 
\bm{G}_{dd}^{a}(0)\,\Bigr]
\,= & \  
\Gamma_N^{2}
\,\bm{G}_{dd}^{r}(0)\,\bm{G}_{dd}^{a}(0) \,.
\label{eq:opritcal_theorem_2}
\end{align}
Here, we have used the property that  
 the imaginary part of the interacting self-energy 
vanishes $\mathrm{Im}\, \bm{\Sigma}_{U}^{r}(0) =0$
and the noninteracting one is given by 
$\bm{\Sigma}_{0}^{r}(\omega) 
-\bm{\Sigma}_{0}^{a}(\omega) 
= -2i \Gamma_N^{} \bm{1}$.
Taking trace of the Nambu matrices,
the left-hand side of Eq.\ \eqref{eq:opritcal_theorem_2} 
can be calculated as 
\begin{align}
&  \!\!\!\! 
- \frac{\Gamma_N^{}}{2i}
 \mathrm{Tr}\, \Bigl[\,
 \bm{G}_{dd}^{r}(0)   - 
\bm{G}_{dd}^{a}(0)
 \,\Bigr] \nonumber \\
& =  
 - \frac{\Gamma_N^{}}{2i}
\sum_{\sigma}\Bigl[
G_{\gamma,\sigma}^{r}(0)   - 
G_{\gamma,\sigma}^{a}(0)   
\Bigr] 
=  
\sum_{\sigma}\sin^2 \delta_{\sigma}^{}.
\end{align}
Similarly, 
the right-hand side of Eq.\ \eqref{eq:opritcal_theorem_2} 
takes the form
\begin{align}
& \Gamma_N^{2}
\mathrm{Tr}\, 
\Bigl[\,\bm{G}_{dd}^{r}(0)\,\bm{G}_{dd}^{a}(0)\,\Bigr] 
\nonumber \\ 
&  =  \   \Gamma_N^{2}
\biggl[ 
\Bigl|\left\{\bm{G}_{dd}^{r}(0)\right\}_{11}^{}\Bigr|^2
+\Bigl|\left\{\bm{G}_{dd}^{r}(0)\right\}_{12}^{}\Bigr|^2
\nonumber \\
& \qquad \quad   
+\Bigl|\left\{\bm{G}_{dd}^{r}(0)\right\}_{21}^{}\Bigr|^2
+\Bigl|\left\{\bm{G}_{dd}^{r}(0)\right\}_{22}^{}\Bigr|^2
\,\biggr] .
\end{align} 
The last line corresponds to  
$2 \bigl(\mathcal{T}_{\mathrm{ET}}^{} + \mathcal{T}_{\mathrm{CP}}^{} \bigr)$ 
defined in Eqs.\ \eqref{eq:T_ET_single} and  \eqref{eq:T_CP_single}, 
and from this Eq.\ \eqref{eq:opritcal_theorem_1} follows.

\section{Derivation of  the spin-current formula}
 \label{sec:spin_current_appendix}

We briefly describe here the linear-response formula 
for the  spin current following between two normal leads at finite magnetic fields.
The current formula  presented  in Appendix \ref{sec:conductance_derivation} 
can be decomposed into the contributions of  
the $\uparrow$-  and $\downarrow$- spin components, 
which can be rearranged as a spin current:  
\begin{align}
I_{R,\mathrm{spin}}^{} \equiv & \  I_{R,\uparrow}^{} 
- I_{R,\downarrow}^{}
\nonumber\\
= & \ 
\frac{e}{h} \int_{-\infty}^{\infty} \! d\omega
\ 
(-i)
\Gamma_R^{} 
\,\mathrm{Tr} \Biggl[\, 
\bm{G}_{dd}^{r}
\bm{\Sigma}_\mathrm{tot}^\mathrm{K} 
\, \bm{G}_{dd}^{a}
\nonumber \\
& \ 
 -( \bm{1}-2\bm{f}_R)\,
\bm{G}_{dd}^{r}
\Bigr\{
\bm{\Sigma}_\mathrm{tot}^{-+} -\bm{\Sigma}_\mathrm{tot}^{+-} 
\Bigr\} 
\, \bm{G}_{dd}^{a} \,\Biggr].
\end{align}
Specifically at $T = 0$, 
the linear-response spin current  can be expressed in the following form, 
\begin{align}
   I_{R,\mathrm{spin}}^{}\,=& \   \frac{4e^2}{h}
\,\Gamma_L\Gamma_R 
\left( \Bigl | \bigl\{\bm{G}_{dd}^{r}(0)\bigl\}_{11}^{}\Bigr|^2 
- \Bigl | \bigl\{\bm{G}_{dd}^{r}(0)\bigl\}_{22}^{}\Bigr|^2 \right)
\nonumber \\
& \times  \left(V_L -V_R \right) 
\nonumber \\
    =  & \  \frac{4e^2}{h} \,
\frac{\Gamma_L\Gamma_R}{\Gamma_N^2} 
\, \mathcal{T}_{\mathrm{spin}}\,
\left(V_L - V_R \right) \, ,
\\
\mathcal{T}_{\mathrm{spin}}^{} 
\,\equiv & \  
 \left( \sin^2\delta_\uparrow - \sin^2\delta_\downarrow\right) \cos\Theta \,. 
\label{eq:Tspin}
\end{align}
Note that  $I_{L,\uparrow} - I_{L,\downarrow} 
=  I_{R,\uparrow} - I_{R,\downarrow}$.

Similarly,  the current polarization $P_R^{}$, 
defined with respect to symmetric voltages  $V_L = -V_R$, 
can be used as a  measure of the spin current relative to the charge current
\cite{PhysRevLett.90.116602,PhysRevB.78.155303,
PhysRevB.74.161301,PhysRevB.91.155302}:
\begin{align}
    &P_R^{} 
\,\equiv \, 
\frac{I_{R,\uparrow}^{} - I_{R,\downarrow}^{}}
{I_{R,\uparrow}^{} + I_{R,\downarrow}^{}} 
    \ = \  
\frac{\Gamma_L^{} \Gamma_R^{} \mathcal{T}_{\mathrm{spin}}^{}}
{2\left[ \Gamma_L^{} \Gamma_R^{}\mathcal{T}_{\mathrm{ET}^{}}
 + \Gamma_R^2\mathcal{T}_{\mathrm{CP}}^{}\right]} 
\nonumber \\
 & \ \ 
\xrightarrow{\,\Gamma_L^{} = \Gamma_R^{}\,} \  
 \frac{\sin^2\delta_\uparrow^{} - \sin^2\delta_\downarrow^{}}
{\sin^2\delta_\uparrow^{} + \sin^2\delta_\downarrow^{}}\, \cos\Theta \, .
    \label{eq:PR}
\end{align}



\end{document}